\documentclass[aps,rmp,twocolumn,floatfix,superscriptaddress,longbibliography]{revtex4-1}

\usepackage[utf8]{inputenc}
\usepackage[T1]{fontenc}
\usepackage{amsmath, amsthm, amssymb}
\usepackage{graphicx}
\usepackage{dcolumn,times}
\usepackage{bm,bbm}
\usepackage{color}

\usepackage{changes}

\definecolor{myurlcolor}{rgb}{0,0,0.7}
\definecolor{myrefcolor}{rgb}{0.8,0,0}

\usepackage[unicode=true,pdfusetitle, bookmarks=true,bookmarksnumbered=false,bookmarksopen=false, breaklinks=false,pdfborder={0 0 0},backref=false,colorlinks=true, linkcolor=myrefcolor,citecolor=myurlcolor,urlcolor=myurlcolor]{hyperref}
\DeclareGraphicsExtensions{.png,.pdf,.eps}
\graphicspath{ {./figs/} }

\newcommand{\pS}{p_{\rm S}}
\newcommand{\pL}{p_{\rm L}}

\newcommand{\noise}{{\rm noise}}
\newcommand{\vc}{V_\noise}

\newcommand{\vrange}{\Delta V}

\newcommand{\vis}{{\cal V}}

\begin{document}

\title{Randomness in Quantum Mechanics: Philosophy, Physics and Technology}

\author{Manabendra Nath Bera}
\affiliation{ICFO-Institut de Ci\`encies Fot\`oniques, The Barcelona Institute of Science and Technology, E-08860 Castelldefels (Barcelona), Spain}
\author{Antonio Ac\'in}
\affiliation{ICFO-Institut de Ci\`encies Fot\`oniques, The Barcelona Institute of Science and Technology, E-08860 Castelldefels (Barcelona), Spain}
\affiliation{ICREA-Instituci\'o Catalana de Recerca i Estudis Avan\c{c}ats, Lluis Companys 23, E-08010 Barcelona, Spain}
\author{Marek Ku\'s}
\affiliation{Center for Theoretical Physics, Polish Academy of Sciences, Aleja Lotnik{\'o}w 32/44, 02-668 Warszawa, Poland}
\author{Morgan W. Mitchell}
\affiliation{ICFO-Institut de Ci\`encies Fot\`oniques, The Barcelona Institute of Science and Technology, E-08860 Castelldefels (Barcelona), Spain}
\affiliation{ICREA-Instituci\'o Catalana de Recerca i Estudis Avan\c{c}ats, Lluis Companys 23, E-08010 Barcelona, Spain}
\author{Maciej Lewenstein}
\affiliation{ICFO-Institut de Ci\`encies Fot\`oniques, The Barcelona Institute of Science and Technology, E-08860 Castelldefels (Barcelona), Spain}
\affiliation{ICREA-Instituci\'o Catalana de Recerca i Estudis Avan\c{c}ats, Lluis Companys 23, E-08010 Barcelona, Spain}

% \date{\today}

\begin{abstract}

This progress report  covers recent developments in the area of quantum randomness, which is an extraordinarily interdisciplinary area that belongs not only  to physics, but also to philosophy, mathematics, computer science, and technology. For this reason the article contains three parts that will be essentially devoted to different aspects of quantum randomness, and even directed, although not restricted, to various audiences: a philosophical part, a physical part, and a technological part. For these reasons the article is written on an elementary level, combining simple and non-technical descriptions with a concise review of  more advanced results. In this way readers of various provenances will be able to gain while reading the article.

\end{abstract}
\maketitle
\tableofcontents

%% SECTION I: Introduction %%%%%%%%%%%%%%%%%%%%%%%%%%%%%%%%%
\section{Introduction}
Randomness is a very important concept finding many applications in modern science and technologies. At the same time it is also quite controversial, and may have different meanings depending on the field of science it concerns. In this short introduction to our report we explain, in a very general manner, why randomness plays such an important  role in various branches of science and technology. In particular we elaborate the concept of {\it ``apparent randomness''}, to contrast it with what we understand under the name {\it ``intrinsic randomness''}.

{\it Apparent randomness} as an element of more efficient description of nature is 
used practically in all sciences, and in physics in particular, cf.
\cite{Penrose79,Schrodinger89,Khinchin14,Tolman10,Halmos13}. This kind
of randomness expresses our lack of full knowledge of the considered
system. Paradigmatic example concerns classical mechanics of many-body
systems that are simply too complex to be considered with all details.
The complexity of the dynamics of systems consisting of many interacting
constituents makes predictions, even assuming perfect knowledge of
initial conditions, practically impossible. This fact motivates the
development of statistical mechanics and thermodynamics. The
descriptions that employ probability distributions  and statistical
ensembles, or even more reduced thermodynamic description, are more
adequate and useful. Another paradigmatic example concerns chaotic
systems. In deterministic chaos  theory, cf.
\cite{Bricmont95,Ivancevic08,Gleick08} even for the small systems
involving few degrees of freedom, the immanent lack of precision in our
knowledge of initial conditions leads to the impossibility of making
long time  predictions. This is  due to an exponential separation of
trajectories, when small differences at start lead to large
end-effects. Also here, intrinsic ergodicity allows one to use the tools of statistical ensembles.

In  quantum mechanics  {\it apparent (a.k.a. epistemic)  randomness}
also plays an important role and reflects our lack of full knowledge of
the state of a system. A state of a system in quantum mechanics corresponds to a {\it vector} in a Hilbert space, and is 
described by the projector operator on that vector. Such states and the corresponding projectors of rank one are termed as {\it pure states}. In general, we never know the actual (pure) state of the system precisely. Such situation may be caused by our own imperfectness in determining the state in question. Even, these may arise from measurements that result in statistical ensembles of many pure states. The appropriate way of describing such states is using {\it a density matrix}, i.e. the probabilistic mixture of the projectors on the pure states. The pure states are, simply, represented by those density matrices that are just rank-one projectors. In fact, expressing quantum systems, with a lack of the full knowledge about the state in question, constitutes the main reason of the introduction of the density matrix formalism \cite{Messiah14,Tannoudji91}.

However, in quantum physics there is a new form of randomness,
which is rather {\it intrinsic} or {\it inherent} to the theory. Namely, even if the state of the system is pure and we know it exactly, the predictions of quantum mechanics could be {\it intrinsically  probabilistic and random}! Accepting quantum mechanics, that is assuming that the previous sentence is true, we should consequently accept that quantum mechanics could be intrinsically random. We adopt this position in this paper.

To summarize the above discussion let us define: \\

\noindent {\bf Def. 1 -- Apparent (a.k.a. epistemic) randomness.}

Apparent randomness is the randomness that results exclusively from a
lack of full knowledge about the state of the system in consideration.
Had we known the initial state of the system exactly, we could have
predicted its future evolution exactly. Probabilities and stochastic
processes are used here as an {\it efficient tool} to describe at least
a partial knowledge about the system and its features. Apparent
randomness implies and requires existence of the, so called, underlying
{\it hidden variable theory}. It is the lack of knowledge of hidden
variables that causes apparent randomness. Had we known them, we could
have make predictions with certainty. \\

\noindent {\bf Def. 2 -- Intrinsic (a.k.a. inherent or ontic)
randomness.} Intrinsic randomness is the randomness that persists even
if we have the full knowledge about the state of the system in
consideration. Even exact knowledge of the initial state does not allow
to predict future evolution exactly: we can only make probabilistic
predictions. Probabilities and stochastic processes are used here as a
{\it necessary and inevitable  tool} to describe our knowledge about
the system and its behavior. Of course,  intrinsic randomness might
coexist with the apparent one -- for instance, in quantum mechanics
when we have only partial knowledge about the state of the system
expressed by the density matrix, the two causes of randomness are
present. Moreover, intrinsic randomness {\it does not} exclude existence of
effective hidden variable theories that could allow for  partial
predictions of the evolution of the systems with certainty. 
As we shall see, in quantum mechanics of composite systems, an effective {\it local} hidden variable
theories in general cannot be used to make predictions about local measurements and the local outcomes are intrinsically random.
\\

Having defined the main concepts, we present here short resumes of the subsequent parts of the report, where our focus would be mostly on {\it quantum randomness}:
\begin{itemize}

\item {\bf Quantum Randomness and Philosophy.} Inquiries concerning
    the nature of randomness accompany European philosophy from its
    beginnings. We give a short review of classical philosophical
    attitudes to the problem and their motivations. Our aim is to
    relate them to contemporary physics and science in general.
    This is intimately connected to discussion of various concepts
    of determinism and its understanding in classical mechanics,
    commonly treated as an exemplary deterministic theory, where
    chance has only an epistemic status and leaves room for
    indeterminism only in form of statistical physics description
    of the world. In this context, we briefly discuss another kind
    of indeterminism in classical mechanics caused by
    the non-uniqueness of solutions of the Newton's equations and
    requiring supplementing the theory with additional unknown
    laws. We argue that this situation shares similarities with
    that of quantum mechanics, where quantum measurement theory
    \textit{\`a la} von Neumann provides such laws. This brings us
    to the heart of the     problem of intrinsic randomness of
    quantum mechanics from the philosophical point of view. We
    discuss it in two quantum aspects: contextuality and
    nonlocality, paying special attention to the question: can quantum randomness be certified in any sense?

\item {\bf Quantum Randomness and Physics.} Unlike in classical
    mechanics, randomness is considered to be inherent in the
    quantum domain. From a scientific point of view, we would raise
    arguments if this randomness is intrinsic or not. We start by
    briefly reviewing standard the approach to randomness in quantum
    theory. We shortly recall the postulates of quantum mechanics
    and the relation between quantum measurement theory and
    randomness. Nonlocality as an important ingredient of the
    contemporary physical approach to randomness is then discussed.
    We proceed then with a more recent approach to randomness generation
	    based on the so called ``device independent'' scenario, in which one talks exclusively about
    correlations and probabilities to characterize randomness and
    nonlocality. We then  describe several problems of classical
    computer science that have recently found elegant quantum
    mechanical solutions, employing the nonlocality of quantum
    mechanics. Before doing this we devote a subsection to describe
    a contemporary information theoretic approach to the 
    characterization of randomness and random bit sources.  In continuation, we discuss the idea of protocols for Bell
    certified randomness generation (i.e. protocols based on Bell
    inequalities to generate and certify randomness in the device
    independent scenario), such as quantum  randomness expansion (i.e.
    generating larger number of random bits from a shorter seed of random bits),
    quantum randomness amplification (i.e. transforming weakly
    random sequences of, say, bits into ``perfectly'' random ones).
    It should be noted that certification, expansion and
    amplification of randomness are classically not possible or
    require extra assumptions in comparison with what quantum mechanics
    offers. Our goal is to review the recent state-of-art results
    in this area, and their relations and applications for
    device independent quantum secret key distribution. We also
    review briefly and analyze critically  the ``special status''
    of quantum mechanics among the so called no-signaling theories. These are the theories, in which the choice of observable to measure by, say, Bob does not influence the outcomes of measurements of Alice and all other parties (for precise definition in terms of conditional probabilities for arbitrary number of parties, observables and outcomes see Eq. (\ref{eq:no-sig}). While quantum mechanical correlation fulfill the no-signaling conditions, correlations resulting from no-signaling theories form a strictly larger set.  No-signalling correlations were first considered in relation to quantum mechanical ones by Popescu and Rohlich \cite{Popescu92}.     In many situations, it is the no-signaling assumption and Bell
    non-locality that permit certification of randomness and
    perhaps novel possibilities of achieving security in
    communication protocols.

\item {\bf  Quantum Randomness and Technology.} We start this part 
    by shortly reminding the readers why random numbers are useful in technology and what they are used for.
The drawbacks of the classical random number generation,  based on
classical computer science ideas,   are also mentioned.  We describe
proof-of-principle experiments, in which certified randomness was
generated using nonlocality.  We then focus on describing existing
``conventional'' methods of quantum random number generation and
certification. We discuss also the current status of detecting
non-locality and Bell violations. We will then review current
status of commercial implementation of quantum protocols for random
number generations, and the first steps toward device independent
or at least partially device independent implementations. A complementary review of quantum random generators may be found in Ref. \cite{HerreroARX2016}

\item {\bf Quantum Randomness and Future.} In the conclusions we
    outline new interesting perspectives for fundamentals of
    quantum mechanics that are stimulated by the studies of quantum
    randomness: What's the relation between randomness,
    entanglement  and non-locality? What are the ultimate limits
    for randomness generation using quantum resources? How does
    quantum physics compare to other theories in terms of
    randomness generation? What's the maximum amount of randomness
    that can be generated in a theory restricted only by the
    no-signaling principle?
\end{itemize}

Randomness in physics has been a subject of extensive studies and our report neither has ambition, nor objective to cover all existing literature on this subject. We stress that there are of course various, highly recommended reviews on randomness in physics, such as for instance the excellent articles by J. Bricmont \cite{Bricmont95}, or the recent book  by Juan C. Vallejo and Miguel A.F. Sanjuan \cite{Sanjuan17}.
The main novelty of our report lies in the incorporation of the contemporary approach to quantum randomness and its relation to quantum nonlocality and quantum correlations, and emerging device independent quantum information processing and quantum technologies.

In fact, our report focuses on certain aspects of randomness that have become particularly relevant in the view of the recent technical (i.e. qualitative and quantitative, theoretical and experimental) developments in quantum physics and quantum information science: quantum randomness certification, amplification and extension are paradigmatic examples of these developments. The technological progress in constructing publicly or even commercially available, highly efficient quantum random number generators is another important aspect: it has in particular led to the first experimental proof of quantum nonlocality, i.e. loophole-free violation of Bell inequalities \cite{Hensen15, Giustina15,
Shalm15}. In particular: 
\begin{itemize}

\item In the philosophical part we concentrate on the distinction between apparent (epistemic) and intrinsic (inherent or ontic) randomness, and on the question whether intrinsic randomness of quantum mechanics can be certified in certain sense. We devote considerable attention to the recent discussion of non-deterministic models in classical physics, in which (in contrast to the standard Newtonian-Laplacian mechanics) similar questions may be posed. Based on the recently proposed protocols, we argue that observation of nonlocality of quantum correlations can be directly use to certify randomness; moreover this can be achieved in a secure device independent way. Similarly, contextuality of quantum mechanics, i.e. results of measurement depend on the context in which they are performed, or, more precisely, which compatible quantities are simultaneously measured, can be used to certify randomness, although not in device independent way. 

\item In the physical part we concentrate on more detailed presentation of the recent protocols of randomness certification, amplification end expansion. 

\item In the technological part we first discuss the certified randomness generation \cite{Pironio10}, accessible as an open source NIST Beacon \cite{NISTBeacon}. Then we concentrate on the recent technological developments that have led to the first loophole free detection of nonlocality, and are triggering important commercial applications. 

\end{itemize}

Here we limit ourselves to the contemporary, but traditional approach to quantum mechanics and its interpretation, as explicated in the books of Messiah or Cohen-Tannoudji \cite{Messiah14,Tannoudji91}. In this sense this review is not complete, and important relevant philosophical aspects are not discussed. Thus, we do not describe other interpretations and approaches, such as pilot wave theory of Bohm \cite{Bohm51} or many-world interpretation (MWI) of Everett \cite{Everett57}, as they are far beyond the scope of this report. Of course, the meanings of randomness and non-locality are completely different in these approaches. 

For instance, one can consider {\it de Broglie–Bohm's} interpretation of quantum theory. This is also known as the {\it pilot-wave theory, Bohmian mechanics, the Bohm (or Bohm's) interpretation}, and {\it the causal interpretation} of quantum mechanics. There a wave function, on the space of all possible configurations, not only captures the epistemic knowledge of system's state but also carry a ``hidden variable'' to encode it's ontic information and this ``hidden variable'' may not be accessible or observed. In addition to the wave function, the Bohmian particle positions also carry information. Thus, the Bohmian QM has two ontological ingredients: the wave function and positions of particles. As we explain below, the theory is non-local and that is why we do not discuss it in the present review in details.

The time evolution of the system (say, the positions of all particles or the configuration of all fields) is guided by Schrödinger's equation. By construction, the theory is deterministic \cite{Bohm52} and explicitly non-local. In other words, the velocity of one particle relies on  the value of the guiding equation, which depends on the configuration of the system given by its wave function. The latter is constrained to the boundary conditions of the system and that could. in principle, be the entire universe.  Thus, as explicitly stated by D. Bohm \cite{Bohm52}: "In contrast to the usual interpretation, this alternative interpretation permits us to conceive of each individual system as being in a precisely definable state, whose changes with time are determined by definite laws, analogous to (but not identical with) the classical equations of motion. Quantum-mechanical probabilities are regarded (like their counterparts in classical statistical mechanics) as only a practical necessity and not as an inherent lack of complete determination in the properties of matter at the quantum level."

So Bohm's theory has to be regarded as non-local hidden variable theory and therefore it does not allow intrinsic randomness; similarly, the many-world interpretation (MWI) suggests that intrinsic randomness is an illusion \cite{Vaidman14}. MWI asserts the objective reality of ``universal'' wave function and denies any possibility of wave function collapse. MWI implies that all possible pasts and futures are elements of reality, each representing an objective "world" (or "universe"). In simpler words, the interpretation states that there is a very large number of universes, and everything that could possibly have occurred in our past, but did not, has occurred in the past of some other universe or universes.
Therefore, MWI indeed does not leave much space for any kind of probability or randomness, since formally, all outcomes take place with certainty. This is already a sufficient reason to not to consider the WMI in the present review. But, obviously, the whole problem is whether one can speak about probabilities within MWI or not. This problem has been extensively discussed by several authors, e.~g. \cite{Saunders1998, Saunders2010, Papineau2010, Albert2010}.

We stress that we adopt in this review the “traditional” interpretation, in which quantum mechanics is intrinsically random, but nonlocal. This adaptation is the result of our free choice. Other readers may freely, or better to say deterministically, but nonlocally adopt the Bohmian point of view.

\section{Quantum Randomness and Philosophy}
\subsection{Epistemic and ontic character of probability}
Randomness is a fundamental resource indispensable in numerous
scientific and practical applications like Monte-Carlo simulations,
taking opinion polls, cryptography etc. In each case one has to
generate a ``random sample'', or simply a random sequence of digits.
Variety of methods to extract such a random sequence from physical
phenomena were proposed and, in general successfully, applied in
practice. But how do we know that a sequence is ``truly random''? Or,
at least. ``random enough'' for all practical purposes? Such problems
become particularly acute for cryptography where provably unbreakable
security systems are based on the possibility to produce a string of
perfectly random, uncorrelated digits used later to encode data. Such a
random sequence must be unpredictable for an adversary wanting to break
the code, and here we touch a fundamental question concerning the
nature of randomness. If all physical processes are uniquely determined
by their initial conditions and the only cause of unpredictability is
our inability to determine them with an arbitrary precision, or lack of
detailed knowledge of actual conditions that can influence their time
evolution, the security can be compromised, if an adversary finds finer
methods to predict outcomes. On the other hand, if there are processes
``intrinsically'' random, i.e. random by their nature and not due to
gaps in our knowledge, unconditional secure coding schemes are
conceivable.

Two attitudes concerning the nature of randomness in the world
mentioned above can be dubbed as epistemic and ontic. Both agree that
we observe randomness (indeterminacy, unpredictability) in nature, but
differ in identifying the source of the phenomenon. The first claims
that the world is basically deterministic, and the only way in which
a random behavior demanding probabilistic description appears is due to
lack of knowledge of the actual state of the observed system or details
of its interaction with the rest of the universe. In contrast, according
to the second, the world is nondeterministic, randomness is its
intrinsic property, independent of our knowledge and resistant to
attempts aiming at circumventing its consequences by improving
precision of our observations. In other words, ``intrinsic'' means that
this kind of randomness cannot be understood in terms of a
deterministic ``hidden variable'' model. The debate on both epistemic
and ontic nature of randomness can be traced back to the pre-Socratic
beginnings of the European philosophy. For early atomists,
Leucippus\footnote{`Nothing happens at random; everything happens out
of reason and by necessity', from the lost work {Per\'i no\~u}
\textit{On Mind}, see \cite{Diels06}, p. 350, \cite{Freeman48}, p. 140,
fr. 2.} and Democritus\footnote{`All things happen by virtue of
necessity', \cite{Laertius25}, IX, 45.}, the world was perfectly
deterministic. Any occurrence of chance is a consequence of our limited
abilities\footnote{`Men have fashioned an image of {\it chance} as an excuse
for their own stupidity', \cite{Diels06}, p. 407, \cite{Freeman48}, p.
158, fr. 119.}. One century later Epicurus took the opposite side. To
accommodate an objective chance the deterministic motion of atoms must
be interrupted, without a cause, by ``swerves''. Such an indeterminacy
propagates then to macroscopic world. The main motivation was to
explain, or at least to give room for human free will, hardly
imaginable in a perfectly deterministic world\footnote{`Epicurus saw
that if the atoms traveled downwards by their own weight, we should
have no freedom of the will, since the motion of the atoms would be
determined by necessity. He therefore invented a device to escape from
determinism (the point had apparently escaped the notice of
Democritus): he said that the atom while traveling vertically downward
by the force of gravity makes a very slight swerve to one side'
\cite{Cicero33}, I, XXV.}. It should be clear, however, that purely
random nature of human actions is as far from free will, as the
latter from a completely deterministic process. A common feature of
both extreme cases of pure randomness and strict determinism is lack of
any possibility to control or influence the course of events. Such a
possibility is definitely indispensable component of the free will. The
ontological status of randomness is thus here irrelevant and the
discussion whether "truly random theories", (as supposedly should
quantum mechanics be), can "explain the phenomenon of the free will" is
pointless. It does not mean that free will and intrinsic randomness
problems are not intertwined. On one side, as we explain later, the
assumption that we may perform experiments in which we can freely
choose what we measure, is an important ingredient in arguing that
violating of Bell-like inequalities implies ``intrinsic randomness'' of
quantum mechanics. On the second side, as strict determinism in fact
precludes the free will, the intrinsic randomness seems to be a
necessary condition for its existence. But, we need more to produce a
condition that is sufficient. An interesting recent discussion of
connections between free will and quantum mechanics may be found in
Part I of \cite{suarezbook13}. In \cite{gisin13} and \cite{brassard13}
the many-world interpretation of quantum mechanics, which is sometimes
treated as a cure against odds of orthodox quantum mechanics, is either
dismissed as a theory that can accommodate free will \cite{gisin13} or,
taken seriously in \cite{brassard13}, as admitting the possibility that
free will might be a mere illusion. In any case it is clear that one
needs much more than any kind of randomness to understand how free will
appears. In \cite{suarez13} the most radical attitude to the problem
(apparently present also in \cite{gisin13}) is that ``not all that
matters for physical phenomena is contained in space-time''.

\subsection{Randomness in classical physics}
A seemingly clear distinction between two possible sources of
randomness outlined in the previous section becomes less obvious if we
try to make the notion of determinism more precise. Historically, its
definition usually had a strong epistemic flavor. Probably the most
famous characterization of determinism is that of Pierre Simon de Laplace \cite{Laplace14}:
`\emph{Une intelligence qui, pour un instant donn\'e, conna\^{i}trait
toutes les forces dont la nature est anim\'ee, et la situation
respective des \^{e}tres qui la composent, si d'ailleurs elle \'{e}tait
assez vaste pour soumettre ces donn\'ees \`{a} l'analyse, embrasserait
dans la m\^{e}me formule les mouvemens des plus grands corps de
l'univers et ceux du plus l\'eger atome : rien ne serait incertain pour
elle, et l'avenir comme le pass\'e, serait pr\'esent \'{a} ses yeux.\footnote{``We may regard the present state of the universe as the effect of its past and the cause of its future. An intellect which at a certain moment would know all forces that set nature in motion, and all positions of all items of which nature is composed, if this intellect were also vast enough to submit these data to analysis, it would embrace in a single formula the movements of the greatest bodies of the universe and those of the tiniest atom; for such an intellect nothing would be uncertain and the future just like the past would be present before its eyes.'' \cite{Laplace51} p. 4}}'
 Two hundred years later Karl Raimund Popper writes
`We can ... define `scientific' determinism as follows: The doctrine of
`scientific' determinism is the doctrine that the state of any closed
physical system at any given future instant of time can be predicted,
even from within the system, with any specified degree of precision, by
deducing the prediction from theories, in conjunction with initial
conditions whose required degree of precision can always be calculated
(in accordance with the principle of accountability) if the prediction
task is given' \cite{Popper82}. By contraposition thus,
unpredictability implies indeterminism. If we now equate indeterminism
with existence of randomness, we see that a sufficient condition for
the latter is the unpredictability. But, unpredictable can be equally
well events about which we do not have enough information, and those
that are ``random by themselves''. Consequently, as it should have been
obvious, Laplacean-like descriptions of determinism are of no help when
we look for sources of randomness.

Let us thus simply say that a course of events is deterministic if
there is only one future way for it to develop. Usually we may also
assume that its past history is also unique. In such cases the only
kind of randomness is the epistemic one.

As an exemplary theory describing such situations one usually invokes
classical mechanics. Arnol'd in his treatise on ordinary differential
equations, after adopting the very definition of determinism advocated
above\footnote{\,``A process is said to be \textit{deterministic} if
its entire future course and its entire past are uniquely determined by
its state at the present instant of time'', \cite{Arnold73}, p. 1},
writes: ``Thus for example, classical mechanics considers the motion of
systems whose past and future are uniquely determined by the initial
positions and velocities of all points of the
system''\footnote{\,\textit{ibid.}}. The same can be found in his
treatise on classical mechanics\footnote{\,''The initial state of a
mechanical system (the totality of positions and velocities of its
points at some moment of time) uniquely determines all of its motion'',
\cite{Arnold89}, p. 4}. He gives also a kind of justification, ``It is
hard to doubt this fact, since we learn it very
early''\footnote{\textit{ibid}}. But, what he really means is that a
mechanical system are uniquely determined by positions and momenta of
its constituents, ``one can imagine a world, in which to determine the
future of a system one must also know the acceleration at the initial
moment, but experience shows us that our world is not like
this''\footnote{\textit{ibid.}}. It is clearly exposed in another
classical mechanics textbook, Landau and Lifschitz's
\textit{Mechanics}: ``If all the co-ordinates and velocities are
simultaneously specified, it is known from experience that the state of
the system is completely determined and that its subsequent motion can,
in principle, be calculated. Mathematically, this means that, if all
the co-ordinates $q$ and velocities $\dot{q}$ are given at some
instant, the accelerations $\ddot{q}$ at that instant are uniquely
defined''\footnote{\cite{Landau60}, p. 1.}. Apparently, also here the
``experience'' concerns only the observation that positions and
velocities, and not higher time-derivatives of them, are sufficient to
determine the future.

In such a theory there are no random processes. Everything is in fact
completely determined and can be predicted with desired accuracy once
we improve our measuring and computing devices. Statistical physics,
which is based on classical mechanics, is a perfect example of
indeterministic theory where measurable quantities like pressure or
temperature are determined by mean values of microscopical `hidden
variables', for example positions and momenta of gas particles. These
hidden variables, however, are completely determined at each instant of
time by the laws of classical mechanics, and with an appropriate effort
can be, in principle, measured and determined. What makes the theory
`indeterministic' is a practical impossibility to follow trajectories
of individual particles because of their number and/or the
sensitiveness to changes of initial conditions. In fact such a
sensitiveness was pointed as a source of chance by
Poincar\'e\footnote{\,``Le premier exemple que nous allons choisir est
celui de l'\'{e}quilibre instable; si un c\^{o}ne repose sur sa pointe,
nous savons bien qu'il va tomber, mais nous ne savons pas de quel
c\^{o}t\'{e}; il nous semble que le hasard seul va en d\'{e}cider.'',
\cite{Poincare12} page 4, (``The first example we select is that of
unstable equilibrium; if a cone rests upon its apex, we know well that
it will fall, but we do not know toward what side; it seems to us
chance alone will decide.'' \cite{Newman56}, vol.\ 2, p. 1382)} and
Smoluchowski\footnote{\,``...ein ganz wesentliches Merkmal desjenigen,
was man im gew{\"o}hnlichen Leben oder in unserer Wissenschaft als
Zufall bezeichnet ... l{\"a}{\ss}t sich ... kurz in die Worte fassen:
{\it kleine Ursache -- gro{\ss}e Wirkung''}, \cite{Smoluchowski18}
(``...fundamental feature of what one calls chance in everyday life or
in science allows a short formulation: {\it small cause -- big
effect}'')} soon after modern statistical physics was born, but it is
hard to argue that this gives to the chance an ontological status. It
is, however, worth mentioning that Poincar\'e was aware that randomness
might have not only epistemic character. In the above cited
Introduction to his {\it Calcul des probabilit\'es} he states `\emph{Il
faut donc bien que le hasard soit autre chose que le nom que nous
donnons \`{a} notre ignorance}'\footnote{\cite{Poincare12} p.\ 3.},
(`So it must be well that chance is something other than the name we
give our ignorance'\footnote{\cite{Newman56}, vol.\ 2, p. 1381}).

Still, the very existence of deterministic chaos implies that classical mechanics is unpredictable in general in any practical sense. The technical question how important this unpredictability can be is, actually, the subject of intensive studies in the last decades (for recent monographs see \cite{Sanjuan17,Sanjuan16}).

It is commonly believed (and consistent with the above cited
descriptions of determinism in mechanical systems) that on the
mathematical level the deterministic character of classical mechanics
takes form of Newton's Second Law
\begin{equation}\label{N2}
m\frac{d^2\mathbf{x}(t)}{dt^2}=\mathbf{F}(\mathbf{x(t)},t),
\end{equation}
where the second derivatives of the positions, $\mathbf{x}(t)$, are
given in terms of some (known) forces $\mathbf{F}(\mathbf{x(t)},t)$.
But, to be able to determine uniquely the fate of the system we need
something more than merely initial positions $\mathbf{x}(0)$ and
velocities $d\mathbf{x}(t)/dt|_{t=0}$. To guarantee uniqueness of the
solutions of the Newton's equations (\ref{N2}), we need some additional
assumptions about the forces $\mathbf{F}(\mathbf{x(t)},t)$. According to the Picard Theorem\footnote{In mathematics of differential equations, the Picard's existance theorem (also known as Cauchy–Lipschitz theorem) is important to ensure existence and uniqueness of solutions to first-order equations with given initial conditions. Consider an initial value problem, say, $y'(t)=f(t,y(t))$ with $y(t_0)=y_0$. Also assume $f(.,.)$ is is uniformly Lipschitz continuous in $y$ (i.e., the Lipschitz constant can be taken independent of $t$) and continuous in $t$. Then for some values of $\varepsilon >0$, there exists a unique solution of $y(t)$, given the initial condition, in the interval $[t_0 - \varepsilon , t_0 + \varepsilon]$.}
\cite{Coddington55}, 
an additional technical condition that is sufficient
for the uniqueness is the Lipschitz condition, limiting the
variability of the forces with respect to the positions. Breaking it
opens possibilities to have initial positions and velocities that do
not determine uniquely the future trajectory. A world, in which there
are systems governed by equations admitting non-unique solutions is not
deterministic according to our definition. We can either defend
determinism in classical mechanics by showing that such pathologies
never occur in our world, or agree that classical mechanics admits, at
least in some cases, a nondeterministic evolution. Each choice is hard
to defend. In fact it is relatively easy to construct examples of more
or less realistic mechanical systems for which the uniqueness is not
guaranteed. Norton \cite{Norton07} (see also \cite{Norton08}) provided
a model of a point particle sliding on a curved surface under the
gravitational force, for which the Newton equation reduces to
$\frac{d^2r}{dt^2}=\sqrt{r}$. For the initial conditions $r(0)=0,
\frac{dr}{dt}|_{t=0}=0$, the equation has not only an obvious solution
$r(t)=0$, but, in addition, a one parameter family given by
\begin{equation}
r(t)=\left\{
\begin{array}{cl}
0, & \mathrm{for\ } t\le T \\
\frac{1}{144}(t-T)^4, & \mathrm{for\ } t\ge T
\end{array}
\right.
\end{equation}
where $T$ is an arbitrary parameter. For a given $T$ the solution
describes the particle staying at rest at $r=0$ until $T$
and starting to accelerate at $T$.
Since $T$ is arbitrary we can not predict when the change from the
state of rest to the one with a non-zero velocity takes place.

The example triggered discussions
\cite{Korolev07,Korolev07a,Kosyakov08,Malament08,Roberts09,Wilson09,Zinkernagel10,Fletcher12,
Laraudogoitia13}, raising several problems, in particular its physical
relevance in connection with simplifications and idealizations made to
construct it. However, they do not seem to be different from ones
commonly adopted in descriptions of similar mechanical situations,
where the answers given by classical mechanics are treated as perfectly
adequate. At this point classical mechanics is not a complete theory of
the part of the physical reality it aspires to describe. We are
confronted with a necessity to supplement it by additional laws dealing
with situations where the Newton's equation do not posses unique
solutions.

The explicit assumption of incompleteness of classical mechanics has
its history, astonishingly longer than one would expect. Possible
consequences of non-uniqueness of solutions attracted attention of
Boussinesq who in his \emph{{M}\'emoire} for the French Academy of
Moral and Political Sciences writes: `\emph{...les ph{\'e}nom{\`e}nes de
mouvement doivent se diviser en deux grandes classes. La premi{\`e}re
comprendra ceux o{\`u} les lois m{\'e}caniques exprim{\'e}es par les
{\'e}quations diff{\'e}rentielles d{\'e}ter\-mi\-ne\-ront {\`a} elles
seules la suite des {\'e}tats par lesquels passera le syst{\`e}me, et
o{\`u}, par cons{\'e}quent, les forces physico-chimiques ne laisseront
aucun r{\^o}le disponible {\`a} des causes d'une autre nature. Dans la
seconde classe se rangeront, au contraire, les mouvements dont les
{\'e}quations admettront des int{\'e}grales singuli{\`e}res, et dans
lesquels il faudra qu'une cause distincte des forces physico-chimiques
intervienne, de temps  en  temps ou d'une mani{\`e}re continue, sans
d'ailleurs apporter aucune part d'action m{\'e}canique, mais simplement
pour diriger le syst{\`e}me a chaque bifurcation d'int{\'e}grales qui
se pr{\'e}sentera.}'\footnote{\, \cite{Boussinesq78}, p. 39. ``The
movement phenomena should be divided into two major classes. The first
one comprises those for which the laws of mechanics expressed as
differential equations will determine by themselves the sequence of
states through which the system will go and, consequently, the
physico-chemical forces will not admit causes of different nature to
play a role. On the other hand, to the second class we will assign
movements for which the equations will admit singular integrals, and
for which one will need a cause distinct from physico-chemical forces
to intervene, from time to time, or continuously, without using any
mechanical action, but simply to direct the system at each bifurcation
point which will appear.'' The
``singular integrals'' mentioned by Boussinesq are the additional
trajectories coexisting with ``regular'' ones when conditions
guaranteeing uniqueness of solutions are broken.}

Boussinesq does not introduce any probabilistic ingredient to the
reasoning, but definitely, there is a room to go from mere
indeterminism to the awaited `intrinsic randomness'. To this end,
however, we need to postulate an additional law supplementing classical
mechanics by attributing probabilities to different solutions of
non-Lipschitzian equations\footnote{Thus in the Norton's model, the new
law of nature should, in particular, ascribe a probability $p(T)$ to
the event that the point staying at rest at $r=0$ starts to move at
time $T$.}. It is hard to see how to discover (or even look for) such a
law, and how to check its validity. What we find interesting is an
explicit introduction to the theory a second kind of motion. It is
strikingly similar to what we encounter in quantum mechanics, where to
explain all observed phenomena one has to introduce two kinds of
kinematics of a perfectly deterministic Schr\"odinger evolution and
indeterministic state reductions during measurements. Similarity
consist in the fact, that deterministic (Schr\"odinger, Newton)
equations are  not sufficient to describe the full evolution: they have
to be completed, for instance  by probabilistic description of the
results of measurements in quantum mechanics, or by probabilistic
choice of non-unique solutions in the Norton's example\footnote{Similar things seem to happen also in so called "general
no-signaling theories" where, in comparison with quantum mechanics,
the only physical assumption concerning the behavior of a system is the
impossibility of transmitting information with an infinite velocity,
see \cite{Tylec15}.}.

It is interesting to note the ideas of Boussinesq have been in fact a subject of intensive discussion in the recent years in philosophy of science within the, so called, ``second Boussinesq debate''. The first Boussinesq debate took place in France between 1874-1880. As stated by T.M. Mueller \cite{Mueller15}: ``In 1877, a young mathematician named Joseph Boussinesq presented a \textit{ m\'emoire} to the \textit{ Acad\'emie des Sciences}, 
which demonstrated that some differential equations may have more than one solution. Boussinesq linked this fact to indeterminism and to a possible solution to the free will versus determinism debate.''. The more recent debate discovered, in fact, that some hints for the Boussinesq ideas can be also found in the works of James Clerk Maxwell \cite{Maxwell}.  The views of Maxwell, important in this debate and not known very much by physicists, show that he was very much influenced by the work of Joseph Boussinesq and Adh\'emar Jean Claude Barr\'e de Saint-Venant \cite{Mueller15}.  What is also quite unknown by many scientists is that Maxwell learn the statistical ideas from Adolphe Quetelet, a Belgian mathematician, considered to be one of the founders of statistics. Excellent account  on the concepts of determinism versus indeterminism, on the notion of uncertainty, also associated to the idea of randomness, as well as on different meanings that randomness has for different audiences may be found in the set of blogs of Miguel A. F. Sanju\'an \cite{blog1,blog2,blog3} and in the outstanding book \cite{Dahan92}. These references cover also a lot of details 
of the the first and recent Boussinesq debate.  A Polish text by Kole\.zy\'nski \cite{Polish} discusses  related quotations from Boussinesq, Maxwell and Poincar\'e  in a philosophical context of the determinism.

Of course, to great extend Boussinesq debate was stimulated by the attempts toward understanding of nonlinear dynamics and hydrodynamics in general, and the phenomenon of turbulence in particular. A nice review of of various approaches and ideas until 1970s is presented in the lecture by Marie Farge \cite{Farge}. The contemporary approach to turbulence is very much related to the Boussinesq suggestions and the use of non-Lipschitzian, i.e. nondeterministic hydrodynamics, has been develop in the recent years by Falkovich, {Gaw\ifmmode \mbox{\c{e}}\else \c{e}\fi{}dzki}, Vargassola and others (for  outstanding reviews see \cite{Falkovich01,Gawedz-rec}). The history of these works is nicely described in the presentation \cite{Gawedz1}, while the most important particular articles include the series of papers by {Gaw\ifmmode \mbox{\c{e}}\else \c{e}\fi{}dzki} and collaborators
\cite{Gawedz1,Gawedz2}, Vanden Eijnden \cite{Vanden1,Vanden2}, and Le Jan and Raimond \cite{Lejan1,Lejan2,Lejan3}.

\subsection{Randomness in quantum physics}

The chances of proving the existence of `intrinsic randomness' in the
world seem to be much higher, when we switch to quantum mechanics. The
Born's interpretation of the wave function implies that we can count
only on a probabilistic description of reality, therefore quantum
mechanics is inherently probabilistic. 

Obviously, one should ask what is the source of randomness in quantum physics. As pointed out by one of the referees: “In my view all the sources of randomness originate because of interaction of the system (and/or the measurement apparatus) with an environment. The randomness that affects pure states due to measurement is, in my view, due to the interaction of the measurement apparatus with an environment. The randomness that affects open systems (those that directly interact with an environment) is again due to environmental effects.” This point of view is, as considered by many physicists, of course,  parallel to the contemporary theory of quantum measurements, and collapse of the wave function
\cite{WheelerZurek83,Zurek03,Zurek09}.

Still, the end result of such approach to randomness and quantum measurements is that the Born's rule and the traditional Copenhagen interpretation is not far from being rigorously correct. At the same time, quantum mechanics viewed from the device independent point of view, i.e. by regarding only probabilities of outcomes of individual or correlated measurements, incorporates randomness,  which cannot be reduced to our lack of knowledge or imperfectness of our measurements (this will be  discussed with more details below). In this sense for the 
purpose of the present discussion, the detailed form of the major source of the randomness is not relevant, as long as this randomness leads to contextual results of measurements, or nonlocal correlations. 

Let us repeat, both the  pure Born's rule and the advanced theory of quantum measurement    imply that the measurement
outcomes (or expectation value of an observable) may have some
randomness. However, \emph{a priori} there are no obvious reasons for
leaving the Democritean ground  and switch to the Epicurean view. It
might be so that  quantum mechanics, just as statistical physics, is an
incomplete theory admitting deterministic hidden variables, values of
which were beyond our control. To be precise, one may ask how
``intrinsic'' this randomness is and if it can be considered as an
epistemic one. To illustrate it further, we consider two different
examples in the following.

\subsubsection{Contextuality and randomness}

Let us consider a case of a spin-$s$ particle. Now if the particle is
measured in the $z$-direction, there could be $2s+1$ possible outcomes
and each appears with certain probability. Say, the outcomes are
labeled by $\{m\}$, where $m \in [-s,-s+1,\ldots,s-1,s]$ and the
corresponding probabilities by $\{p_m\}$. It means that, with many
repetitions, the experimenter will observe an outcome $m$ with the
frequency approaching $p_m$, as predicted by the Born's rule of quantum
mechanics. The outcomes contain some randomness as they appear
probabilistically. Moreover, these probabilities  are indistinguishable
from classical probabilities. Therefore, the randomness here could be
explained with the help of a deterministic hidden-variable
model\footnote{Note, here we do not impose any constraint on the hidden
variables and these could be even {\it nonlocal}. In fact, the quantum
theory becomes deterministic if one assumes the hidden variables to be
{\it nonlocal} \cite{Gudder70}.} and it is simply a consequence of the
ignorance of the hidden-variable(s).

But, as we stress in the definition in the Introduction: intrinsic
randomness of quantum mechanics  does not exclude existence of hidden
variable models that  can describe outcomes of
measurements. Obviously, if the system is in the pure state
corresponding to $m_0$, the outcome of the measurement of $z$-component
of the spin will be deterministic: $m_0$ with certainty. If we measured
$x$-component of the spin, however, the result would be again
non-deterministic and described only probabilistically. In fact, this
is an instance  of the existence of the, so called, non-commuting
observables in quantum mechanics that cannot be measured simultaneously
with certainty. Uncertainty of measurements of such non-commuting
observables is quantitatively bounded from below by generalized
Heisenberg Uncertainty Principle \cite{Messiah14,Tannoudji91}.

One of important consequences of the existence of
non-commuting observables is the fact that quantum mechanics is {\it
contextual}, as demonstrated in the famous Kochen-Specker theorem
(\cite{Kochen67}, for philosophical discussion see
\cite{Bub99,Isham98}). The Kochen–Specker (KS) theorem \cite{Kochen67},
also known as the Bell-Kochen-Specker theorem \cite{Bell66}, is a "no
go" theorem \cite{Bub99}, proved by J.S. Bell in 1966 and by S.B.
Kochen and E. Specker in 1967. KS theorem  places certain constraints
on the permissible types of hidden variable theories, which try to
explain the randomness of quantum mechanics as an apparent randomness,
resulting from lack of knowledge of hidden variables in an underlying
deterministic model.  The version of the theorem proved by Kochen and
Specker also gave an explicit example for this constraint in terms of a
finite number of state vectors (cf. \cite{Peres95}). The KS theorem
deals with single quantum systems and is thus   a complement to Bell's
theorem that deals with composite systems.

As proved by the KS theorem, there is a contradiction
between two basic assumptions of the hidden variable theories, which is intended
to reproduce the results of quantum mechanics where all hidden
variables corresponding to quantum mechanical observables have definite
values at any given time, and that the values of those variables are
intrinsic and independent of the measurement devices. An immediate contradiction can be caused by non-commutative observables, that are allowed by quantum mechanics. If the Hilbert space
dimension is at least three, it turns out to be impossible to
simultaneously embed all the non-commuting sub-algebras of the algebra of
these observables in one commutative algebra, which is expected to represent the
classical structure of the hidden variable theory\footnote{In fact, it was A. Gleason
\cite{Gleason75}, who pointed out first that quantum contextuality may
exist in dimensions greater than two.  For a single
qubit, i.e. for the especially simple  case of two dimensional Hilbert
space, one can explicitly construct the non-contextual hidden variable
models that describes all measurements (cf.
\cite{Wodkiewicz85,Wodkiewicz95,Scully89}). In this sense, a single qubit does not exhibit intrinsic randomness. 
For the consistency of approach, we should thus consider that intrinsic randomness  could appear in all quantum mechanics, with exception  of quantum mechanics of single qubits. In this report we will neglect this subtlety, and talk about  intrinsic randomness for the whole quantum mechanics without  exceptions, remembering, however, Gleason’s result.}.

The Kochen–Specker theorem excludes hidden variable theories that
require elements of physical reality to be non-contextual (i.e.
independent of the measurement arrangement). As succinctly worded by
Isham and Butterfield \cite{Isham98}, the Kochen–Specker theorem
"asserts the impossibility of assigning values to all physical
quantities whilst, at the same time, preserving the functional
relations between them."

In a more recent approach to contextuality, i.e. where the measurement results depend on the context with which they are measured, 
one proves that non-contextual hidden variable
theories lead to probabilities of measurement outcomes that fulfill
certain inequalities~\cite{Cabello08}, similar to Bell's inequalities for composite
systems. More specifically there are Bell-type inequalities for
non-contextual theories that are violated by any quantum state. Many of
these inequalities between the correlations of compatible measurements
are particularly suitable for testing this state-independent violation
in an experiment, and indeed violations have been experimentally
demonstrated \cite{Kirchmair09,Bartosik09}. Quantifying and
characterizing  contextuality of different physical theories is
particularly elegant in  a general graph-theoretic framework
\cite{Cabello14,Acin15a}.

This novel approach to contextuality is on hand parallel to the earlier observation by N. Bohr \cite{Bohr35} that EPR-like paradoxes may occur in the quantum systems without the need for entangled composite systems. On the other hand it offers a way to certify intrinsic randomness of quantum mechanics. If Cabello-like inequalities are violated in an experiment, it implies that there exist no non-contextual deterministic hidden variable theory that can reproduce results of this experiment, {\it ergo} the results are intrinsically random. Unfortunately, this kind of randomness certification is not very secure, since it explicitly depends on the non-commuting observables that are measured, and in effect is not device independent.

\subsubsection{Nonlocality and randomness}

It is important to extend  the situation beyond the one mentioned above
to multi-party systems and local measurements.  For example, consider
multi-particle system with each particle  placed in a 
separated region. Now, instead of observing the system as a whole, one
may get interested to observe only a part of it, i.e. perform local
measurements.  Given two important facts that QM allows superposition
and no quantum system can be absolutely isolated, spatially separated
quantum systems can be non-trivially correlated, beyond classical
correlations allowed by classical mechanics. In such situation, the
information contained in the whole system is certainly larger than that
of sum of individual local systems. The information residing in the
nonlocal correlations cannot be accessed by observing locally
individual particles of the systems. It means that local descriptions
cannot completely describe the total state of the system. Therefore,
outcomes due to any local observation are bound to incorporate a
randomness in the presence of nonlocal correlation, as long as we do
not have access to the global system or ignore the nonlocal
correlation. In technical terms, the randomness appearing in the {\it
local} measurement outcomes cannot be understood in terms of
deterministic {\it local} hidden variable model and a ``true'' local
indeterminacy is present\footnote{Of course one could  argue that such randomness appears only to be
``intrinsic'', since it is essentially epistemic in nature and arises
due to the inaccessibility or ignorance of the information that resides
in the nonlocal correlations. In another words, this kind of randomness
on the local level is caused by our lack of knowledge of the global
state, and further, it can be explained  using deterministic {\it nonlocal} hidden
variable models.}. Moreover,  randomness on the local level appears even if the
global state of the system is pure and completely known -- the
necessary condition for this is just entanglement of the pure state in
question. That is typically  referred as ``intrinsic'' randomness in the literature, and that is
the point of view we adopt in this report.

\begin{figure}
 \includegraphics[width=35 mm]{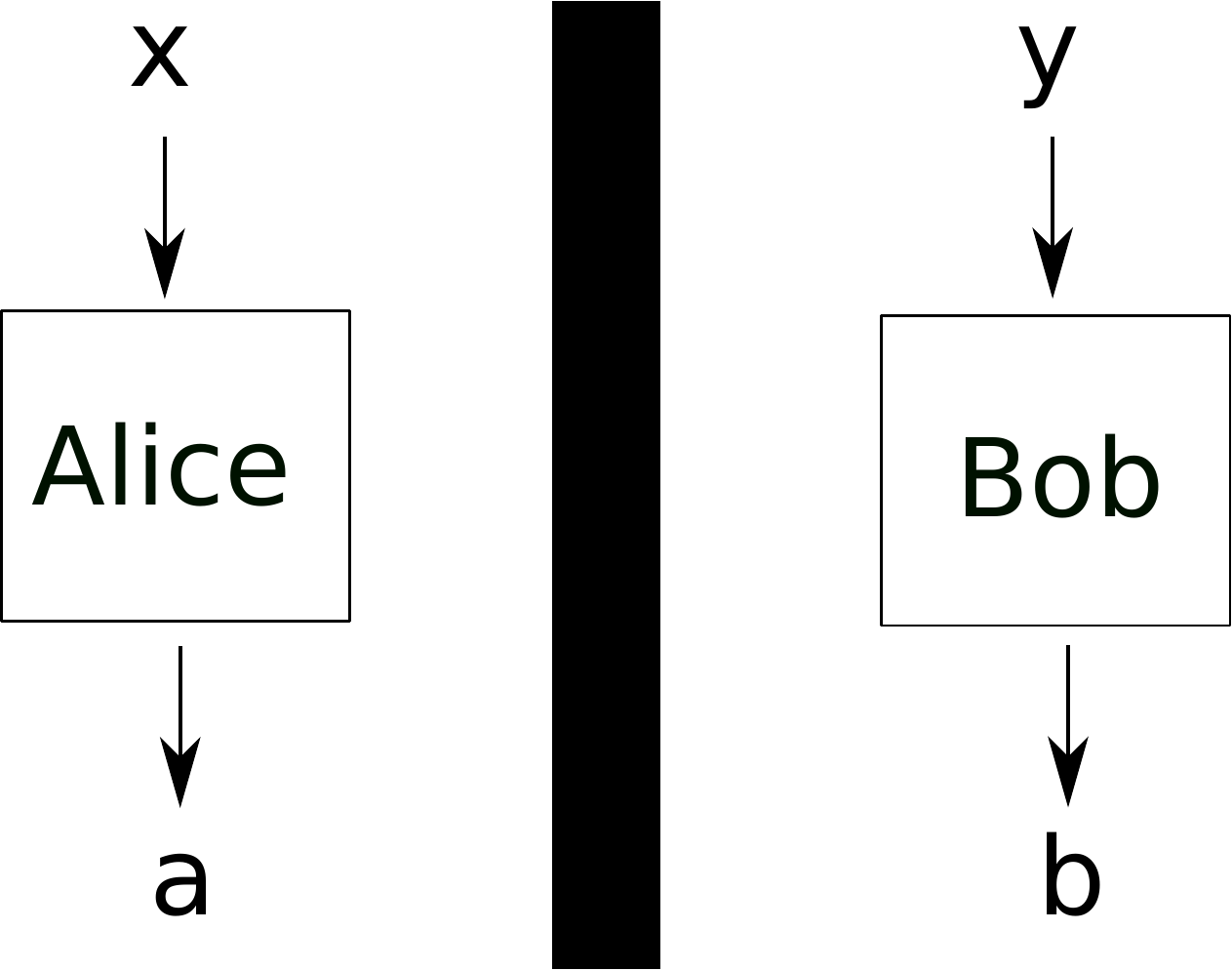}
 \caption{\label{fig:BellEx} Schematic of a two-party Bell-like experiment. The experimenters Alice and Bob are separated and cannot communicate as indicated by the black barrier. The measurements settings and outcomes, of Alice and Bob, are denoted by $x, \ y$ and $a, \ b$ respectively.}
\end{figure}

Before we move further in discussing quantum randomness in the presence
of quantum correlation, let us make a short detour through the history
of foundation of quantum mechanics. The possibility of nonlocal
correlation, also known as quantum entanglement, was initially put
forward with the question if quantum mechanics respects local realism,
by Einstein, Podolsky and Rosen (EPR) \cite{EPR35}. According to EPR,
two main properties any reasonable physical theory should satisfy are
realism and locality. The first one states that if a measurement outcome of a physical
quantity, pertaining to some system, is predicted with unit probability, then there must exits `an element of
reality'' correspond to this physical quantity having a value equal
to the predicted one, at the moment of measurement. 
In other words, the
values of observables, revealed in measurements, are intrinsic
properties of the measured system. The second one, locality, demands
that elements of reality pertaining to one system cannot be affected by
measurement performed on another sufficiently far system. Based on
these two essential ingredients, EPR studied the measurement
correlations between two entangled particles and concluded that
the wave function describing the quantum state ``does not provide a
complete description of physical reality''. Thereby they argued that
quantum mechanics is an incomplete but effective theory and conjectured
that a complete theory describing the physical reality is possible.

In these discussions, one needs to clearly understand what locality and
realism mean. In fact, they could be replaced with  no-signaling and
determinism, respectively. The no-signaling principle states that
infinitely fast communication is impossible. The relativistic
limitation of the speed, by the velocity of light, is just a special
case of no-signaling principle. If two observers are 
separated and perform space-like separated measurements 
(as depicted in Fig. \ref{fig:BellEx}), then the principle
ascertains that the statistics seen by one observer, when measuring her
particle, is completely independent of the measurement choice made on the space-like separated other particle. Clearly, if it
were not the case, one observer could, by changing her measurement
choice, make a noticeable change on the other and thereby
instantaneously communicate with an infinite speed.

Determinism, the other important ingredient, implies that correlations
observed in an experiment can be decomposed as mixtures of
deterministic ones i.e., occurring in situations where all measurements
have deterministic outcomes. A deterministic theory accepts the fact
that the apparent random outcomes in an experiment, like in coin
tossing, are only consequences of ignorance of the actual state of the
system. Therefore, each run of the experiment does have an \textit{a
priori} definite result, but we have only an access to averages.

In 1964, Bell showed that all theories that satisfy locality and
realism (in the sense of EPR) are incompatible with quantum mechanics
\cite{Bell64, Bell66}. In a simple experiment, mimicking Bell's
scenario, two correlated quantum particles are sent to two spatially
separated measuring devices (see Fig. \ref{fig:BellEx}), and each device
can perform two different measurements with two possible outcomes. The
measurement processes are 
space-like separated and no communication is
possible when these are performed. With this configuration a
local-realistic model gives bounds on the correlation between the
outcomes observed in the two measurement devices, called
Bell inequalities \cite{Bell64}. In other words, impossibility of
instantaneous communication (no-signaling) between spatially separated
systems together with full local determinism imply that all correlations
between measurement results must obey the Bell inequalities.

Strikingly, these inequalities are violated with correlated (entangled)
quantum particles, and therefore have no explanations in terms of
deterministic local hidden variables. In fact, the correlations
predicted by the no-signaling and determinism are exactly the same as
predicted by EPR model, and they are equivalent. The experimental
violations of the Bell inequalities in 1972 \cite{FreedmanPRL1972}, in
1981 \cite{Aspect81} and in 1982 \cite{Aspect82}, along with the recent
loophole-free Bell-inequality violations \cite{Hensen15, Giustina15,
Shalm15} confirm that any local-realistic theory is unable to predict
the correlations observed in quantum mechanics.
It immediately implies that either no-signaling or local determinism has to be abandoned. For the most physicists, it is favorable to dump local determinism and save no-signaling. Assuming that the nature
respects no-signaling principle, any violation of Bell inequality
implies thus that the outcomes could not be predetermined in advance.

Thus, once the no-signaling principle is
accepted to be true, the experimental outcomes, due to local
measurements, cannot be deterministic and therefore are random.
Of course, a valid alternative is to abandon the no-signaling principle, allow for non-local hidden variables, but 
save the determinism, as for instance is done in Bohm's theory \cite{Bohm51,Bohm52}. In any case, some kind of non-locality is needed to explain Bell correlations.
One can, also, abandon both no-signaling and  local determinism: such sacrifice is, however, hard to be accepted by majority of physicists, and scientists in general.

Another crucial assumption is considered for Bell experiments, that is
the measurements performed with the local measurement devices have to
be chosen ``freely''. In other words, the measurement choices cannot, in principle, be predicted in advance. If the
free-choice condition is relaxed and the chosen measurements could be
predicted in advance, then it is easy to construct a no-signaling, but
deterministic theory that leads to Bell violations. It has been shown
in \cite{Hall10, Koh12} that one does not have to give up measurement
independence completely to violate Bell inequalities. Even, relaxing
free-choice condition to a certain degree, the Bell inequities could be
maximally violated using no-signaling and deterministic model
\cite{Hall10}. However, in the Bell-like experiment scenarios where the
local observers are  separated, it is very natural to assume
that the choices of the experiments are completely free (this is often
referred to free-will assumption). Therefore, the Bell-inequality
violation in the quantum regime, with the no-signaling principle,
implies that local measurement outcomes are ``intrinsically'' random.

The lesson that we should learn from the above discussion is that the question raised by Einstein, Rosen, Podolsky found its operational meaning in  Bell's theorem that showed incompatibility of hidden-variable theories with quantum mechanics \cite{Bell64}, \cite{Bell66}. Experiment could now decide about existence or non-existence of nonlocal correlations. Exhibiting non-local correlations in an experiment gave, under the assumption of no-signaling, a proof of a non-deterministic nature of quantum mechanical reality, and allowed certifying the existence of truly random processes. These experiments require, however, random adjustments of measuring devices \cite{Bell64}. There must exist a truly random process controlling their choice. This, ironically, closes an unavoidable \textit{circulus vitiosus}. We can check the indeterministic character of the physical reality only assuming that it is, in fact, indeterministic.

\section{Quantum Randomness and Physics}

In this section we consider randomness form the point of view of
physics or in particular, quantum physics. In doing so, first we
briefly introduce  quantum measurements, nonlocality and information
theoretic measures of randomness. Then we turn to outline, how the
quantum feature such as nonlocality can be exploited not only to
generate ``true'' randomness but also to certify, expand and amplify
randomness.

\subsection{Quantum measurements}
According to standard textbook approach, quantum mechanics (QM)  is an
inherently  probabilistic theory (cf.
\cite{Messiah14,Tannoudji91,WheelerZurek83}-- the prediction of QM
concerning results of measurements are typically probabilistic. Only in
very rare instances measurements give deterministic outcomes -- this
happens when the systems is in an eigenstate of an observable to be
measured. Note, that in general, even if we have full information about
the quantum mechanical state of the system, the outcome of the
measurements is in principle random. The paradigmatic example is
provided a $d$-state system (a qudit), whose space of states is spanned
by the states$|1\rangle$, $|2\rangle$,..., $|d\rangle$. Suppose that we
know  the system is in the superposition state
$|\phi\rangle=\sum_{j=1}^d\alpha_j|j\rangle$, where $\alpha_j$ are
complex probability amplitudes and $\sum_{j=1}^d|\alpha_j|^2 =1$, and
we ask whether it is in a state $|i\rangle$. To find out, we measure an
observable $\hat P=|i\rangle \langle i|$ that projects on the state
$|i\rangle$.
The result of such measurement will be one (yes, the system is in the
state $|i\rangle$) with probability $|\alpha_i|^2$ and zero  with
probability $1-\sum_{j \ne i}^d|\alpha_j|^2$.

We do not want to enter here deeply into the subject of the
foundations of QM, but we want to remind the readers  the "standard" approach to QM.

\subsubsection{Postulates of QM}

The postulates of QM for simple mechanical systems (single or many
particle), as given in \cite{Tannoudji91}, read: 
\begin{itemize}
\item {\bf First Postulate.} At a fixed time $t_0$, the state of a
    physical system is defined by specifying a wave function
    $\psi(x; t_0)$, where $x$ represents collection of parameters to specify the state.

\item {\bf Second Postulate.} Every measurable physical quantity
    $Q$ is described by an operator $\hat Q$; this operator is
    called an observable.

\item{\bf Third Postulate.} The only possible result of the
    measurement of a physical quantity $Q$ is one of the
    eigenvalues of the corresponding observable $\hat Q$.

\item {\bf Fourth Postulate (non-degenerate case).} When the
    physical quantity $Q$ is measured on a system in the normalized
    state  $\psi$, the probability $P(q_n)$ of obtaining the
    non-degenerate eigenvalue $q_n$ of the corresponding observable
    $\hat Q$ is
$$P(q_n) = |\int dx \ \varphi_n(x)\psi(x)|^2, $$
where $ \varphi_n$  is the normalized eigenvector of $\hat Q$
associated with the eigenvalue $q_n$.

\item {\bf Fifth Postulate (collapse).} If the measurement of the
    physical quantity $Q$ on the system in the state  $\psi$ gives
    the result $q_n$, the state of the system immediately after the
    measurement is $\varphi_n$.

\item{\bf Sixth Postulate (time evolution).} The time evolution of
    the wave function $\psi(x; t)$ is governed by the
    Schr\"odinger equation
$$i\hbar \frac{\partial \psi}{\partial t}={\hat H}\psi,$$
where $\hat H$ is the observable associated with the total energy
of the system.

\item{\bf Seventh Postulate (symmetrization).} When a system
    includes several identical particles, only certain wave
    functions can describe its physical states (leads to the
    concept of bosons and fermions). For electrons (which are
    fermions), the wave function must change sign whenever the
    coordinates of two electrons are interchanged. For hydrogen
    atoms (regarded as composite bosons) the wave function must not
    change whenever the coordinates of two bosons are interchanged.

\end{itemize}

\subsubsection{Measurement theory}
Evidently, the inherent randomness of QM is associated with the
measurement processes (Fourth and Fifth Postulates). The quantum
measurement theory has been a subject of intensive studies and long
debate, see e.g., \cite{WheelerZurek83}. In particular the question of
the physical meaning of the wave function collapse has been partially
solved only in the last 30 years by analyzing interactions of the
measured system with the environment (reservoir), describing the
measuring apparatus (see seminal works of Zurek
\cite{Zurek03,Zurek09})

In the abstract formulation in the early days of QM, one has considered
von Neumann measurements \cite{Neumann55}, defined in the following
way. Let the observable $\hat Q$ has (possibly degenerated) eigenvalues
$q_n$ and let $\hat E_n$ denote projectors on the corresponding
invariant subspaces (one dimensional for non-degenerate eigenvalues,
$k$-dimensional for $k$-fold degenerated eigenvalues). Since the
invariant subspace are orthogonal, we have $\hat E_n\hat
E_m=\delta_{nm}\hat E_n$, where $\delta_{mn}$ is the Kronecker delta.
If $\hat P_\psi$ denotes the projector, which describes a state of a
system, the measurement outcome corresponds to the eigenvalue $q_n$ of
the observable will appear with probability $p_ n={\rm Tr}(\hat
P_\psi\hat E_n)$, where ${\rm Tr}(.)$ denotes the matrix trace
operation.
Also, after the measurement, the systems is found in the state $\hat E_n\hat P_\psi\hat E_n/p_n$ with probability $p_n$.

In the contemporary quantum measurement theory the measurements are
generalized beyond the von Neumann projective ones.
To define the, so called, positive-operator valued measures (POVM),
one still considers von Neumann measurements, but on a system plus an
additional ancilla system \cite{Peres95}. POVMs
are defined by a set of Hermitian positive semidefinite operators
$\{F_i\}$ on a Hilbert space $\mathcal{H}$ that sum to the identity
operator,

    $$\sum_{i=1}^K F_i = \mathbb{I}_H.$$

This is a generalization of the decomposition of a (finite-dimensional)
Hilbert space by a set of orthogonal projectors, $\{E_i\}$, defined for
an orthogonal basis $\{\left|\phi_{i}\right\rangle\}$ by

$$E_i=\left|\phi_{i}\right\rangle \left\langle\phi_{i}\right|, $$

hence,

$$\sum_{i=1}^N E_i = \mathbb{I}_H, \quad E_i E_j = \delta_{i j}
E_i $$

An important difference is that the elements of POVM are not
necessarily orthogonal, with the consequence that the number $K$ of
elements in the POVM can be larger than the dimension $N$ of the
Hilbert space they act on.

The post-measurement state depends on the way the system plus  ancilla
are measured. For instance, consider the case where the ancilla is
initially a pure state $|0\rangle_B$. We entangle the ancilla with the
system, taking

$$  |\psi\rangle_A |0\rangle_B \rightarrow \sum_i M_i |\psi\rangle_A
|i\rangle_B,$$ and perform a projective measurement on the ancilla in
the $\{|i\rangle_B\}$ basis. The operators of the resulting POVM are
given by

$$F_i = M_i ^\dagger M_i .$$

Since the $M_i$ are not required to be positive, there are an infinite
number of solutions to this equation. This means that there are
infinitely many different experimental apparatus giving the same
probabilities for the outcomes. Since the post-measurement state of the
system (expressed now as a density matrix)

    $$\rho_i = {M_i \rho M_i^\dagger \over {\rm tr}(M_i \rho M_i^\dagger)}$$
depends on the $M_i$, in general it cannot be inferred from the POVM alone.

If we accept quantum mechanics and its inherent randomness, then it is possible in principle to implement measurements of an observable on copies of a state that is not an eigenstate of this observable, to generate a set of  perfect random numbers. Early experiments and commercial devices attempted to mimic a perfect coin with probability 1/2 of getting head and tail. To this aim quantum two-level systems were used, for instance single photons of two orthogonal circular polarizations. If such photons are transmitted through a linear polarizer of arbitrary direction then they pass (do not pass) with probability 1/2.  In practice, the generated numbers are never perfect, and randomness extraction is required to generate good random output.  The challenges of sufficiently approximating the ideal two-level scenario, and the complexity of detectors for single quantum systems, have motivated the development of other randomness generation strategies. In particular, continuous-variable techniques are now several orders of magnitude faster, and allow for randomness extraction based on known predictability bounds. See Section \ref{sec:RandTech}.

It is worth mentioning that the Heisenberg uncertainty relation
\cite{Heisenberg27} and its generalized version, i.e., the
Robertson-Scr\"odinger relation \cite{Robertson29, Schrodinger30,
WheelerZurek83}, often mentioned in the context of quantum
measurements, signify how precisely two non-commuting observables can
be measured on a quantum state. Quantitatively, for a given state
$\rho$ and observables $X$ and $Y$, it gives a lower bound on the
uncertainty when they are measures simultaneously, as
\begin{align}
\delta X^2 \delta Y^2 \geq \frac{1}{4} |\mbox{Tr} \rho \left[X,Y \right]|^2,
\label{eq:RSUR}
\end{align}
where $\delta X^2=\mbox{Tr}\rho X^2-(\mbox{Tr} \rho X)^2$ is the variance
and $\left[X,Y \right]=XY-YX$ is the commutator. A non-vanishing
$\delta X$ represents a randomness in the measurement process and that
may arise from non-commutativity (misalignment in the eigenbases) between
state and observable, or even it may appear due to classical
uncertainty present in the state (i.e., not due to quantum
superposition). In fact Eq. (\ref{eq:RSUR}) does allow to have either
$\delta X$ or $\delta Y$ vanishing, but not simultaneously for a given
state $\rho$ and $\left[X,Y \right]\neq 0$. However, when $\delta X$
vanishes, it is nontrivial to infer on $\delta Y$ and vice versa. To
overcome this drawback, the uncertainty relation is extended to
sum-uncertainty relations, both in terms of variance \cite{Maccone14}
and entropic quantities \cite{Beckner75,Birula75,Deutsch83,Maassen88}.
We refer to \cite{Coles15} for an excellent review on this subject. The
entropic uncertainty relation was also considered in the presence of
quantum memory \cite{Berta10}. It has been shown that, in the presence
of quantum memory, any two observables can simultaneously be measured
with arbitrary precision. Therefore the  randomness appearing in the
measurements can be compensated by the side information stored in the
quantum memory. As we mentioned in the previous section, Heisenberg
uncertainty relations are closely related to the contextuality of
quantum mechanics at the level of single systems. Non-commuting
observables are indeed responsible for the fact that there does not exist
non-contextual hidden variable theories that can explain all results of
quantum mechanical measurements on a given system.

The inherent randomness considered in this work is steaming out the
Born's rule in quantum mechanics, irrespective of the fact if there is
more than one observable being simultaneously measured or not. Furthermore
the existence of nonlocal correlations (and quantum correlations) in
the quantum domain give rise to possibility of, in  a sense, a new
form of randomness. In the following we consider such randomness and
its connection to nonlocal correlations. Before we do so, we shall
discuss nonlocal correlations in more detail.

\subsection{Nonlocality}

\subsubsection{Two-party nonlocality}
Let us now turn to  nonlocality, i.e. property of correlations that violate  Bell inequalities \cite{Bell64, Brunner14}. As we will see below, nonlocality is intimately connected to the intrinsic quantum randomness. In the traditional scenario a Bell nonlocality test relies
on two spatially separated observers, say Alice and Bob, who perform space-like measurements on a bipartite system possibly produced by a common source. For a schematic see Fig. \ref{fig:BellEx}. Suppose Alice's measurement choices are $x\in \mathcal{X}=\{1,\ldots,M_A\}$ and Bob's choices $y\in \mathcal{Y}=\{1,\ldots,M_B\}$ and the corresponding outcomes $a\in \mathcal{A}=\{1,\ldots,m_A\}$ and $b\in \mathcal{B}=\{1,\ldots,m_B\}$ respectively. After repeating many times, Alice and Bob communicate their measurement settings and outcomes to each other and estimate the joint probability
$p(a,b|x,y)=p(A=a,B=b|X=x,Y=y)$ where $X$, $Y$ are the random variables that govern the inputs and $A$, $B$ are the random variables that govern the outputs. The outcomes are considered to be correlated, for some $x,y,a,b$, if
\begin{align}
 p(a,b|x,y)\neq p(a|x)p(b|y).
\end{align}
Observing such correlations is not surprising as there are many classical sources and  natural processes that lead to correlated statistics. These can be modeled with the help of another random variable $\Lambda$ with the outcomes $\lambda$, which has a causal influence on both the measurement outcomes and is inaccessible to the observers or ignored. 

In a {\it local hidden-variable model}, considering all possible causes $\Lambda$, the joint probability can then be expressed as
\begin{align}
 p(a,b|x,y,\lambda) = p(\lambda) p(a|x, \lambda)p(b|y,\lambda ).
\end{align}
One, thereby, could explain any observed correlation in accordance with
the fact that Alice's outcomes solely depends on her local measurement
settings $x$, on the common cause $\lambda$,  and are independent of
Bob's measurement settings. Similarly Bob's outcomes are independent of
Alice's choices. This assumption -- the no-signaling condition -- is
crucial -- it is required by the theory of relativity, where nonlocal
causal influence between space-like separated events is forbidden.
Therefore, any joint probability, under the {\it local hidden-variable
model}, becomes
\begin{align}
 p(a,b|x,y) = \int_{\Lambda} d\lambda p(\lambda) p(a|x, \lambda)p(b|y,\lambda ),
 \label{eq:lhv}
\end{align}
with the implicit assumption that the measurement settings $x$ and $y$
could be chosen independently of $\lambda$, i.e.,
$p(\lambda|x,y)=p(\lambda)$. Note that so far we have  not assumed
anything about the nature of the local measurements, whether they are
deterministic or not. In a {\it deterministic local hidden-variable
model}, Alice's outcomes are completely determined by the choice $x$
and the $\lambda$. In other words, for an outcome $a$, given input $x$
and hidden cause $\lambda$, the probability $p(a|x,\lambda)$ is either
$1$ or $0$ and so as for Bob's outcomes. Importantly, the {\it
deterministic local hidden-variable model} has been shown to be fully
equivalent to the {\it local hidden-variable model} \cite{Fine82}.
Consequently, the observed correlations that admit a join probability
distribution as in (\ref{eq:lhv}), can have an explanation based on a
{\it deterministic local hidden-variable model}. 

In 1964, Bell showed
that any {\it local hidden-variable model} is bound to respect a set of
linear inequalities, which are commonly know as Bell inequalities. In
terms of  joint probabilities they can be expressed  as
\begin{align}
 \sum_{a,b,x,y} \alpha^{xy}_{ab} \ p(a,b|x,y) \leq \mathcal{S}_L,
 \label{eq:bi}
\end{align}
where $\alpha^{xy}_{ab}$ are some coefficients and $\mathcal{S}_L$ is
the classical bound. Any violation of Bell inequalities (\ref{eq:bi})
implies a presence of correlations that cannot be explained by a  {\it
local hidden-variable model}, and therefore have a nonlocal character.
Remarkably, there indeed exists correlations
violating Bell inequalities that could be observed with certain choices of
local measurements on a quantum system, and hence do not admit a {\it
deterministic local hidden-variable model}.

To understand it better, let us consider an example of  the most
studied two-party Bell inequalities, also known as Clauser-Horne-Shimony-Holt (CHSH) inequalities,
introduced in \cite{Clauser69}. Assume the simplest scenario (as in
Fig.~\ref{fig:BellEx}) in which  Alice and Bob both choose one of two
local measurements $x,y \in \{0,1\}$ and obtain one of two measurement
outcomes $a,b \in \{-1,1\}$. Let the expectation values of the local
measurements are $\langle a_x b_y \rangle = \sum_{a,b} a\cdot b \cdot
p(a,b|x,y)$, then the CHSH-inequality reads:
\begin{align}
 I_{CHSH}=\langle a_0 b_0 \rangle + \langle a_0 b_1 \rangle + \langle
 a_1 b_0 \rangle - \langle a_1 b_1 \rangle \leq 2.
 \label{eq:CHSHprob}
\end{align}
One can maximize $I_{CHSH}$ using local deterministic strategy and to
do so one needs to achieve the highest possible values of $\langle a_0 b_0
\rangle, \ \langle a_0 b_1 \rangle, \ \langle a_1 b_0 \rangle$ and the
lowest possible value of $ \langle a_1 b_1 \rangle$. By choosing
$p(1,1|0,0)=p(1,1|0,1)=p(1,1|1,0)=1$, the first three expectation
values can be maximized. However, in such situation the $p(1,1|1,1)=1$
and $I_{CHSH}$ could be saturated to 2. Thus the inequality is
respected. However, it can be violated in a quantum setting. For
example, considering a quantum state
$|\Psi^+\rangle=\frac{1}{\sqrt{2}}(|00\rangle + |11\rangle)$ and
measurement choices $A_0=\sigma_z$, $A_1=\sigma_x$,
$B_0=\frac{1}{\sqrt{2}}(\sigma_z+\sigma_x)$,
$B_1=\frac{1}{\sqrt{2}}(\sigma_z-\sigma_x)$
one could check that for the quantum expectation values $\langle
a_\alpha b_\beta \rangle = \langle \Psi^+ | A_\alpha \otimes B_\beta
|\Psi^+ \rangle $ we get $I_{CHSH}=2\sqrt{2}$. Here $\sigma_z$ and
$\sigma_x$ are the Pauli spin matrices and $|0\rangle$ and $|1\rangle$
are two eigenvectors of $\sigma_z$. Therefore the joint probability
distribution $p(a,b|x,y)$ cannot be explained in terms of local
deterministic model.

\subsubsection{Multi-party nonlocality and device independent approach}
Bell-type inequalities can be also constructed in multi-party scenario.
Their violation signifies nonlocal correlations distributed over many
parties. A detailed account may be found in \cite{Brunner14}.

Here we introduce the concept of nonlocality using the contemporary
language of device independent approach (DIA) \cite{Brunner14}. Recent
successful hacking attacks on quantum cryptographic devices
\cite{Lydersen10} triggered this novel approach to quantum information
theory in which protocols are defined independently of the inner
working of the devices used in the implementation. That leads to avalanches of works in the field of
device independent quantum information processing and technology
\cite{Brunner14a, Pironio15}.

\begin{figure}
\begin{center}
\includegraphics[width=0.25\textwidth]{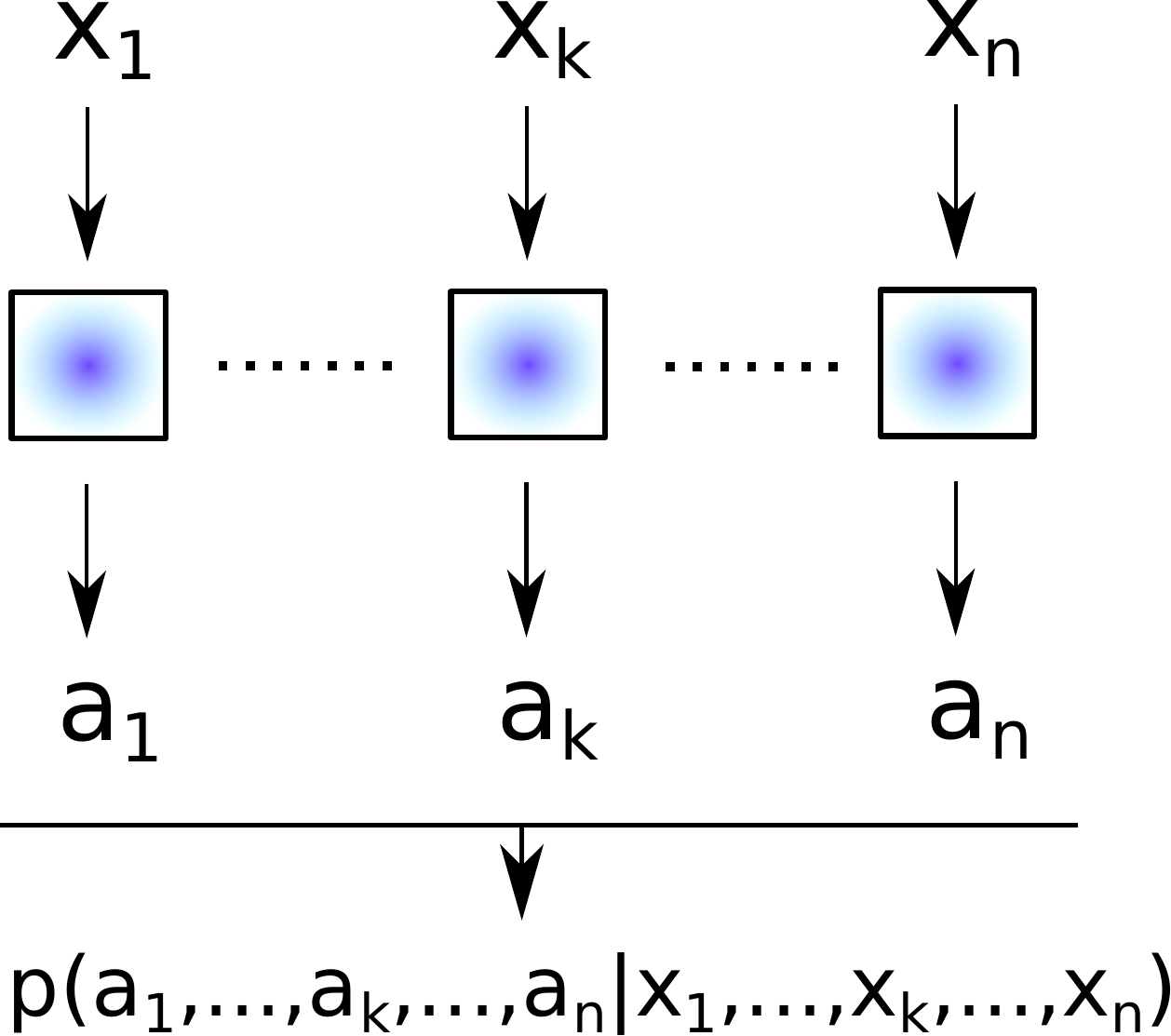}
\end{center}
\caption{\label{fig:Bell1} Schematic representation of device independent approach. In this approach several users could access to uncharacterized black-boxes (shown as squares) possibly prepared by an adversary. The users are allowed to choose inputs $\left(x_1, \hdots, x_k, \hdots, x_n\right)$ for the boxes and acquire outputs $\left(a_1, \hdots, a_k, \hdots, a_n\right)$ as results. The joint probability with which the outputs appear is $p\left(a_1, \hdots, a_k, \hdots, a_n |  x_1, \hdots, x_k, \hdots, x_n\right)$.}
\end{figure}

The idea of DIA is schematically given in Fig. \ref{fig:Bell1}. We
consider here the following scenario, usually referred to as the
\textit{$(n,m,d)$ scenario}. Suppose $n$ spatially separated parties
$A_1,\ldots,A_n$. Each of them possesses a black box with $m$
measurement choices (or observables) and $d$  measurement outcomes.
Now, in each round of the experiment every party is allowed to perform one
choice of measurement and acquires one outcome. The accessible
information, after the measurements, is contained in a set of $(md)^n$
conditional probabilities $p(a_1,\ldots, a_n|x_1,\ldots, x_n)$ of
obtaining outputs $a_1, a_2, \ldots, a_n$, provided observables $x_1,
x_2, \ldots, x_n$ were measured.
The set of all such probability distributions forms a convex set; in fact, it is a polytope in the probability manifold. From the physical point of view (causality, special relativity) the probabilities must fulfill the {\it no-signaling condition}, i.e., the choice of measurement by the $k$-th party, cannot be instantaneously signalled to the others. Mathematically it means that for any $k=1,\ldots,n$, the following condition
\begin{align}
\label{eq:no-sig}
&\sum_{a_k}p(a_1, \ldots,  a_k, \ldots, a_n|x_1,\ldots, x_k,\ldots, x_n)\nonumber\\
& = p(a_1,\ldots,  a_{k-1}, a_{k+1}\ldots, a_n|x_1, \ldots, x_{k-1},x_{k+1}\ldots, x_n), 
\end{align}
is fulfilled.

The \textit{local correlations} are defined via the concept of a local
hidden variable $\lambda$ with the associated probability
$q_{\lambda}$.
The correlations that the parties are able to establish in such case
are of the form

\begin{align}
p(a_1,  & \ldots, a_n|x_1,  \ldots,  x_n) \nonumber \\
 & = \sum_{\lambda} q_\lambda D(a_1|x_1,\lambda)  \ldots D(a_n|x_n,\lambda),
\end{align}
where $D(a_k|x_k,\lambda)$ are deterministic probabilities, i.e., for
any $\lambda$, $D(a_k|x_k,\lambda)$ equals one for some outcome, and
zero for all others. What is important in this expression is that
measurements of different parties are independent, so that the
probability is a product of terms corresponding to different parties.
In this $n$-party scenario the local hidden variable model bounds the
joint probabilities to follow the Bell inequalities, given as
\begin{align}
 \sum_{a_1, \ldots, a_n,x_1,  \ldots,  x_n} \alpha_{a_1, \ldots,
 a_n}^{x_1,  \ldots,  x_n} \ p(a_1,\ldots, a_n|x_1, & \ldots,  x_n)
 \nonumber \\
  &\leq  \mathcal{S}_L^n,
\end{align}
where $\alpha_{a_1, \ldots, a_n}^{x_1,  \ldots,  x_n}$ are some
coefficients and $\mathcal{S}_L^n$ is the classical bound.

\begin{figure}
\begin{center}
\includegraphics[width=0.48\textwidth]{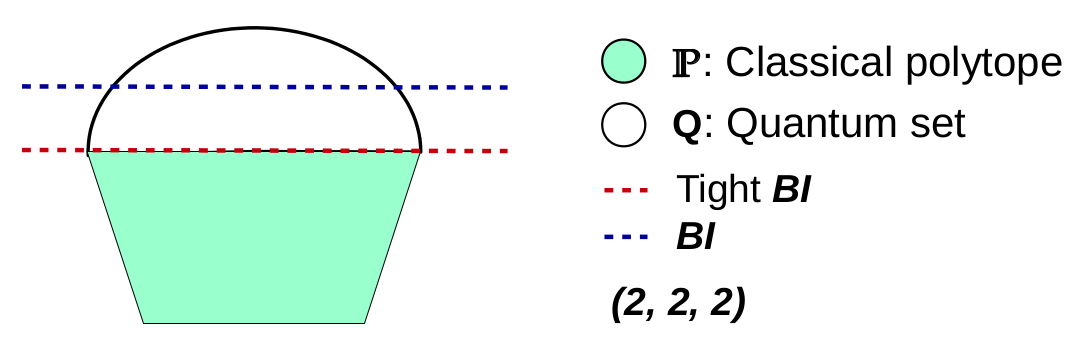}
\end{center}
\caption{Schematic representation of different sets of correlations:
classical (grey area) and quantum (the area bounded by the thick line).
Clearly, the former is the subset of the latter and, as has been shown
by Bell \cite{Bell64}, they are not equal -- there are quantum
correlations that do not fall into the grey area. The black dots
represent the vertices of the classical polytope $\mathbbm{P}$ --
deterministic classical correlations -- satisfying deterministic local
hidden variable models. The dashed lines represent Bell inequalities.
In particular, the black dashed line is tight and it corresponds to the
facet of the classical set.\label{fig:zbiory}}
\end{figure}

The probabilities that follow local (classical) correlations form a
convex set that is also a polytope, denoted $\mathbbm{P}$ (cf.
Fig.~\ref{fig:zbiory}). Its extremal points (or vertices) are given by
$\prod_{i=1}^n D(a_i|x_i,\lambda)$ with fixed $\lambda$.
The Bell theorem states that the quantum-mechanical probabilities,
which also form a convex set $\mathcal{Q}$, may stick out of the
classical polytope \cite{Bell64, Fine82}. The quantum probabilities are
given by the trace formula for the set of local measurements
\begin{equation}
p(a_1, \ldots, a_n | x_1, \ldots, x_n)={\rm Tr}(\rho \otimes_{i=1}^n
M_{a_i}^{x_i}),
\end{equation}
where $\rho$ is some $n$-partite state and $M_{a_i}^{x_i}$ denote the
measurement operators (POVMs) for any choice of the measurement $x_i$
and party $i$.
As we do not impose any constraint on the local dimension, we can
always choose the measurements to be projective, i.e., the measurement
operators additionally satisfy
$M_{a'_i}^{x_i}M_{a_i}^{x_i}=\delta_{a'_i,a_i} M_{a_i}^{x_i}$.

This approach towards the Bell inequalities is explained in
Fig.~\ref{fig:zbiory}. Any hyperplane in the space of probabilities
that separates the classical polytope from the rest determines a Bell
inequality: everything that is above the upper horizontal dashed line
is obviously nonlocal. But the most useful are the {\it tight Bell
inequalities} corresponding to the facets of the classical polytope,
i.e. its walls of maximal dimensions (lower horizontal dashed line).

In general $(n,m,d)$ scenarios, the complexity of characterizing the
corresponding classical polytope is enormous. It is fairly easy to see
that, even for $(n,2,2)$, the number of its vertices (extremal points)
is equal to $2^{2n}$, hence it grows exponentially with $n$.
Nevertheless, a considerable effort has been made in recent time to
characterize multi-party nonlocality \cite{Brunner14, Tura14a, Tura14,
Liang15, Tura15, Rosicka16}.

Among the many other device independent applications, the nonlocality
appears to be a valuable resource in random number generation,
certification, expansion and amplification, which we outline in the
following subsections. In fact, it has been shown that Bell
nonlocal correlation is a genuine resource, in the framework of a
resource theory, where the allowed operations are restricted to device
independent local operations \cite{GallegoPRL2012, Vicente14}.

\subsection{Randomness: information theoretic approach\label{sec:Randomness}}

Before turning to the quantum protocols involving randomness, we
discuss in this section randomness from the  information theory
standpoint. It is worth mentioning the role of randomness in various
applications, beyond its fundamental implications. In fact 
randomness is a  resource in many different areas -- for a good
overview see Refs.  \cite{Motwani95,Menezes96}. Random numbers play
important role in cryptographic applications, in  numerical simulations
of complex physical, chemical, economical, social and biological
systems, not to mention gambling.  That is why, much efforts were put
forward to (1) develop good, reliable sources of random numbers, and
(2) to design reliable certification tests for a given source of random
numbers. 

In general, there exists three types of random number
generators (RNG). They are ``true'' RNGs, pseudo-RNGs and the
quantum-RNGs. The true RNGs are based on some physical processes that
are hard to predict, like noise in electrical circuits, thermal or
atmospheric noises, radioactive decays etc. The pseudo-RNGs rely on the
output of a deterministic function with a shorter random seed possibly
generated by a true RNG. Finally, quantum RNGs use genuine quantum
features to generate random bits.

We consider here  a finite sample space and denote it  by the set
$\Omega$. The notions of ideal and weak random strings
describe distributions over $\Omega$ with certain properties. When a
distribution is {\it uniform} over $\Omega$, we say that it has ideal
randomness. A uniform distribution over $n$-bit strings is denoted by
$U_n$. The uniform distributions are very natural to work with.
However, when we are working with physical systems, the processes or
measurements are usually biased. Then the bit strings resulting from
such sources are not uniform. A string with nonuniform distribution,
due to some bias (could be unknown), is referred to have {\it weak}
randomness and the sources of such strings are termed as weak sources.

Consider the random variables denoted by the letters $(X,Y,\ldots)$.
Their values will be denoted by $(x,y,\ldots)$. The probability of a
random variable $X$ with a value $x$ is denoted as $p(X=x)$ and when
the random variable in question is clear we use the shorthand notation
$p(x)$.
Here we briefly introduce the operational approach to define randomness
of a random variable. In general, the degree of randomness or bias of a
source is unknown and it is insufficient to define a weak source by a
random variable $X$ with a probability distribution $P(X)$. Instead one
needs to model the weak randomness by a random variable with an unknown
probability distribution. In another words, one need to characterize a
set of probability distributions with desired properties. If we suppose
that the probability distribution $P(X)$ of the variable $X$ comes from
a set $\Omega$, then the degree of randomness is determined  by the
properties of the set, or more specifically, by the least random
probability distribution(s) in the set. The types of weak randomness
differ with the types of distribution $P(X)$ on  $\Omega$ and the set
$\Omega$ itself -- they are determined  by the allowed distributions
motivated by a physical source. There are many ways to classify the
weak random sources, and an interested reader may go through Ref.
\cite{Pivoluska14}. Here we shall consider two types
of weak random sources, Santha-Vazirani (SV) and Min-Entropy (ME)
sources, which will be sufficient for our later discussions.

A  Santha-Vazirani (SV) source \cite{Santha86} is defined as a sequence
of binary random variables $(X_1, X_2,\ldots,X_n)$, such that
\begin{align} \label{eq:SVsource}
  \frac{1}{2} - \epsilon \leq & p(x_i=1|x_1,\ldots,x_{i-1}) \leq \frac{1}{2} + \epsilon, \\
 &  \forall i \in \mathbb{N}, \forall x_1,\ldots,x_{i-1} \in \{0,1\} \nonumber,
\end{align}
where the conditional probability $p(x_i=1|x_1,\ldots,x_{i-1})$ is the
probability of the value $x_i=1$ conditioned on the values
$x_1,\ldots,x_{i-1}$. The $0 \leq \epsilon \leq \frac{1}{2}$ represents
bias of the source. For fixed $\epsilon$ and $n$, the SV-source
represents a set of probability distributions over $n$-bit strings. If
a random string satisfies (\ref{eq:SVsource}), then we say that it is
$\epsilon$-free.  For $\epsilon=0$ the string is perfectly random --
uniformly distributed sequence of bits $U_n$. For
$\epsilon=\frac{1}{2}$, nothing can be inferred about the string and it
can be even deterministic.  Note that in SV sources the bias can not
only change for each bit $X_i$, but it also can depend on the
previously generated bits. It requires that each produced bit must have
some amount of randomness, when $\epsilon \neq \frac{1}{2}$, and even
be conditioned on the previous one.

In order the generalize it further one considers {\it block} source
\cite{Chor88}, where the randomness is not guaranteed in every single
bit, but rather for a block of $n$-bits. Here, in general, the
randomness is quantified by the min-entropy, which is defined as:
\begin{align}
 H_{\infty}(Y)=\mbox{min}_{y} [-\mbox{log}_2(p(Y=y))],
\end{align}
for an $n$-bit random variable $Y$. For a block source, the randomness
is guaranteed by the most probable  $n$-bit string appearing in the
outcome of the variable -- simply by guessing the most probable element
-- provided that the probability is less than one. A {\it block}
$(n,k)$ source can now be modeled, for $n$-bit random variables
$(X_1,X_2,...,X_n)$, such that
\begin{align}
 H_{\infty}&(X_i|X_{i-1}=x_{i-1},...,X_1=x_1)\geq k, \\
 & \forall i \in \mathbb{N}, \forall x_1,...,x_{i-1} \in \{0,1\}^n.\nonumber
\end{align}
These block sources are generalizations of SV-sources; the latter are
recovered with $n=1$ and $\epsilon = 2^{-H_{\infty}(X)}-\frac{1}{2}$.
The block sources can  be further generalized to sources of randomness
of finite output size, where no internal structure is given, e.g.,
guaranteed randomness in every bit (SV-sources) or every block of
certain size (block sources). The randomness is only guaranteed by its
overall min-entropy. Such sources are termed as the {\it min-entropy}
sources \cite{Chor88} and are defined, for an $n$-bit random variable
$X$,  such that
\begin{align}
 H_{\infty}(X)\geq k.
\end{align}
Therefore, a min-entropy source represents a set of probability
distributions where the randomness is upper-bounded by the probability
of the most probable element, measured by min-entropy.

Let us now briefly outline the {\it randomness extraction (RE)}, as it
is one of the most common operations that is applied in the
post-processing of weak random sources. The randomness extractors are
the algorithms that produce nearly perfect (ideal) randomness useful
for potential applications. The aim of RE is to convert randomness of a
 string from a weak source into a possibly shorter string of bits
that is {\it close} to a perfectly random one. The closeness is defined
as follows. The random variables $X$ and $Y$ over a same domain
$\Omega$ are $\varepsilon$-close, if:
\begin{align}
 \Delta(X,Y)=\frac{1}{2}\sum_{x \in \Omega}| p(X=x)-p(Y=x)| \leq \varepsilon.
\end{align}
With respect to RE, the weak sources can be divided into two classes --
extractable sources and non-extractable sources. Only from extractable
sources a perfectly random string can be extracted by a deterministic
procedure. Though there exist many non-trivial extractable sources (see
for example \cite{Kamp11}), most of the natural sources, defined by
entropies, are non-extractable and in such cases non-deterministic
({\it stochastic}) randomness extractors are necessary.

Deterministic extraction fails for the random strings from SV-sources,
but it is possible to post-process them with a help of an additional
random string. As shown in \cite{Vazirani87}, for any $\epsilon$ and
two mutually independent $\epsilon$-free strings from SV-sources, it is
possible to efficiently extract a single almost perfect bit
($\epsilon^\prime \rightarrow 0$). For two $n$-bit independent strings
$X=(X_1,...,X_n)$ and $Y=(Y_1,...Y_n)$, the post-processing function,
{\it Ex}, has been defined as
\begin{align}
 Ex(X,Y)=(X_1\cdot Y_1)\oplus(X_2 \cdot Y_2)\oplus \cdots \oplus (X_n \cdot Y_n),
\end{align}
where $\oplus$ denotes the sum modulo 2. The function {\it Ex} is the
inner product between the $n$-bit strings $X$ and $Y$ modulo 2. 
Randomness extraction of SV-sources are sometime referred as {\it
randomness amplification} as two $\epsilon$-free strings from
SV-sources are converted to one-bit string of  $\epsilon^\prime$-free
and with $\epsilon^\prime < \epsilon$.

Deterministic extraction is also impossible for the min-entropy
sources. Nevertheless, an extraction might be possible with the help of
{\it seeded extractor} in which an extra resource of uniformly
distributed string, called the seed, is exploited. A function, $Ex: \
\{0,1\}^n \ \times \ \{0,1\}^r \mapsto \{0,1\}^m $ is seeded
$(k,\varepsilon)$-extractor, for every string from block $(n,k)$-source
of random variable $X$, if
\begin{align}
 \Delta (Ex(X,U_r),U_m) \leq \varepsilon.
\end{align}
Here $U_r$ ($U_m$) is the uniformly distributed $r$-bit ($m$-bit)
string. In fact, for a variable $X$, min-entropy gives the upper bound
on the number of extractable perfectly random bits \cite{Shaltiel02}.
Randomness extraction is well developed area of research in classical
information theory. There are many  randomness extraction techniques
using multiple strings \cite{Dodis04,Raz05,Barak10,
Nisan99,Shaltiel02,Gabizon08}, such as universal hashing extractor,
Hadamard extractor, DEOR extractor, BPP extractor etc.,  useful for
different post-processing.

\subsection{Nonlocality, random number generation and certification}
Here we link the new form of randomness, i.e the   presence of
nonlocality (in terms of Bell violation) in the quantum regime, to
random number generation and certification. To do so, we outline how
nonlocal correlations can be used to generate new types of random
numbers, what has been experimentally demonstrated in \cite{Pironio10}.
Consider the Bell-experiment scenario (Fig.~(\ref{fig:BellEx})), as
explained before. Two  separated observers perform different
measurements, labeled as $x$ and $y$, on two quantum particles in their
possession and get measurement outcomes $a$ and $b$, respectively. With
many repetitions they can estimate the joint probability, $p(a,b|x,y)$,
for the outcomes $a$ and $b$ with the measurement choices $x$ and $y$.
With the joint probabilities the observers could check if the
Bell inequalities are respected. If a violation is observed the
outcomes are guaranteed to be random. The generation of these random
numbers is independent of working principles of the measurement
devices. Hence, this is a  device independent random number generation.
In fact, there is a quantitative relation between the amount of
Bell-inequality violation and the observed randomness. Therefore,
these random numbers could be (a) certifiable, (b) private, and (c)
device independent \cite{Colbeck07,Pironio10}. The resulting string of random
bits, obtained by $N$ uses of measurement device, would be made up of
$N$ pair of outcomes, $(a_1,b_1,...,a_N,b_N)$, and their randomness
could be guaranteed by the violation of Bell inequalities. 

There is however an important point to be noted. \textit{A priori}, the
observers do not know whether the measurement devices violate
Bell inequalities or not. To confirm they need to execute statistical
tests, but such tests cannot be carried out in predetermined way. Of
course, if the measurement settings are known in advance, then an
external agent could prepare the devices that are completely
deterministic and the Bell-inequality violations could be achieved
even in the absence of nonlocal correlations. Apparently there is a
contradiction between the aim of making random number generator and the
requirement of random choices to test the nonlocal nature of the
devices. However, it is natural to assume that the observers can make
free choices when they are  separated.

Initially it was speculated that the more particles are nonlocally
correlated (in the sense of Bell violation), the stronger would be the
observed randomness. However, this intuition is not entirely correct,
as shown in \cite{Acin12} -- a maximum production of random bits could
be achieved with a non-maximal CHSH violation. To establish a
quantitative relation between the nonlocal correlation and generated
randomness, let us assume that the devices follow quantum mechanics.
There exist thus  a quantum state $\rho$ and measurement operators
(POVMs) of each device $M^x_a$ and $M^y_b$ such that the joint
probability distribution $P(a,b|x,y)$ could be expressed, through Born
rule, as
\begin{equation}
 P_Q(a,b|x,y)=\mbox{Tr}(\rho M^x_a \otimes M^y_b),
\end{equation}
where the tensor product reflects the fact that the measurements are
local, i.e., there are  no interactions between the devices, while the
measurement takes place. The set of quantum correlations consists of
all such probability distributions. Consider a linear combinations of
them,
\begin{equation}
 \sum_{x,y,a,b} \alpha^{xy}_{ab} P_Q(a,b|x,y) = \mathcal{S},
\end{equation}
specified by real coefficients $\alpha^{xy}_{ab}$. For local
hidden-variable model, with certain coefficients $\alpha^{xy}_{ab}$,
the Bell inequalities can be then expressed as
\begin{equation}
 \mathcal{S} \leq \mathcal{S}_L.
 \label{eq:BI}
\end{equation}
This bound could be violated ($\mathcal{S} > \mathcal{S}_L$) for some
quantum states and measurements indicating that the state contains
nonlocal correlation.

Let us consider the the measure of randomness quantified by
min-entropy. For a $d$-dimensional probability distribution $P(X)$,
describing a random variable $X$, the min-entropy is defined as
$H_{\infty}(X)=-\mbox{log}_2\left[ \mbox{max}_x p(x) \right]$.
%which measures in bits.
Clearly, for the perfectly deterministic distribution this maximum
equals one and the min-entropy is zero. On the other hand, for a
perfectly random (uniform) distribution, the entropy acquires the
maximum value, $\mbox{log}_2 d$. In the Bell scenario, the randomness
in the outcomes, generated by the pair of measurements $x$ and $y$,
reads $H_{\infty}(A,B|x,y)=-\mbox{log}_2 c_{xy}$, where
$c_{xy}=\mbox{max}_{ab}P_Q(a,b|x,y)$. For a given observed value
$\mathcal{S} > \mathcal{S}_L$, violating Bell inequality, one could
find a quantum realization, i.e., the quantum states and set of
measurements, that minimizes the min-entropy of the outcomes
$H_{\infty}(A,B|x,y)$ \cite{Navascues08}. Thus, for any violation of
Bell inequalities, the randomness of a pair of outcomes satisfies
\begin{equation}
 H_{\infty}(A,B|x,y) \geq f(\mathcal{S}),
 \label{eq:RandBound1}
\end{equation}
where $f$ is a convex function and vanishes for the case of no
Bell-inequality violation, $\mathcal{S} \leq \mathcal{S}_L$. Hence,
the (\ref{eq:RandBound1}) quantitatively states that {\it a violation
of Bell inequalities guarantees some amount randomness}. Intuitively,
if the joint probabilities admit (\ref{eq:BI}), then for each setting
$x, \ y$ and a hidden cause $\lambda$, the outcomes $a$ and $b$ can be
deterministically assigned. However, the violation of (\ref{eq:BI})
rules out such possibility. As a consequence, the observed correlation
cannot be understood with a deterministic model and the outcomes are
fundamentally undetermined at the local level.

Although there are many different approaches to generate random
numbers \cite{Marsaglia08, Bassham10}, the certification of randomness is highly non-trivial. 
However, this
problem could be solved, in one stroke, if the random sequence shows a
Bell violation, as it certifies a new form of ``true'' randomness that
has no deterministic classical analogue.

\subsection{Nonlocality and randomness expansion}
Nonlocal correlations can be also used for the construction of novel types of 
randomness expansion protocols. In these protocols, a user 
expands an initial random string, known as seed, into a larger string of
random bits. Here, we focus on protocols that achieve this expansion by using 
randomness certified by a Bell inequality violation. Since the first proposals in 
Refs.~\cite{Colbeck07,Pironio10}, there have been 
several different works studying Bell-based randomness expansion protocols, see for 
instance~\cite{Colbeck07, Pironio10, Colbeck10, Vazirani12, coudron_yuen, miller_shi, Chung14, miller_shi2,EATQKD}.
It is not the scope of this section to review the contributions of all these works, which
in any case should be interpreted as a representative but non-exhaustive list. However,
most of them consider protocols that have the structure described in what follows.
Note that the description aims at providing the main intuitive steps in a general randomness
expansion protocols and technicalities are deliberately omitted (for details see the 
references above).

The general scenario consists of a user who is able to run a Bell test. He thus has
access to $n\geq 2$ devices where he can implement $m$ local measurements of $d$ outputs.
For simplicity, we restrict the description in what follows to protocols involving only
two devices, which are also more practical form an implementation point of view.
The initial seed is used to randomly choose the local measurement settings in the Bell experiment. 
The choice of settings does not need to be uniformly random. In fact, in many situations, there 
is a combination of settings in the Bell test that produces more randomness than the rest. It is then convenient
to bias the choice of measurement towards these settings so that (i) the amount of random bits
consumed from the seed, denoted by $N_s$, is minimize and 
(ii) the amount of randomness produced during the Bell test is maximized.

The choice of settings is then used to perform the Bell test. After $N$ repetitions of the Bell
test, the user acquires enough statistics to have a proper estimation of the non-locality of the generated 
data. If not enough confidence about a Bell violation is obtained in this process, the protocol is aborted 
or more data are generated. From the observed Bell violation, it is possible to bound the amount 
of randomness in the generated bits.
This is often done by means of the so-called min-entropy, $H_{\infty}$. In general,
for a random variable $X$, the min-entropy is expressed in bits and is
equal to $H_{\infty}=-\log_2\max_x P(X=x)$.
The observed Bell violation is used to establish a lower bound on the min-entropy
of the generated measurement outputs. Usually, after this process, the user concludes
that with high confidence the $N_G\leq N$ generated bits have an entropy 
at least equal to $R\leq N_g$, that is $H_{\infty}\geq R$.

This type of bounds is useful to run the last step in the protocol: the final randomness
distillation using a randomness extractor~\cite{randextr,Nisan99}. This consists of classical post-processing 
of the measurement outcomes, 
in general assisted by some extra $N_e$ random bits from the seed, which map
the $N_g$ bits with entropy at least $R$ to $R$ bits with the same entropy, that is,
$R$ random bits. Putting all the things together, the final expansion of the protocols is 
given by the ration $R/(N_s+N_e)$.

Every protocol comes with a security proof, which guarantees that the final list of 
$R$ bits is unpredictable to any possible observer, or eavesdropper, who could share correlated
quantum information with the devices used in the Bell test. Security proofs are also
considered in the case of eavesdroppers who can go beyond the quantum formalism,
yet without violating the no-signaling principle. All the works mentioned above represent
important advances in the design of randomness expansion protocols. At the moment, it is for instance known that
(i) any Bell violation is enough to run a randomness expansion protocol~\cite{miller_shi} and (ii)
in the previous simplest configuration presented here, there exist protocol attaining an 
exponential randomness expansion~\cite{Vazirani12}. More complicated variants, where devices
are nested, even attain an unbounded expansion~\cite{coudron_yuen, Chung14}.
Before concluding, it is worth mentioning that another interesting scenario 
consists of the case in which a trusted source of public randomness is available. 
Even if public, this trusted randomness can safely be used to run the Bell test and 
does not need to be taken into account in the final rate.

\subsection{Nonlocality and randomness amplification}
Here we discuss the usefulness of nonlocal correlation for randomness
amplification, a task related but in a way complementary to randomness expansion. 
While in randomness expansion one assumes the existence of an initial list
of perfect random bits and the goal is to generate a longer list, in randomness
amplification the user has access to a list of imperfect randomness and
the goal is to generate a source of higher, ideally arbitrarily good, quality.
As above, the goal is to solve this information task by exploiting Bell violating correlations.

\begin{figure}[h]
 \includegraphics[width=70 mm]{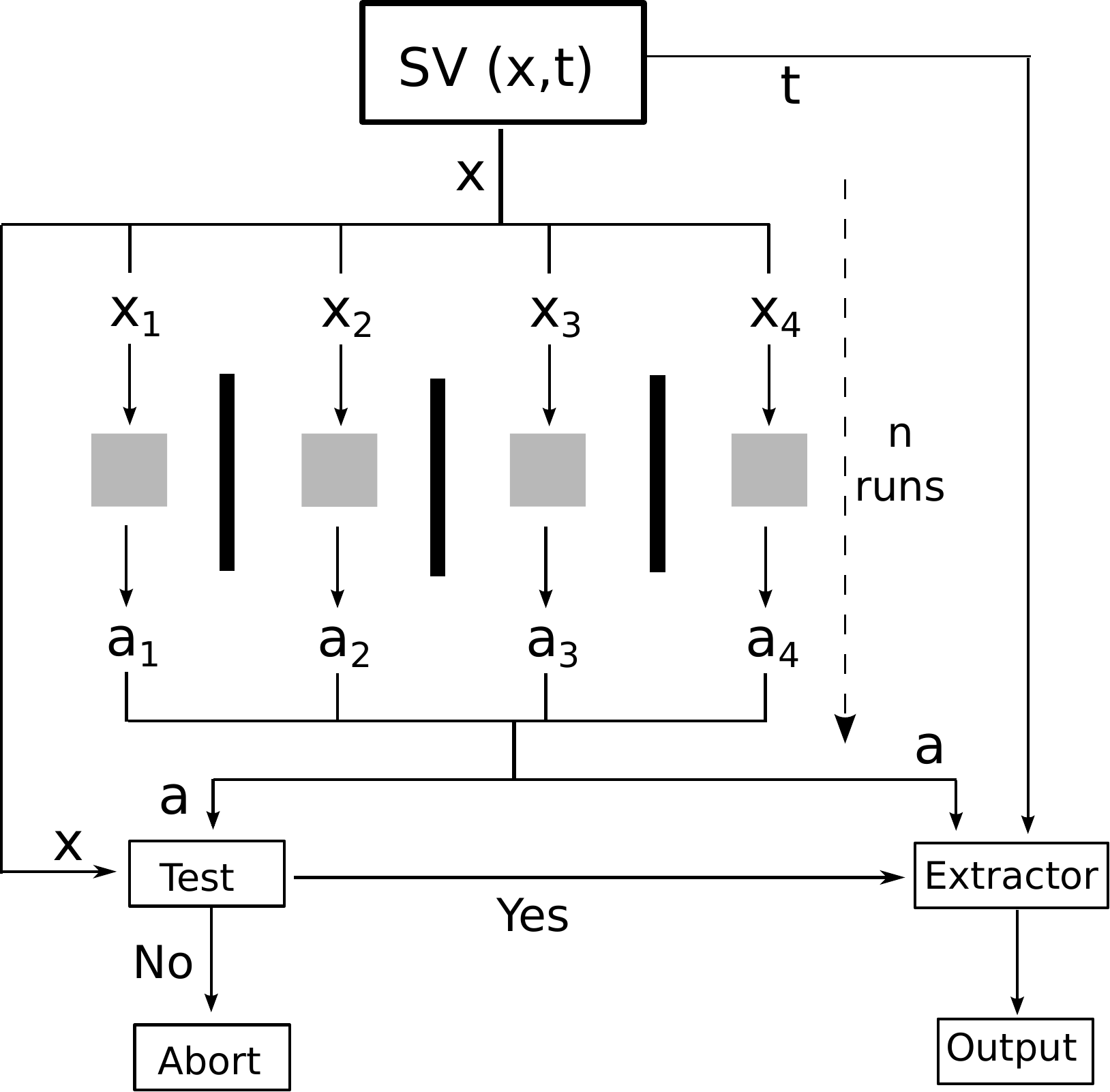}
 \caption{(Color online.)  \label{fig:RandAmp} Scheme of randomness
 amplification using four devices, as in \cite{Brandao16}. The devices are shielded from each
 other as indicated with black barriers. The local measurement choices,
 in each run, are governed by the part of the SV-source, $x$, and
 corresponding output forms random bits $a$. After n-runs Bell test is
 performed (denoted by -- Test). If the test violates Bell inequalities,
then the outputs and rest of the initial
 SV-source $t$ are fed into an extractor (denoted by -- Extractor) in
 order to obtain final outputs. If the test doesn't violate Bell
 inequalities, the protocol is aborted.}
\end{figure}

Randomness amplification based on non-locality was introduced in~\cite{Colbeck12}.
There, the initial source of imperfect randomness consisted of a SV source.
Recall that the amplification of SV sources is impossible by classical means.
A protocol was constructed based on the two-party chained Bell inequalities
that was able to map an SV source with parameter $\epsilon < 0.058$ 
into a new source with $\epsilon$ arbitrarily close to zero.
This result
is only valid in an asymptotic regime in which the user implements the chained
Bell inequality with an infinite number of measurements.
Soon after, a more complicated protocol attained full randomness amplification~\cite{Gallego13}, 
that is, it was able to map SV sources of arbitrarily weak randomness,
$\epsilon<1/2$, to arbitrarily good sources of randomness, $\epsilon\rightarrow 0$.
The final result was again asymptotic, in the sense that to attain full randomness
amplification the user now requires an infinite number of devices.
Randomness amplification protocols have been studied by several other works, see for 
instance~\cite{Grudka14,Mironowicz15,Brandao16,Bouda14, Chung14,coudron_yuen,WBGHHHPR16, Ravi16a}.
As above, the scope of this section is not to provide a complete description of all
the works studying the problem of randomness amplification, but rather to provide 
a general framework that encompasses most of them. In fact, randomness amplification 
protocols (see e.g., Fig.~\ref{fig:RandAmp}) have a structure similar to randomness expansion ones.

The starting point of a protocol consists of a source of imperfect randomness. This is often modelled
by an SV source, although some works consider a weaker source of randomness, known as 
min-entropy source, in which the user only knows a lower bound on the min-entropy of the 
symbols generated by the source~\cite{Bouda14,Chung14}. The bits of imperfect randomness
generated by the user are used to perform $N$ repetitions of the Bell test. If the observed Bell violation 
is large enough, with enough statistical confidence, bits defining the final source are constructed 
from the measurement outputs, possibly assisted by new random bits from the imperfect source.
Note that contrarily to the previous case, the extraction process cannot be assisted with
a seed of perfect random numbers, as this seed could be trivially be used to produce the final source.
As in the case of expansion protocols, any protocol should be accompanied by a security proof
showing that the final bits are 
unpredictable to any observer sharing a system correlated with the devices in the user's hands.

\section{Quantum randomness and technology \label{sec:RandTech}}

Random numbers have been a part of human technology since ancient
times. If Julius Caesar indeed said ``Alea iacta est'' (``the die is
cast,'') when he crossed the Rubicon, he referred to a technology that
had already been in use for thousands of years. Modern uses for random
numbers include cryptography, computer simulations, dispute resolution,
and gaming. The importance of random numbers in politics, social
science and medicine should also not be underestimated; randomized
polling and randomized trials are essential methodology in these areas.

A major challenge for any modern randomness technology is
quantification of the degree to which the output could be predicted or
controlled by an adversary.  A common misconception is that the {\em
output} of a random number generator can be tested for randomness, for
example using statistical tests such as Diehard/Dieharder
\cite{Marsaglia08,BrownWEB2004}, NIST SP800-22 \cite{RukhinNIST2010},
or TestU01 \cite{LEcuyerACM2007}. While it is true that failing these
tests indicates the presence of patterns in the output, and thus a
degree of predictability, passing the tests does not indicate
randomness. This becomes clear if you imagine a device that on its
first run outputs a truly random sequence, perhaps from ideal
measurements of radioactive decay, and on subsequent runs replays this
same sequence from a recording it kept in memory. Any of these
identical output sequences will pass the statistical tests, but only
the first one is random; the others are completely predictable. We can
summarize this situation with the words of John von Neumann: ``there is
no such thing as a random number -- there are only methods to produce
random numbers'' \cite{VonNeumannAMS1951}.

How can we know that a process does indeed produce random numbers?  In
light of the difficulties in determining the predictability of the
apparent randomness seen in thermal fluctuations and other classical
phenomena, using the intrinsic randomness of quantum processes is very
attractive. One approach, described in earlier sections, is to use
device-independent methods.   In principle, device-independent
randomness protocols can be implemented with any technology capable of
a strong Bell-inequality violation, including  ions \cite{Pironio10},
photons \cite{Giustina15,Shalm15}, nitrogen-vacancy centres
\cite{Hensen15}, neutral atoms \cite{RosenfeldOS2011} and
superconducting qubits \cite{JergerARX2016}.

Device-independent randomness expansion based on Bell inequality
violations was first demonstrated using a pair of  Yb$^+$ ions held in
spatially-separated traps \cite{Pironio10}. In this protocol, each ion
is made to emit a photon which, due to the availability of multiple
decay channels with orthogonal photon polarizations, emerges from the
trap entangled with the internal state of the ion.  When the two
photons are combined on a beamsplitter, the Hong-Ou-Mandel effect
causes a coalescence of the two photons into a single output channel,
except in the case that the photons are in a polarization-antisymmetric
Bell state.  Detection of a photon pair, one at each beamsplitter
output, thus accomplishes a projective measurement onto this
antisymmetric Bell state, and this projection in turn causes an
entanglement swapping that leaves the ions entangled.  Their internal
states can then be detected with high efficiency using fluorescence
readout.  This experiment strongly resembles a loophole-free Bell test,
with the exception that the spatial separation of about one meter is
too short to achieve space-like separation.  Due to the low probability
that both photons were collected and  registered on the detectors, the
experiment had a very low success rate, but this does not reduce the
degree of Bell  inequality violation or the quality of the randomness
produced.  The experiment generated  42 random bits in about one month
of continuous running, or $1.6 \times 10^{-5}$ bits/s.

A second experiment, in principle similar but using very different
technologies, was performed with entangled photons and high-efficiency
detectors \cite{ChristensenPRL2013} to achieve a randomness extraction
rate of 0.4 bits/s. While further improvements in speed can be expected
in the near future \cite{NISTBeacon}, at present device-independent
techniques are quite slow, and nearly all applications must still use
traditional quantum randomness techniques.

It is also worth noting that device-independent experiments consume a
large quantity of random bits in choosing the measurement settings in
the Bell test.  Pironio et al. used publicly available randomness
sources drawn from radioactive decay, atmospheric noise, and remote
network activity. Christensen et al. used photon-arrival-time random
number generators to choose the measurement settings. Although it has
been argued that no additional physical randomness is needed in Bell
tests \cite{PironioARX2015}, there does not appear to be agreement on
this point. At least in practice if not in principle, it seems likely
that there will be a need for traditional quantum randomness technology
also in device-independent protocols.

If one does not stick to the device-independent approach, 
it is in fact fairly easy to obtain signals from quantum processes, and
devices to harness the intrinsic randomness of quantum mechanics have
existed since the 1950s. This began with devices to observe the timing
of nuclear decay \cite{Isida1956},  followed by a long list of quantum
physical processes including electron shot noise in semiconductors,
splitting of photons on beamsplitters, timing of photon arrivals,
vacuum fluctuations, laser phase diffusion, amplified spontaneous
emission, Raman scattering, atomic spin diffusion, and others.  See
\cite{HerreroARX2016} for a thorough review.

While measurements on many physical processes can give signals that
contain some intrinsic randomness, any real measurement will also be
contaminated by other signal sources, which might be predictable or of
unknown origin. For example, one could make a simple random number
generator by counting the number of electrons that pass through a Zener
diode in a given amount of time.  Although electron shot noise will
make an intrinsically-random contribution, there will also be an
apparently-random contribution from thermal fluctuations
(Johnson-Nyquist noise), and a quite non-random contribution due to
technical noises from the environment.  If the physical understanding
of the device permits a description in terms of the conditional
min-entropy (see Section \ref{sec:Randomness})
\begin{align}
 H_{\infty}&(X_i|h_i)\geq k,
\forall i \in \mathbb{N}, \forall h_i
\end{align}
where $X_i$ is the $i$'th output string and $h_i$ is the ``history'' of
the the device at that moment, including all fluctuating quantities not
ascribable to intrinsic randomness, then randomness extraction
techniques can be used to produce arbitrarily-good output bits from
this source.  Establishing this min-entropy level can be an important
challenge, however.

\newcommand{\var}{{\rm var}}
The prevalence of optical technologies in recent work on quantum random
number generators is in part in response to this challenge.  The high
coherence and relative purity of optical phenomena allows experimental
systems to closely approximate idealized quantum measurement scenarios.
For example, fluorescence detection of the state of a single trapped
atom is reasonably close to an ideal von Neumann projective
measurement, with fidelity errors at the part-per-thousand level
\cite{MyersonPRL2008}.  Some statistical characterizations can also be
carried out directly using the known statistical properties of quantum
systems. For example, in linear optical systems  shot noise can be
distinguished from other noise sources based purely on scaling
considerations, and provides a very direct calibration of the quantum
versus thermal and technical contributions, without need for detailed
modeling of the devices used. Considering an optical power measurement,
the photocurrent $I_1$ that passes in unit time will obey
\begin{equation}
\var(I_1) = A + B \langle I_1 \rangle + C \langle I_1 \rangle^2
\end{equation}
where $A$ is the ``electronic noise'' contribution, typically of
thermal origin, $C  \langle I_1 \rangle^2$ is the technical noise
contribution, and $B \langle I_1 \rangle$ is the shot-noise
contribution.  Measuring $\var(I_1)$ as a function of $\langle I_1
\rangle$ then provides a direct quantification of the noise contributed
by each of these distinct sources. This methodology has been used to
estimate entropies in continuous-wave phase diffusion random number
generators \cite{XuOE2012}. %\cite{GabrielNPhot2010}.

\begin{figure*}
 \includegraphics[width=1.8 \columnwidth]{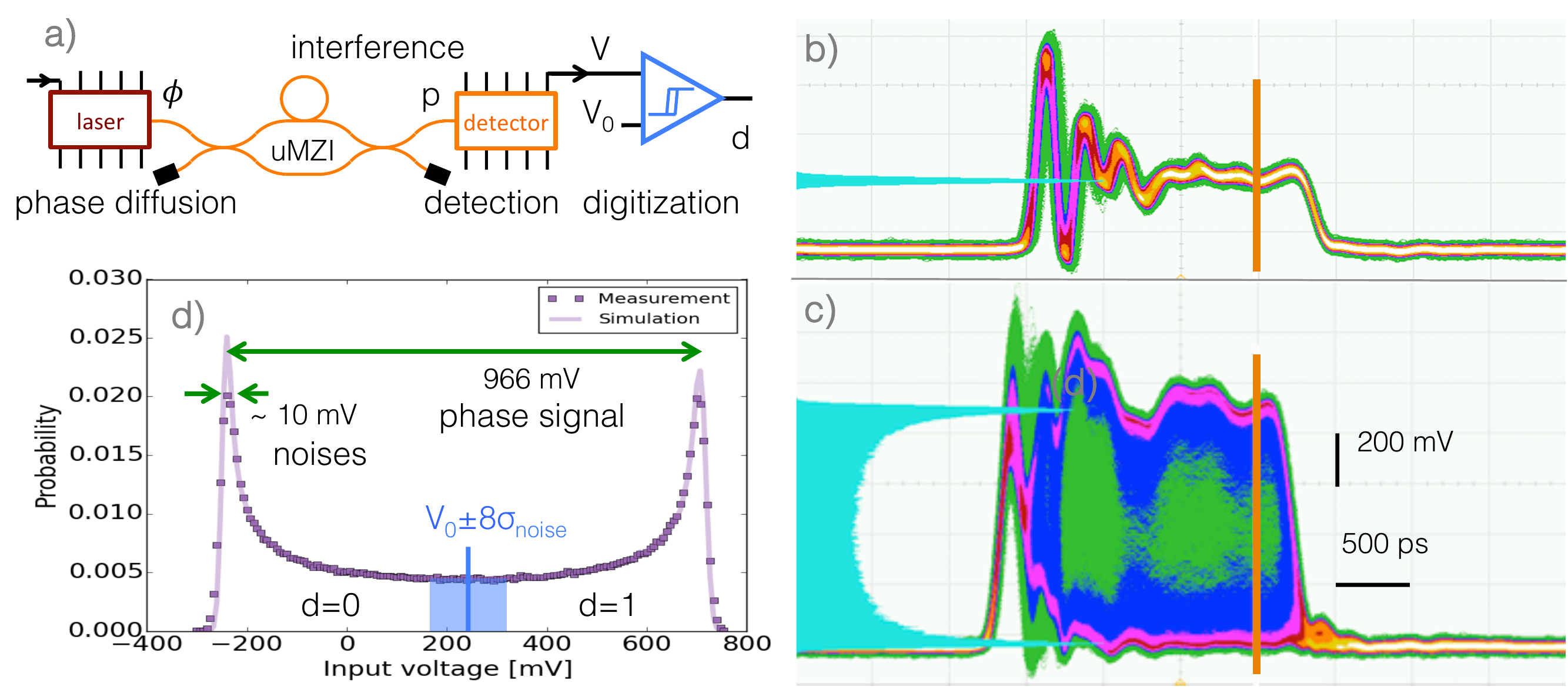}
 \caption{(Color online.) \label{fig:PDQRNG} Laser phase-diffusion
 quantum random number generator (LPD-QRNG). a) schematic diagram
 showing components of a LPD-QRNG using a pulsed laser and single-bit
 digitization.  A laser, driven with pulses of injection current,
 produces optical pulses with very similar wave-forms and with relative
 phases randomized due to phase diffusion between the pulses.   The
 pulses enter a single-mode fiber unbalanced Mach-Zehnder
 interferometer (uMZI), which produces interference between subsequent
 pulses, converting the random relative phase into a strong variation
 in the power reaching the detector, a linear photodiode. A comparator
 produces a digital output in function of the detector signal. b)
 Statistics of the pulse shapes produced by the laser, obtained by
 blocking one arm of the interferometer and recording on an
 oscilloscope. Main image shows the distribution of the pulse shapes
 warmer colors show higher probability density. Strong ``relaxation
 oscillations'' are seen, but are highly reproducible; all traces show
 the same behavior.  Side image shows histogram taken in the region
 labelled in orange, showing a narrow peak indicating very small
 variation in power from one pulse to the next. c) Same acquisition
 strategy, but using both arms of the interferometer and thus producing
 interference.  The variation due to the random phase $\Delta \phi$ is
 orders of magnitude stronger that the noise, and the minimum values
 approach zero power, indicating high interference visibility. The
 histogram shows the classic ``arcsine'' distribution expected for the
 cosine of a random phase. d) illustration of the digitization process.
 Curve and points show expected and observed frequencies for the input
 voltage, approximating an arcsine distribution. The finite width of
 the peaks is a result of convolving the ideal arcsine distribution
 with the noise distribution, of order 10 mV.  The comparator splits
 assigns a value $d=0$ or $d=1$ in function of the input voltage. The
 probability of a noise-produced error can be bounded by considering
 the effect of noise on digitization, giving an
 experimentally-guaranteed min-entropy for the output bits.   }
\end{figure*}

To date, the fastest quantum random number generators are based on
laser phase diffusion \cite{JofreOE2011b,
XuOE2012,AbellanOE2014,YuanAPL2014}, with the record at the time of
writing being 68 Gbits/second \cite{Nie2015}.  These devices,
illustrated in Fig. \ref{fig:PDQRNG} work entirely with macroscopic
optical signals (the output of lasers), which greatly enhances their
speed and signal-to-noise ratios. It is perhaps surprising that
intrinsic randomness can be observed in the macroscopic regime, but in
fact laser phase diffusion (and before it maser phase diffusion) was
one of the first predicted quantum-optical signals, described by
Schawlow and Townes in 1958 \cite{SchawlowPR1958}.

 Because stimulated emission is always accompanied by spontaneous
 emission, the light inside a laser cavity experiences random
 phase-kicks due to spontaneous emission. The laser itself has no
 phase-restoring mechanism; its governing equations are
 phase-invariant, and the phase diffuses in a random walk. As the kicks
 from spontaneous emission accumulate, the phase distribution  rapidly
 approaches a uniform distribution on $[0,2\pi)$, making the laser
 field a macroscopic variable with one degree of freedom fully
 randomized by intrinsic randomness.  The phase diffusion accumulated
 in a given time can be detected simply by introducing an optical delay
 and interfering earlier output with later output in an unbalanced
 Mach-Zehnder interferometer.

It is worth noting that the phase distribution is fully insensitive to
technical and thermal contributions; it is irrelevant if the
environment or an adversary introduces an additional phase shift if the
phase, a cyclic variable, is already fully randomized, i.e. uniformly distributed
on $[0,2\pi)$. 

Considerable effort has gone into determining the min-entropy due to
intrinsic randomness of laser phase-diffusion random number generators
\cite{MitchellPRA2015}, especially in the context of Bell tests
\cite{AbellanPRL2015}.  To date, laser phase diffusion random number
generators have been used to choose the settings for all loophole-free
Bell tests \cite{ Hensen15, Giustina15, Shalm15}. Here we outline the
modeling and measurement considerations used to bound the min-entropy
of the output of these devices.

Considering an interferometer with two paths, short (S) and long (L)
with relative delay $\tau$, fed by the laser output $E(t) = |E(t)|
\exp[i \phi(t)]$, the instantaneous power that reaches the detector is
\begin{equation}
\label{eq:Interference}
p_I(t) = \pS(t) +  \pL(t) +  2\vis \sqrt{\pS(t) \pL(t)} \cos \Delta \phi(t),
\end{equation}
where $\pS(t) \equiv \frac{1}{4} |E(t)|^2$, $\pL(t) \equiv \frac{1}{4}
|E(t-\tau)|^2$,  $\Delta \phi(t) = \phi(t) - \phi(t-\tau)$ and $\vis$
is the interference visibility. Assuming $\tau$ gives sufficient time
for a full phase diffusion, $\Delta \phi(t)$ is uniformly distributed
on $[0,2\pi)$ due to intrinsic quantum randomness.  The contributions
of $\pS(t)$ and $\pL(t)$, however, may reflect technical or thermal
fluctuations, and constitute a contamination of the measurable signal
$p_I(t)$. The process of detection converts this to a voltage $V(t)$,
and in doing so adds other technical and thermal noises.  Also, the
necessarily finite speed of the detection system implies that $V(t)$ is
a function of not only of $p_I(t)$, but also to a lesser extent of
prior values $p_I(t'), t'<t$. This ``hangover,'' which is predictable
based on the previous values, must be accounted for so as to not
overestimate the entropy in $p_I(t)$.

Digitization is the conversion from the continuous signal $V$ to a
digital value $d$. Considering only the simplest case of binary
digitization, we have
\begin{eqnarray}
d_i & = & \left\{
\begin{array}{rl}
0 & V(t_i) < V_0 \\
1 & V(t_i) \ge V_0
\end{array}
\right.
\end{eqnarray}
where $V_0$ is the threshold voltage, itself a random variable
influenced by a combination of thermal and technical noises.

We can now express the distribution of $d$ in function of the total
noise $\vc$
\begin{eqnarray}
\label{eq:PredFromVc}
P(d=1|\vc) &=&   \frac{2}{\pi} \arcsin \sqrt{\frac{1}{2} + \frac{\vc}{2\vrange}}
\end{eqnarray}
where $2 \vrange \propto 4 \vis \sqrt{\pS \pL}$ is the peak-to-peak
range of the signal due to the random $\Delta \phi$. This  derives from
the ``arcsin'' distribution that describes the cumulative distribution
function of the cosine of a uniformly-random phase.

The noise contributions can all be measured in ways that conservatively
estimate their variation; for example interrupting one or the other
path of the interferometer we can measure the distribution of $\pS$ and
$\pL$, and comparing digitizer input to output we can upper bound the
variation in $V_0$. With the measured distributions in hand, we can
assign probabilities to $\vc$ and thus to the min-entropy of $d$. For
example, if the total noise $\vc$ is described by a normal distribution
with  zero mean and width $\sigma_{\noise} = $ 10 mV, and  $\vrange =
0.5$ V, a probability $P(d|\vc) > P(d=1| 8 \sigma_{\noise}) \approx
\frac{1}{2} + 0.0511$ will occur as often as $\vc$ exceeds $8
\sigma_{\noise}$, which is to say with probability $\approx 6 \times
10^{-16}$. Since $P(d|\vc) \le \frac{1}{2} + 0.0511$ implies a
single-bit min entropy $H_{\infty}(d|\vc) > 0.86$, a randomness
extraction based on this min-entropy can then be applied to give
fully-random output strings.

It is worth emphasizing that the characterizations used to guarantee
randomness of this kind of source are not measurements of the digital
output of the source, which as mentioned already can never demonstrate
randomness.  Rather, they are arguments based on physical principles,
backed by measurements of the internal workings of the device. To
summarize the argument, the trusted random variable $\Delta \phi(t)$ is
known to be fully randomized by the intrinsic quantum randomness of
spontaneous emission.  This statement relies on general principles that
concern laser physics, such as Einstein's A and B coefficient argument
linking spontaneous emission to stimulated emission and the fact that
lasers have no preferred phase, due to the time-translation invariance
of physical law.  The next step of the guarantee follows from a model
of the interference process, Eq. (\ref{eq:Interference}), whose
simplicity mirrors the simplicity of the experimental situation, in
which single-mode devices (fibres) are used to ensure a low-dimensional
field characterized only by the time variable. Finally there is an
experimental characterization of the noises and a simple computation to
bound their effects on the distribution of outputs. The computation can
and should be performed with worst-case assumptions, assuming for
example that all noise contributions are maximally correlated, unless
the contrary has been experimentally demonstrated.

\section{Quantum Randomness and Future}

Randomness is a fascinating concept that absorbs human attention since centuries. Nowadays we are witnessing a novel situation, when the theoretical and experimental developments of  quantum physics allow to investigate  quantum randomness from completely novel points of view. The present review  provides an overview of the problem of quantum randomness, and covers
the implications and new directions emerging in the studies of this problem. 

From a philosophical and fundamental perspective, the recent results have 
significantly improved our understanding of what can and cannot be said about randomness in nature
using quantum physics. While the presence of randomness cannot be proven
without making some assumptions about the systems, theses assumptions
are constantly weakened and it is an interesting open research problem to identify
the weakest set of assumption sufficient to certify the presence of randomness.

From a theoretical physics perspective, the recent results have provided
a much better understanding of the relation between non-locality and
randomness in quantum theory. Still, the exact relation between 
these two fundamental concepts is not fully 
understood. For instance, small amount of non-locality, or even entanglement,
sometimes suffice to certify the presence of maximal randomness in the measurement
outputs of a Bell experiment~\cite{Acin12}. The relation between non-locality and randomness
can also be studied in the larger framework of no-signaling theories, that is theories
only limited by the no-signaling principle, which can go beyond quantum physics~\cite{PR}.
For instance it is known that in these theories maximal randomness certification is impossible,
while it is in quantum physics~\cite{delatorre}.

From a more applied perspective, quantum protocols for randomness generation follow
different approaches and require different assumptions. Until very recently,
all quantum protocol required a precise knowledge of the devices used in the protocol 
and certified the presence of randomness by means of standard statistical tests.
The resulting protocols are cheap, feasible to implement in practice, including
the development of commercial products, and lead
to reasonably high randomness generation rates. Device-independent solutions provide a 
completely different approach, in which no modeling of the devices is needed
and the certification comes from a Bell inequality violation. Their implementation
is however more challenging and only few much slower experimental 
realizations have until now been reported
\footnote{An important discussion of the commercial and practical aspects of quantum random number generation, and cryptography based on device dependent and independent protocols can be found in the lecture No. 7 by Alain Aspect and Michel Brune \cite{moon}}.

Due to the importance and need of random numbers in our information society,
we expect important advances in all these approaches, resulting in a large 
variety of quantum empowered solutions for randomness generation.

\section{Acknowledgements}
We thank anonymous referees for constructive criticism and valuable suggestions. We are very grateful to Krzysztof Gaw\k edzki, Alain Aspect, Philippe Grangier and Miguel A.F. Sanjuan for enlightening discussions about non-deterministic theories and unpredictability in classical physics. 
We acknowledge financial support from the John Templeton Foundation, the European Commission (FETPRO QUIC, STREP EQuaM and RAQUEL), the European Research Council (AdG OSYRIS, AdG IRQUAT, StG AQUMET, CoG QITBOX and PoC ERIDIAN), the AXA Chair in Quantum Information Science,
the Spanish MINECO (Grants No. FIS2008-01236, No. FIS2013-40627-P, No. FIS2013-46768-P FOQUS, FIS2014-62181-EXP, FIS2015-68039-P, FIS2016-80773-P, and Severo Ochoa Excellence Grant SEV-2015-0522) with the support of FEDER funds, the Generalitat de Catalunya (Grants No. 2014-SGR-874, No. 2014-SGR-875, No. 2014-SGR-966 and No. 2014-SGR-1295 and CERCA Program), and Fundaci\'o Privada Cellex.

% \bibliography{bib_filo}

\begin{thebibliography}{180}%
\makeatletter
\providecommand \@ifxundefined [1]{%
 \@ifx{#1\undefined}
}%
\providecommand \@ifnum [1]{%
 \ifnum #1\expandafter \@firstoftwo
 \else \expandafter \@secondoftwo
 \fi
}%
\providecommand \@ifx [1]{%
 \ifx #1\expandafter \@firstoftwo
 \else \expandafter \@secondoftwo
 \fi
}%
\providecommand \natexlab [1]{#1}%
\providecommand \enquote  [1]{``#1''}%
\providecommand \bibnamefont  [1]{#1}%
\providecommand \bibfnamefont [1]{#1}%
\providecommand \citenamefont [1]{#1}%
\providecommand \href@noop [0]{\@secondoftwo}%
\providecommand \href [0]{\begingroup \@sanitize@url \@href}%
\providecommand \@href[1]{\@@startlink{#1}\@@href}%
\providecommand \@@href[1]{\endgroup#1\@@endlink}%
\providecommand \@sanitize@url [0]{\catcode `\\12\catcode `\$12\catcode
  `\&12\catcode `\#12\catcode `\^12\catcode `\_12\catcode `\%12\relax}%
\providecommand \@@startlink[1]{}%
\providecommand \@@endlink[0]{}%
\providecommand \url  [0]{\begingroup\@sanitize@url \@url }%
\providecommand \@url [1]{\endgroup\@href {#1}{\urlprefix }}%
\providecommand \urlprefix  [0]{URL }%
\providecommand \Eprint [0]{\href }%
\providecommand \doibase [0]{http://dx.doi.org/}%
\providecommand \selectlanguage [0]{\@gobble}%
\providecommand \bibinfo  [0]{\@secondoftwo}%
\providecommand \bibfield  [0]{\@secondoftwo}%
\providecommand \translation [1]{[#1]}%
\providecommand \BibitemOpen [0]{}%
\providecommand \bibitemStop [0]{}%
\providecommand \bibitemNoStop [0]{.\EOS\space}%
\providecommand \EOS [0]{\spacefactor3000\relax}%
\providecommand \BibitemShut  [1]{\csname bibitem#1\endcsname}%
\let\auto@bib@innerbib\@empty
%</preamble>
\bibitem [{\citenamefont {Abell\'{a}n}\ \emph {et~al.}(2014)\citenamefont
  {Abell\'{a}n}, \citenamefont {Amaya}, \citenamefont {Jofre}, \citenamefont
  {Curty}, \citenamefont {Ac\'{i}n}, \citenamefont {Capmany}, \citenamefont
  {Pruneri},\ and\ \citenamefont {Mitchell}}]{AbellanOE2014}%
  \BibitemOpen
  \bibfield  {author} {\bibinfo {author} {\bibnamefont {Abell\'{a}n},
  \bibfnamefont {C}}, \bibinfo {author} {\bibfnamefont {W.}~\bibnamefont
  {Amaya}}, \bibinfo {author} {\bibfnamefont {M.}~\bibnamefont {Jofre}},
  \bibinfo {author} {\bibfnamefont {M.}~\bibnamefont {Curty}}, \bibinfo
  {author} {\bibfnamefont {A.}~\bibnamefont {Ac\'{i}n}}, \bibinfo {author}
  {\bibfnamefont {J.}~\bibnamefont {Capmany}}, \bibinfo {author} {\bibfnamefont
  {V.}~\bibnamefont {Pruneri}}, \ and\ \bibinfo {author} {\bibfnamefont
  {M.~W.}\ \bibnamefont {Mitchell}}} (\bibinfo {year} {2014}),\ \bibfield
  {title} {\enquote {\bibinfo {title} {Ultra-fast quantum randomness generation
  by accelerated phase diffusion in a pulsed laser diode},}\ }\href {\doibase
  10.1364/OE.22.001645} {\bibfield  {journal} {\bibinfo  {journal} {Opt.
  Express}\ }\textbf {\bibinfo {volume} {22}}~(\bibinfo {number} {2}),\
  \bibinfo {pages} {1645--1654}}\BibitemShut {NoStop}%
\bibitem [{\citenamefont {Abell\'an}\ \emph {et~al.}(2015)\citenamefont
  {Abell\'an}, \citenamefont {Amaya}, \citenamefont {Mitrani}, \citenamefont
  {Pruneri},\ and\ \citenamefont {Mitchell}}]{AbellanPRL2015}%
  \BibitemOpen
  \bibfield  {author} {\bibinfo {author} {\bibnamefont {Abell\'an},
  \bibfnamefont {Carlos}}, \bibinfo {author} {\bibfnamefont {Waldimar}\
  \bibnamefont {Amaya}}, \bibinfo {author} {\bibfnamefont {Daniel}\
  \bibnamefont {Mitrani}}, \bibinfo {author} {\bibfnamefont {Valerio}\
  \bibnamefont {Pruneri}}, \ and\ \bibinfo {author} {\bibfnamefont {Morgan~W.}\
  \bibnamefont {Mitchell}}} (\bibinfo {year} {2015}),\ \bibfield  {title}
  {\enquote {\bibinfo {title} {Generation of fresh and pure random numbers for
  loophole-free {Bell} tests},}\ }\href {\doibase
  10.1103/PhysRevLett.115.250403} {\bibfield  {journal} {\bibinfo  {journal}
  {Phys. Rev. Lett.}\ }\textbf {\bibinfo {volume} {115}},\ \bibinfo {pages}
  {250403}}\BibitemShut {NoStop}%
\bibitem [{\citenamefont {Ac{\'i}n}\ \emph {et~al.}(2015)\citenamefont
  {Ac{\'i}n}, \citenamefont {Fritz}, \citenamefont {Leverrier},\ and\
  \citenamefont {Sainz}}]{Acin15a}%
  \BibitemOpen
  \bibfield  {author} {\bibinfo {author} {\bibnamefont {Ac{\'i}n},
  \bibfnamefont {Antonio}}, \bibinfo {author} {\bibfnamefont {Tobias}\
  \bibnamefont {Fritz}}, \bibinfo {author} {\bibfnamefont {Anthony}\
  \bibnamefont {Leverrier}}, \ and\ \bibinfo {author} {\bibfnamefont
  {Ana~Bel{\'e}n}\ \bibnamefont {Sainz}}} (\bibinfo {year} {2015}),\ \bibfield
  {title} {\enquote {\bibinfo {title} {A combinatorial approach to nonlocality
  and contextuality},}\ }\href {\doibase 10.1007/s00220-014-2260-1} {\bibfield
  {journal} {\bibinfo  {journal} {Communications in Mathematical Physics}\
  }\textbf {\bibinfo {volume} {334}}~(\bibinfo {number} {2}),\ \bibinfo {pages}
  {533--628}}\BibitemShut {NoStop}%
\bibitem [{\citenamefont {Ac\'{\i}n}\ \emph {et~al.}(2012)\citenamefont
  {Ac\'{\i}n}, \citenamefont {Massar},\ and\ \citenamefont {Pironio}}]{Acin12}%
  \BibitemOpen
  \bibfield  {author} {\bibinfo {author} {\bibnamefont {Ac\'{\i}n},
  \bibfnamefont {Antonio}}, \bibinfo {author} {\bibfnamefont {Serge}\
  \bibnamefont {Massar}}, \ and\ \bibinfo {author} {\bibfnamefont {Stefano}\
  \bibnamefont {Pironio}}} (\bibinfo {year} {2012}),\ \bibfield  {title}
  {\enquote {\bibinfo {title} {Randomness versus nonlocality and
  entanglement},}\ }\href {\doibase 10.1103/PhysRevLett.108.100402} {\bibfield
  {journal} {\bibinfo  {journal} {Phys. Rev. Lett.}\ }\textbf {\bibinfo
  {volume} {108}},\ \bibinfo {pages} {100402}}\BibitemShut {NoStop}%
\bibitem [{\citenamefont {Albert}(2010)}]{Albert2010}%
  \BibitemOpen
  \bibfield  {author} {\bibinfo {author} {\bibnamefont {Albert}, \bibfnamefont
  {David}}} (\bibinfo {year} {2010}),\ \bibfield  {title} {\enquote {\bibinfo
  {title} {{Probability in the Everett Picture}},}\ }in\ \href
  {http://www.oxfordscholarship.com/view/10.1093/acprof:oso/9780199560561.001.0001/acprof-9780199560561-chapter-13}
  {\emph {\bibinfo {booktitle} {{Many Worlds?: Everett, Quantum Theory, and
  Reality}}}},\ \bibinfo {editor} {edited by\ \bibinfo {editor} {\bibfnamefont
  {Simon}\ \bibnamefont {Saunders}}, \bibinfo {editor} {\bibfnamefont
  {Jonathan}\ \bibnamefont {Barrett}}, \bibinfo {editor} {\bibfnamefont
  {Adrian}\ \bibnamefont {Kent}}, \ and\ \bibinfo {editor} {\bibfnamefont
  {David}\ \bibnamefont {Wallace}}},\ Chap.~\bibinfo {chapter} {13}\ (\bibinfo
  {publisher} {Oxford Publisher},\ \bibinfo {address} {Oxford})\ pp.\ \bibinfo
  {pages} {355--368}\BibitemShut {NoStop}%
\bibitem [{\citenamefont {Arnol'd}(1973)}]{Arnold73}%
  \BibitemOpen
  \bibfield  {author} {\bibinfo {author} {\bibnamefont {Arnol'd}, \bibfnamefont
  {V~I}}} (\bibinfo {year} {1973}),\ \href@noop {} {\emph {\bibinfo {title}
  {{O}rdinary differential equations}}}\ (\bibinfo  {publisher} {{MIT} Press,
  Cambridge, Massachusetts})\BibitemShut {NoStop}%
\bibitem [{\citenamefont {Arnol'd}(1989)}]{Arnold89}%
  \BibitemOpen
  \bibfield  {author} {\bibinfo {author} {\bibnamefont {Arnol'd}, \bibfnamefont
  {V~I}}} (\bibinfo {year} {1989}),\ \href@noop {} {\emph {\bibinfo {title}
  {{M}athematical methods of classical mechanics}}}\ (\bibinfo  {publisher}
  {Springer, New York})\BibitemShut {NoStop}%
\bibitem [{\citenamefont {Arnon-Friedman}\ \emph {et~al.}(2016)\citenamefont
  {Arnon-Friedman}, \citenamefont {Renner},\ and\ \citenamefont
  {Vidick}}]{EATQKD}%
  \BibitemOpen
  \bibfield  {author} {\bibinfo {author} {\bibnamefont {Arnon-Friedman},
  \bibfnamefont {R}}, \bibinfo {author} {\bibfnamefont {R.}~\bibnamefont
  {Renner}}, \ and\ \bibinfo {author} {\bibfnamefont {T.}~\bibnamefont
  {Vidick}}} (\bibinfo {year} {2016}),\ \bibfield  {title} {\enquote {\bibinfo
  {title} {{Simple and tight device-independent security proofs}},}\
  }\href@noop {} {\bibfield  {journal} {\bibinfo  {journal} {ArXiv e-prints}\
  }}\Eprint {http://arxiv.org/abs/1607.01797} {arXiv:1607.01797 [quant-ph]}
  \BibitemShut {NoStop}%
\bibitem [{\citenamefont {Aspect}\ and\ \citenamefont {Brune}(2016)}]{moon}%
  \BibitemOpen
  \bibfield  {author} {\bibinfo {author} {\bibnamefont {Aspect}, \bibfnamefont
  {Alain}}, \ and\ \bibinfo {author} {\bibfnamefont {Michel}\ \bibnamefont
  {Brune}}} (\bibinfo {year} {2016}),\ \href
  {https://www.coursera.org/learn/quantum-optics-single-photon} {\enquote
  {\bibinfo {title} {Course in {Q}uantum {O}ptics; {L}ecture {N}o. 7},}\
  }\bibinfo {howpublished} {{\'E}cole {P}olytechnique}\BibitemShut {NoStop}%
\bibitem [{\citenamefont {Aspect}\ \emph {et~al.}(1981)\citenamefont {Aspect},
  \citenamefont {Grangier},\ and\ \citenamefont {Roger}}]{Aspect81}%
  \BibitemOpen
  \bibfield  {author} {\bibinfo {author} {\bibnamefont {Aspect}, \bibfnamefont
  {Alain}}, \bibinfo {author} {\bibfnamefont {Philippe}\ \bibnamefont
  {Grangier}}, \ and\ \bibinfo {author} {\bibfnamefont {G\'erard}\ \bibnamefont
  {Roger}}} (\bibinfo {year} {1981}),\ \bibfield  {title} {\enquote {\bibinfo
  {title} {Experimental tests of realistic local theories via {B}ell's
  theorem},}\ }\href {\doibase 10.1103/PhysRevLett.47.460} {\bibfield
  {journal} {\bibinfo  {journal} {Phys. Rev. Lett.}\ }\textbf {\bibinfo
  {volume} {47}},\ \bibinfo {pages} {460--463}}\BibitemShut {NoStop}%
\bibitem [{\citenamefont {Aspect}\ \emph {et~al.}(1982)\citenamefont {Aspect},
  \citenamefont {Grangier},\ and\ \citenamefont {Roger}}]{Aspect82}%
  \BibitemOpen
  \bibfield  {author} {\bibinfo {author} {\bibnamefont {Aspect}, \bibfnamefont
  {Alain}}, \bibinfo {author} {\bibfnamefont {Philippe}\ \bibnamefont
  {Grangier}}, \ and\ \bibinfo {author} {\bibfnamefont {G\'erard}\ \bibnamefont
  {Roger}}} (\bibinfo {year} {1982}),\ \bibfield  {title} {\enquote {\bibinfo
  {title} {Experimental realization of {E}instein-{P}odolsky-{R}osen-{B}ohm
  \textit{Gedankenexperiment} : A new violation of {B}ell's inequalities},}\
  }\href {\doibase 10.1103/PhysRevLett.49.91} {\bibfield  {journal} {\bibinfo
  {journal} {Phys. Rev. Lett.}\ }\textbf {\bibinfo {volume} {49}},\ \bibinfo
  {pages} {91--94}}\BibitemShut {NoStop}%
\bibitem [{\citenamefont {Barak}\ \emph {et~al.}(2010)\citenamefont {Barak},
  \citenamefont {Kindler}, \citenamefont {Shaltiel}, \citenamefont {Sudakov},\
  and\ \citenamefont {Wigderson}}]{Barak10}%
  \BibitemOpen
  \bibfield  {author} {\bibinfo {author} {\bibnamefont {Barak}, \bibfnamefont
  {B}}, \bibinfo {author} {\bibfnamefont {G.}~\bibnamefont {Kindler}}, \bibinfo
  {author} {\bibfnamefont {R.}~\bibnamefont {Shaltiel}}, \bibinfo {author}
  {\bibfnamefont {B.}~\bibnamefont {Sudakov}}, \ and\ \bibinfo {author}
  {\bibfnamefont {A.}~\bibnamefont {Wigderson}}} (\bibinfo {year} {2010}),\
  \bibfield  {title} {\enquote {\bibinfo {title} {Simulating independence: New
  constructions of condensers, {R}amsey graphs, dispersers, and extractors},}\
  }\href {\doibase 10.1145/1734213.1734214} {\bibfield  {journal} {\bibinfo
  {journal} {J. ACM}\ }\textbf {\bibinfo {volume} {57}}~(\bibinfo {number}
  {4}),\ \bibinfo {pages} {20:1--20:52}}\BibitemShut {NoStop}%
\bibitem [{\citenamefont {Bartosik}\ \emph {et~al.}(2009)\citenamefont
  {Bartosik}, \citenamefont {Klepp}, \citenamefont {Schmitzer}, \citenamefont
  {Sponar}, \citenamefont {Cabello}, \citenamefont {Rauch},\ and\ \citenamefont
  {Hasegawa}}]{Bartosik09}%
  \BibitemOpen
  \bibfield  {author} {\bibinfo {author} {\bibnamefont {Bartosik},
  \bibfnamefont {H}}, \bibinfo {author} {\bibfnamefont {J.}~\bibnamefont
  {Klepp}}, \bibinfo {author} {\bibfnamefont {C.}~\bibnamefont {Schmitzer}},
  \bibinfo {author} {\bibfnamefont {S.}~\bibnamefont {Sponar}}, \bibinfo
  {author} {\bibfnamefont {A.}~\bibnamefont {Cabello}}, \bibinfo {author}
  {\bibfnamefont {H.}~\bibnamefont {Rauch}}, \ and\ \bibinfo {author}
  {\bibfnamefont {Y.}~\bibnamefont {Hasegawa}}} (\bibinfo {year} {2009}),\
  \bibfield  {title} {\enquote {\bibinfo {title} {Experimental test of quantum
  contextuality in neutron interferometry},}\ }\href {\doibase
  10.1103/PhysRevLett.103.040403} {\bibfield  {journal} {\bibinfo  {journal}
  {Phys. Rev. Lett.}\ }\textbf {\bibinfo {volume} {103}},\ \bibinfo {pages}
  {040403}}\BibitemShut {NoStop}%
\bibitem [{\citenamefont {Bassham}\ \emph {et~al.}(2010)\citenamefont
  {Bassham}, \citenamefont {Rukhin}, \citenamefont {Soto}, \citenamefont
  {Nechvatal}, \citenamefont {Smid}, \citenamefont {Barker}, \citenamefont
  {Leigh}, \citenamefont {Levenson}, \citenamefont {Vangel}, \citenamefont
  {Banks}, \citenamefont {Heckert}, \citenamefont {Dray},\ and\ \citenamefont
  {Vo}}]{Bassham10}%
  \BibitemOpen
  \bibfield  {author} {\bibinfo {author} {\bibnamefont {Bassham}, \bibfnamefont
  {III, Lawrence~E}}, \bibinfo {author} {\bibfnamefont {Andrew~L.}\
  \bibnamefont {Rukhin}}, \bibinfo {author} {\bibfnamefont {Juan}\ \bibnamefont
  {Soto}}, \bibinfo {author} {\bibfnamefont {James~R.}\ \bibnamefont
  {Nechvatal}}, \bibinfo {author} {\bibfnamefont {Miles~E.}\ \bibnamefont
  {Smid}}, \bibinfo {author} {\bibfnamefont {Elaine~B.}\ \bibnamefont
  {Barker}}, \bibinfo {author} {\bibfnamefont {Stefan~D.}\ \bibnamefont
  {Leigh}}, \bibinfo {author} {\bibfnamefont {Mark}\ \bibnamefont {Levenson}},
  \bibinfo {author} {\bibfnamefont {Mark}\ \bibnamefont {Vangel}}, \bibinfo
  {author} {\bibfnamefont {David~L.}\ \bibnamefont {Banks}}, \bibinfo {author}
  {\bibfnamefont {Nathanael~Alan}\ \bibnamefont {Heckert}}, \bibinfo {author}
  {\bibfnamefont {James~F.}\ \bibnamefont {Dray}}, \ and\ \bibinfo {author}
  {\bibfnamefont {San}\ \bibnamefont {Vo}}} (\bibinfo {year} {2010}),\
  \href@noop {} {\emph {\bibinfo {title} {SP 800-22 Rev. 1a. A Statistical Test
  Suite for Random and Pseudorandom Number Generators for Cryptographic
  Applications}}},\ \bibinfo {type} {Tech. Rep.}\ (\bibinfo {address}
  {Gaithersburg, MD, United States})\BibitemShut {NoStop}%
\bibitem [{\citenamefont {Beckner}(1975)}]{Beckner75}%
  \BibitemOpen
  \bibfield  {author} {\bibinfo {author} {\bibnamefont {Beckner}, \bibfnamefont
  {William}}} (\bibinfo {year} {1975}),\ \bibfield  {title} {\enquote {\bibinfo
  {title} {Inequalities in fourier analysis},}\ }\href
  {http://www.jstor.org/stable/1970980} {\bibfield  {journal} {\bibinfo
  {journal} {Annals of Mathematics}\ }\textbf {\bibinfo {volume}
  {102}}~(\bibinfo {number} {1}),\ \bibinfo {pages} {159--182}}\BibitemShut
  {NoStop}%
\bibitem [{\citenamefont {Bell}(1964)}]{Bell64}%
  \BibitemOpen
  \bibfield  {author} {\bibinfo {author} {\bibnamefont {Bell}, \bibfnamefont
  {J~S}}} (\bibinfo {year} {1964}),\ \bibfield  {title} {\enquote {\bibinfo
  {title} {{O}n the {E}instein-{P}odolsky-{R}osen paradox},}\ }\href@noop {}
  {\bibfield  {journal} {\bibinfo  {journal} {Physics}\ }\textbf {\bibinfo
  {volume} {1}},\ \bibinfo {pages} {195}}\BibitemShut {NoStop}%
\bibitem [{\citenamefont {Bell}(1966)}]{Bell66}%
  \BibitemOpen
  \bibfield  {author} {\bibinfo {author} {\bibnamefont {Bell}, \bibfnamefont
  {J~S}}} (\bibinfo {year} {1966}),\ \bibfield  {title} {\enquote {\bibinfo
  {title} {On the problem of hidden variables in quantum mechanics},}\ }\href
  {\doibase 10.1103/RevModPhys.38.447} {\bibfield  {journal} {\bibinfo
  {journal} {Rev. Mod. Phys.}\ }\textbf {\bibinfo {volume} {38}},\ \bibinfo
  {pages} {447--452}}\BibitemShut {NoStop}%
\bibitem [{\citenamefont {Bernard}\ \emph {et~al.}(1998)\citenamefont
  {Bernard}, \citenamefont {Gawedzki},\ and\ \citenamefont
  {Kupiainen}}]{Gawedz1}%
  \BibitemOpen
  \bibfield  {author} {\bibinfo {author} {\bibnamefont {Bernard}, \bibfnamefont
  {Denis}}, \bibinfo {author} {\bibfnamefont {Krzysztof}\ \bibnamefont
  {Gawedzki}}, \ and\ \bibinfo {author} {\bibfnamefont {Antti}\ \bibnamefont
  {Kupiainen}}} (\bibinfo {year} {1998}),\ \bibfield  {title} {\enquote
  {\bibinfo {title} {{S}low {M}odes in {P}assive {A}dvection},}\ }\href
  {\doibase 10.1023/A:1023212600779} {\bibfield  {journal} {\bibinfo  {journal}
  {Journal of Statistical Physics}\ }\textbf {\bibinfo {volume} {90}}~(\bibinfo
  {number} {3}),\ \bibinfo {pages} {519--569}}\BibitemShut {NoStop}%
\bibitem [{\citenamefont {Berta}\ \emph {et~al.}(2010)\citenamefont {Berta},
  \citenamefont {Christandl}, \citenamefont {Colbeck}, \citenamefont {Renes},\
  and\ \citenamefont {Renner}}]{Berta10}%
  \BibitemOpen
  \bibfield  {author} {\bibinfo {author} {\bibnamefont {Berta}, \bibfnamefont
  {Mario}}, \bibinfo {author} {\bibfnamefont {Matthias}\ \bibnamefont
  {Christandl}}, \bibinfo {author} {\bibfnamefont {Roger}\ \bibnamefont
  {Colbeck}}, \bibinfo {author} {\bibfnamefont {Joseph~M.}\ \bibnamefont
  {Renes}}, \ and\ \bibinfo {author} {\bibfnamefont {Renato}\ \bibnamefont
  {Renner}}} (\bibinfo {year} {2010}),\ \bibfield  {title} {\enquote {\bibinfo
  {title} {The uncertainty principle in the presence of quantum memory},}\
  }\href {\doibase 10.1038/nphys1734} {\bibfield  {journal} {\bibinfo
  {journal} {Nat Phys}\ }\textbf {\bibinfo {volume} {6}},\ \bibinfo {pages}
  {659--662}}\BibitemShut {NoStop}%
\bibitem [{\citenamefont {Bia{\l}ynicki-Birula}\ and\ \citenamefont
  {Mycielski}(1975)}]{Birula75}%
  \BibitemOpen
  \bibfield  {author} {\bibinfo {author} {\bibnamefont {Bia{\l}ynicki-Birula},
  \bibfnamefont {Iwo}}, \ and\ \bibinfo {author} {\bibfnamefont {Jerzy}\
  \bibnamefont {Mycielski}}} (\bibinfo {year} {1975}),\ \bibfield  {title}
  {\enquote {\bibinfo {title} {Uncertainty relations for information entropy in
  wave mechanics},}\ }\href {\doibase 10.1007/BF01608825} {\bibfield  {journal}
  {\bibinfo  {journal} {Communications in Mathematical Physics}\ }\textbf
  {\bibinfo {volume} {44}}~(\bibinfo {number} {2}),\ \bibinfo {pages}
  {129--132}}\BibitemShut {NoStop}%
\bibitem [{\citenamefont {Bohm}(1951)}]{Bohm51}%
  \BibitemOpen
  \bibfield  {author} {\bibinfo {author} {\bibnamefont {Bohm}, \bibfnamefont
  {David}}} (\bibinfo {year} {1951}),\ \href@noop {} {\emph {\bibinfo {title}
  {Quantum Theory}}}\ (\bibinfo  {publisher} {Prentice-Hall, Inc., New
  York})\BibitemShut {NoStop}%
\bibitem [{\citenamefont {Bohm}(1952)}]{Bohm52}%
  \BibitemOpen
  \bibfield  {author} {\bibinfo {author} {\bibnamefont {Bohm}, \bibfnamefont
  {David}}} (\bibinfo {year} {1952}),\ \bibfield  {title} {\enquote {\bibinfo
  {title} {A suggested interpretation of the quantum theory in terms of
  "hidden" variables. i},}\ }\href {\doibase 10.1103/PhysRev.85.166} {\bibfield
   {journal} {\bibinfo  {journal} {Phys. Rev.}\ }\textbf {\bibinfo {volume}
  {85}},\ \bibinfo {pages} {166--179}}\BibitemShut {NoStop}%
\bibitem [{\citenamefont {Bohr}(1935)}]{Bohr35}%
  \BibitemOpen
  \bibfield  {author} {\bibinfo {author} {\bibnamefont {Bohr}, \bibfnamefont
  {N}}} (\bibinfo {year} {1935}),\ \bibfield  {title} {\enquote {\bibinfo
  {title} {Can quantum-mechanical description of physical reality be considered
  complete?}}\ }\href {\doibase 10.1103/PhysRev.48.696} {\bibfield  {journal}
  {\bibinfo  {journal} {Phys. Rev.}\ }\textbf {\bibinfo {volume} {48}},\
  \bibinfo {pages} {696--702}}\BibitemShut {NoStop}%
\bibitem [{\citenamefont {Bouda}\ \emph {et~al.}(2014)\citenamefont {Bouda},
  \citenamefont {Paw\l{}owski}, \citenamefont {Pivoluska},\ and\ \citenamefont
  {Plesch}}]{Bouda14}%
  \BibitemOpen
  \bibfield  {author} {\bibinfo {author} {\bibnamefont {Bouda}, \bibfnamefont
  {Jan}}, \bibinfo {author} {\bibfnamefont {Marcin}\ \bibnamefont
  {Paw\l{}owski}}, \bibinfo {author} {\bibfnamefont {Matej}\ \bibnamefont
  {Pivoluska}}, \ and\ \bibinfo {author} {\bibfnamefont {Martin}\ \bibnamefont
  {Plesch}}} (\bibinfo {year} {2014}),\ \bibfield  {title} {\enquote {\bibinfo
  {title} {Device-independent randomness extraction from an arbitrarily weak
  min-entropy source},}\ }\href {\doibase 10.1103/PhysRevA.90.032313}
  {\bibfield  {journal} {\bibinfo  {journal} {Phys. Rev. A}\ }\textbf {\bibinfo
  {volume} {90}},\ \bibinfo {pages} {032313}}\BibitemShut {NoStop}%
\bibitem [{\citenamefont {Boussinesq}(1878)}]{Boussinesq78}%
  \BibitemOpen
  \bibfield  {author} {\bibinfo {author} {\bibnamefont {Boussinesq},
  \bibfnamefont {M~J}}} (\bibinfo {year} {1878}),\ \href@noop {} {\emph
  {\bibinfo {title} {{C}onciliation du v\'eritable d\'eterminisme m\'ecanique
  avec l'existence de la vie et de la libert\'e morale}}}\ (\bibinfo
  {publisher} {{M}\'emoire de {M}. {J}. {B}oussinesq pr\'ec\'ed\'e d'un rapport
  \'a {l'Acad\'emie} des {S}ciences {M}orales et {P}olitiques, par {M}. {P}aul
  {J}anet, Paris})\BibitemShut {NoStop}%
\bibitem [{\citenamefont {Brand{\~a}o}\ \emph {et~al.}(2016)\citenamefont
  {Brand{\~a}o}, \citenamefont {Ramanathan}, \citenamefont {Grudka},
  \citenamefont {Horodecki}, \citenamefont {Horodecki}, \citenamefont
  {Horodecki}, \citenamefont {Szarek},\ and\ \citenamefont
  {Wojewodka}}]{Brandao16}%
  \BibitemOpen
  \bibfield  {author} {\bibinfo {author} {\bibnamefont {Brand{\~a}o},
  \bibfnamefont {Fernando G S~L}}, \bibinfo {author} {\bibfnamefont
  {Ravishankar}\ \bibnamefont {Ramanathan}}, \bibinfo {author} {\bibfnamefont
  {Andrzej}\ \bibnamefont {Grudka}}, \bibinfo {author} {\bibfnamefont {Karol}\
  \bibnamefont {Horodecki}}, \bibinfo {author} {\bibfnamefont {Micha{\l}}\
  \bibnamefont {Horodecki}}, \bibinfo {author} {\bibfnamefont {Pawe{\l}}\
  \bibnamefont {Horodecki}}, \bibinfo {author} {\bibfnamefont {Tomasz}\
  \bibnamefont {Szarek}}, \ and\ \bibinfo {author} {\bibfnamefont {Hanna}\
  \bibnamefont {Wojewodka}}} (\bibinfo {year} {2016}),\ \bibfield  {title}
  {\enquote {\bibinfo {title} {Realistic noise-tolerant randomness
  amplification using finite number of devices},}\ }\href {\doibase
  10.1038/ncomms11345} {\bibfield  {journal} {\bibinfo  {journal} {Nat Commun}\
  }\textbf {\bibinfo {volume} {7}},\ 10.1038/ncomms11345}\BibitemShut {NoStop}%
\bibitem [{\citenamefont {Brassard}\ and\ \citenamefont
  {Robichaud}(2013)}]{brassard13}%
  \BibitemOpen
  \bibfield  {author} {\bibinfo {author} {\bibnamefont {Brassard},
  \bibfnamefont {Gilles}}, \ and\ \bibinfo {author} {\bibfnamefont
  {Paul~Raymond}\ \bibnamefont {Robichaud}}} (\bibinfo {year} {2013}),\
  \bibfield  {title} {\enquote {\bibinfo {title} {{C}an {F}ree {W}ill {E}merge
  from {D}eterminism in {Q}uantum {T}heory?}}\ }in\ \href@noop {} {\emph
  {\bibinfo {booktitle} {{I}s {S}cience {C}ompatible with {F}ree {W}ill?
  {E}xploring {F}ree {W}ill and {C}onsciousness in the {L}ight of {Q}uantum
  {P}hysics and {N}euroscience}}},\ Chap.~\bibinfo {chapter} {4}\ (\bibinfo
  {publisher} {Springer},\ \bibinfo {address} {New York})\ pp.\ \bibinfo
  {pages} {41--61}\BibitemShut {NoStop}%
\bibitem [{\citenamefont {Bricmont}(1995)}]{Bricmont95}%
  \BibitemOpen
  \bibfield  {author} {\bibinfo {author} {\bibnamefont {Bricmont},
  \bibfnamefont {J}}} (\bibinfo {year} {1995}),\ \bibfield  {title} {\enquote
  {\bibinfo {title} {{SCIENCE OF CHAOS OR CHAOS IN SCIENCE?}}}\ }\href
  {\doibase 10.1111/j.1749-6632.1996.tb23135.x} {\bibfield  {journal} {\bibinfo
   {journal} {Annals of the New York Academy of Sciences}\ }\textbf {\bibinfo
  {volume} {775}}~(\bibinfo {number} {1}),\ \bibinfo {pages}
  {131--175}}\BibitemShut {NoStop}%
\bibitem [{\citenamefont {Brown}(2004)}]{BrownWEB2004}%
  \BibitemOpen
  \bibfield  {author} {\bibinfo {author} {\bibnamefont {Brown}, \bibfnamefont
  {Robert~G}}} (\bibinfo {year} {2004}),\ \bibfield  {title} {\enquote
  {\bibinfo {title} {Dieharder: A random number test suite},}\ }\href
  {http://www.phy.duke.edu/~rgb/General/dieharder.php} {\bibfield  {journal}
  {\bibinfo  {journal}
  {http://www.phy.duke.edu/{$\sim$}rgb/General/dieharder.php}\ }\textbf
  {\bibinfo {volume} {~}}}\BibitemShut {NoStop}%
\bibitem [{\citenamefont {Brunner}(2014)}]{Brunner14a}%
  \BibitemOpen
  \bibfield  {author} {\bibinfo {author} {\bibnamefont {Brunner}, \bibfnamefont
  {Nicolas}}} (\bibinfo {year} {2014}),\ \bibfield  {title} {\enquote {\bibinfo
  {title} {Device-independent quantum information processing},}\ }in\ \href
  {\doibase 10.1364/QIM.2014.QW3A.2} {\emph {\bibinfo {booktitle} {Research in
  Optical Sciences}}}\ (\bibinfo  {publisher} {Optical Society of America})\
  p.\ \bibinfo {pages} {QW3A.2}\BibitemShut {NoStop}%
\bibitem [{\citenamefont {Brunner}\ \emph {et~al.}(2014)\citenamefont
  {Brunner}, \citenamefont {Cavalcanti}, \citenamefont {Pironio}, \citenamefont
  {Scarani},\ and\ \citenamefont {Wehner}}]{Brunner14}%
  \BibitemOpen
  \bibfield  {author} {\bibinfo {author} {\bibnamefont {Brunner}, \bibfnamefont
  {Nicolas}}, \bibinfo {author} {\bibfnamefont {Daniel}\ \bibnamefont
  {Cavalcanti}}, \bibinfo {author} {\bibfnamefont {Stefano}\ \bibnamefont
  {Pironio}}, \bibinfo {author} {\bibfnamefont {Valerio}\ \bibnamefont
  {Scarani}}, \ and\ \bibinfo {author} {\bibfnamefont {Stephanie}\ \bibnamefont
  {Wehner}}} (\bibinfo {year} {2014}),\ \bibfield  {title} {\enquote {\bibinfo
  {title} {Bell nonlocality},}\ }\href {\doibase 10.1103/RevModPhys.86.419}
  {\bibfield  {journal} {\bibinfo  {journal} {Rev. Mod. Phys.}\ }\textbf
  {\bibinfo {volume} {86}},\ \bibinfo {pages} {419--478}}\BibitemShut {NoStop}%
\bibitem [{\citenamefont {Bub}(1999)}]{Bub99}%
  \BibitemOpen
  \bibfield  {author} {\bibinfo {author} {\bibnamefont {Bub}, \bibfnamefont
  {Jeffrey}}} (\bibinfo {year} {1999}),\ \href
  {http://www.cambridge.org/us/academic/subjects/physics/quantum-physics-quantum-information-and-quantum-computation/interpreting-quantum-world}
  {\emph {\bibinfo {title} {Interpreting the Quantum World}}}\ (\bibinfo
  {publisher} {Cambridge University Press})\BibitemShut {NoStop}%
\bibitem [{\citenamefont {Cabello}(2008)}]{Cabello08}%
  \BibitemOpen
  \bibfield  {author} {\bibinfo {author} {\bibnamefont {Cabello}, \bibfnamefont
  {Ad\'an}}} (\bibinfo {year} {2008}),\ \bibfield  {title} {\enquote {\bibinfo
  {title} {Experimentally testable state-independent quantum contextuality},}\
  }\href {\doibase 10.1103/PhysRevLett.101.210401} {\bibfield  {journal}
  {\bibinfo  {journal} {Phys. Rev. Lett.}\ }\textbf {\bibinfo {volume} {101}},\
  \bibinfo {pages} {210401}}\BibitemShut {NoStop}%
\bibitem [{\citenamefont {Cabello}\ \emph {et~al.}(2014)\citenamefont
  {Cabello}, \citenamefont {Severini},\ and\ \citenamefont
  {Winter}}]{Cabello14}%
  \BibitemOpen
  \bibfield  {author} {\bibinfo {author} {\bibnamefont {Cabello}, \bibfnamefont
  {Ad\'an}}, \bibinfo {author} {\bibfnamefont {Simone}\ \bibnamefont
  {Severini}}, \ and\ \bibinfo {author} {\bibfnamefont {Andreas}\ \bibnamefont
  {Winter}}} (\bibinfo {year} {2014}),\ \bibfield  {title} {\enquote {\bibinfo
  {title} {Graph-theoretic approach to quantum correlations},}\ }\href
  {\doibase 10.1103/PhysRevLett.112.040401} {\bibfield  {journal} {\bibinfo
  {journal} {Phys. Rev. Lett.}\ }\textbf {\bibinfo {volume} {112}},\ \bibinfo
  {pages} {040401}}\BibitemShut {NoStop}%
\bibitem [{\citenamefont {Chor}\ and\ \citenamefont
  {Goldreich}(1988)}]{Chor88}%
  \BibitemOpen
  \bibfield  {author} {\bibinfo {author} {\bibnamefont {Chor}, \bibfnamefont
  {Benny}}, \ and\ \bibinfo {author} {\bibfnamefont {Oded}\ \bibnamefont
  {Goldreich}}} (\bibinfo {year} {1988}),\ \bibfield  {title} {\enquote
  {\bibinfo {title} {Unbiased bits from sources of weak randomness and
  probabilistic communication complexity},}\ }\href {\doibase 10.1137/0217015}
  {\bibfield  {journal} {\bibinfo  {journal} {SIAM Journal on Computing}\
  }\textbf {\bibinfo {volume} {17}}~(\bibinfo {number} {2}),\ \bibinfo {pages}
  {230--261}}\BibitemShut {NoStop}%
\bibitem [{\citenamefont {Christensen}\ \emph {et~al.}(2013)\citenamefont
  {Christensen}, \citenamefont {McCusker}, \citenamefont {Altepeter},
  \citenamefont {Calkins}, \citenamefont {Gerrits}, \citenamefont {Lita},
  \citenamefont {Miller}, \citenamefont {Shalm}, \citenamefont {Zhang},
  \citenamefont {Nam}, \citenamefont {Brunner}, \citenamefont {Lim},
  \citenamefont {Gisin},\ and\ \citenamefont {Kwiat}}]{ChristensenPRL2013}%
  \BibitemOpen
  \bibfield  {author} {\bibinfo {author} {\bibnamefont {Christensen},
  \bibfnamefont {B~G}}, \bibinfo {author} {\bibfnamefont {K.~T.}\ \bibnamefont
  {McCusker}}, \bibinfo {author} {\bibfnamefont {J.~B.}\ \bibnamefont
  {Altepeter}}, \bibinfo {author} {\bibfnamefont {B.}~\bibnamefont {Calkins}},
  \bibinfo {author} {\bibfnamefont {T.}~\bibnamefont {Gerrits}}, \bibinfo
  {author} {\bibfnamefont {A.~E.}\ \bibnamefont {Lita}}, \bibinfo {author}
  {\bibfnamefont {A.}~\bibnamefont {Miller}}, \bibinfo {author} {\bibfnamefont
  {L.~K.}\ \bibnamefont {Shalm}}, \bibinfo {author} {\bibfnamefont
  {Y.}~\bibnamefont {Zhang}}, \bibinfo {author} {\bibfnamefont {S.~W.}\
  \bibnamefont {Nam}}, \bibinfo {author} {\bibfnamefont {N.}~\bibnamefont
  {Brunner}}, \bibinfo {author} {\bibfnamefont {C.~C.~W.}\ \bibnamefont {Lim}},
  \bibinfo {author} {\bibfnamefont {N.}~\bibnamefont {Gisin}}, \ and\ \bibinfo
  {author} {\bibfnamefont {P.~G.}\ \bibnamefont {Kwiat}}} (\bibinfo {year}
  {2013}),\ \bibfield  {title} {\enquote {\bibinfo {title}
  {Detection-loophole-free test of quantum nonlocality, and applications},}\
  }\href {\doibase 10.1103/PhysRevLett.111.130406} {\bibfield  {journal}
  {\bibinfo  {journal} {Phys. Rev. Lett.}\ }\textbf {\bibinfo {volume} {111}},\
  \bibinfo {pages} {130406}}\BibitemShut {NoStop}%
\bibitem [{\citenamefont {Chung}\ \emph {et~al.}(2014)\citenamefont {Chung},
  \citenamefont {Shi},\ and\ \citenamefont {Wu}}]{Chung14}%
  \BibitemOpen
  \bibfield  {author} {\bibinfo {author} {\bibnamefont {Chung}, \bibfnamefont
  {Kai-Min}}, \bibinfo {author} {\bibfnamefont {Yaoyun}\ \bibnamefont {Shi}}, \
  and\ \bibinfo {author} {\bibfnamefont {Xiaodi}\ \bibnamefont {Wu}}} (\bibinfo
  {year} {2014}),\ \href@noop {} {\enquote {\bibinfo {title} {Physical
  randomness extractors: Generating random numbers with minimal assumptions},}\
  }\Eprint {http://arxiv.org/abs/1402.4797} {arXiv:1402.4797} \BibitemShut
  {NoStop}%
\bibitem [{\citenamefont {Cicero}(1933)}]{Cicero33}%
  \BibitemOpen
  \bibfield  {author} {\bibinfo {author} {\bibnamefont {Cicero}, \bibfnamefont
  {Marcus~Tullius}}} (\bibinfo {year} {1933}),\ \href@noop {} {\emph {\bibinfo
  {title} {{{D}e {N}atura {D}eorum}, {E}nglish translation by H. Rackham}}}\
  (\bibinfo  {publisher} {Harvard {U}niversity {P}ress, {C}ambridge,
  {M}assachusetts})\BibitemShut {NoStop}%
\bibitem [{\citenamefont {Clauser}\ \emph {et~al.}(1969)\citenamefont
  {Clauser}, \citenamefont {Horne}, \citenamefont {Shimony},\ and\
  \citenamefont {Holt}}]{Clauser69}%
  \BibitemOpen
  \bibfield  {author} {\bibinfo {author} {\bibnamefont {Clauser}, \bibfnamefont
  {John~F}}, \bibinfo {author} {\bibfnamefont {Michael~A.}\ \bibnamefont
  {Horne}}, \bibinfo {author} {\bibfnamefont {Abner}\ \bibnamefont {Shimony}},
  \ and\ \bibinfo {author} {\bibfnamefont {Richard~A.}\ \bibnamefont {Holt}}}
  (\bibinfo {year} {1969}),\ \bibfield  {title} {\enquote {\bibinfo {title}
  {Proposed experiment to test local hidden-variable theories},}\ }\href
  {\doibase 10.1103/PhysRevLett.23.880} {\bibfield  {journal} {\bibinfo
  {journal} {Phys. Rev. Lett.}\ }\textbf {\bibinfo {volume} {23}},\ \bibinfo
  {pages} {880--884}}\BibitemShut {NoStop}%
\bibitem [{\citenamefont {Coddington}\ and\ \citenamefont
  {Levinson}(1955)}]{Coddington55}%
  \BibitemOpen
  \bibfield  {author} {\bibinfo {author} {\bibnamefont {Coddington},
  \bibfnamefont {Earl~A}}, \ and\ \bibinfo {author} {\bibfnamefont
  {N}~\bibnamefont {Levinson}}} (\bibinfo {year} {1955}),\ \href
  {http://trove.nla.gov.au/work/10556041} {\emph {\bibinfo {title} {Theory of
  ordinary differential equations}}}\ (\bibinfo  {publisher} {New York :
  McGraw-Hill ; New Delhi : Tata McGraw-Hill})\BibitemShut {NoStop}%
\bibitem [{\citenamefont {Cohen-Tannoudji}\ \emph {et~al.}(1991)\citenamefont
  {Cohen-Tannoudji}, \citenamefont {Diu},\ and\ \citenamefont
  {Laloe}}]{Tannoudji91}%
  \BibitemOpen
  \bibfield  {author} {\bibinfo {author} {\bibnamefont {Cohen-Tannoudji},
  \bibfnamefont {Claude}}, \bibinfo {author} {\bibfnamefont {Bernard}\
  \bibnamefont {Diu}}, \ and\ \bibinfo {author} {\bibfnamefont {Frank}\
  \bibnamefont {Laloe}}} (\bibinfo {year} {1991}),\ \href@noop {} {\emph
  {\bibinfo {title} {{Q}uantum {M}echanics, Vol. 1}}}\ (\bibinfo  {publisher}
  {Wiley})\BibitemShut {NoStop}%
\bibitem [{\citenamefont {Colbeck}(2007)}]{Colbeck07}%
  \BibitemOpen
  \bibfield  {author} {\bibinfo {author} {\bibnamefont {Colbeck}, \bibfnamefont
  {R}}} (\bibinfo {year} {2007}),\ \href@noop {} {\emph {\bibinfo {title}
  {{Q}uantum and {R}elativistic {P}rotocols for {S}ecure {M}ultiparty
  {C}omputation, {PhD Thesis}}}}\ (\bibinfo  {publisher} {University of
  Cambridge})\BibitemShut {NoStop}%
\bibitem [{\citenamefont {Colbeck}\ and\ \citenamefont
  {Kent}(2011)}]{Colbeck10}%
  \BibitemOpen
  \bibfield  {author} {\bibinfo {author} {\bibnamefont {Colbeck}, \bibfnamefont
  {Roger}}, \ and\ \bibinfo {author} {\bibfnamefont {Adrian}\ \bibnamefont
  {Kent}}} (\bibinfo {year} {2011}),\ \bibfield  {title} {\enquote {\bibinfo
  {title} {Private randomness expansion with untrusted devices},}\ }\href
  {http://stacks.iop.org/1751-8121/44/i=9/a=095305} {\bibfield  {journal}
  {\bibinfo  {journal} {Journal of Physics A: Mathematical and Theoretical}\
  }\textbf {\bibinfo {volume} {44}}~(\bibinfo {number} {9}),\ \bibinfo {pages}
  {095305}}\BibitemShut {NoStop}%
\bibitem [{\citenamefont {Colbeck}\ and\ \citenamefont
  {Renner}(2012)}]{Colbeck12}%
  \BibitemOpen
  \bibfield  {author} {\bibinfo {author} {\bibnamefont {Colbeck}, \bibfnamefont
  {Roger}}, \ and\ \bibinfo {author} {\bibfnamefont {Renato}\ \bibnamefont
  {Renner}}} (\bibinfo {year} {2012}),\ \bibfield  {title} {\enquote {\bibinfo
  {title} {Free randomness can be amplified},}\ }\href {\doibase
  10.1038/nphys2300} {\bibfield  {journal} {\bibinfo  {journal} {Nature
  Physics}\ }\textbf {\bibinfo {volume} {8}},\ \bibinfo {pages}
  {450}}\BibitemShut {NoStop}%
\bibitem [{\citenamefont {Coles}\ \emph {et~al.}(2015)\citenamefont {Coles},
  \citenamefont {Berta}, \citenamefont {Tomamichel},\ and\ \citenamefont
  {Wehner}}]{Coles15}%
  \BibitemOpen
  \bibfield  {author} {\bibinfo {author} {\bibnamefont {Coles}, \bibfnamefont
  {Patrick~J}}, \bibinfo {author} {\bibfnamefont {Mario}\ \bibnamefont
  {Berta}}, \bibinfo {author} {\bibfnamefont {Marco}\ \bibnamefont
  {Tomamichel}}, \ and\ \bibinfo {author} {\bibfnamefont {Stephanie}\
  \bibnamefont {Wehner}}} (\bibinfo {year} {2015}),\ \bibfield  {title}
  {\enquote {\bibinfo {title} {Entropic uncertainty relations and their
  applications},}\ }\href {http://arxiv.org/abs/1511.04857} {\bibinfo
  {journal} {arXiv:1511.04857}\ }\BibitemShut {NoStop}%
\bibitem [{\citenamefont {Coudron}\ and\ \citenamefont
  {Yuen}(2013)}]{coudron_yuen}%
  \BibitemOpen
\bibfield  {journal} {  }\bibfield  {author} {\bibinfo {author} {\bibnamefont
  {Coudron}, \bibfnamefont {M}}, \ and\ \bibinfo {author} {\bibfnamefont
  {H.}~\bibnamefont {Yuen}}} (\bibinfo {year} {2013}),\ \bibfield  {title}
  {\enquote {\bibinfo {title} {{Infinite Randomness Expansion and Amplification
  with a Constant Number of Devices}},}\ }\href@noop {} {\bibfield  {journal}
  {\bibinfo  {journal} {ArXiv e-prints}\ }}\Eprint
  {http://arxiv.org/abs/1310.6755} {arXiv:1310.6755 [quant-ph]} \BibitemShut
  {NoStop}%
\bibitem [{\citenamefont {Dahan-Dalmedico}\ \emph {et~al.}(1992)\citenamefont
  {Dahan-Dalmedico}, \citenamefont {Chabert},\ and\ \citenamefont
  {Chemla}}]{Dahan92}%
  \BibitemOpen
  \bibfield  {author} {\bibinfo {author} {\bibnamefont {Dahan-Dalmedico},
  \bibfnamefont {Amy}}, \bibinfo {author} {\bibfnamefont {Jean-Luc}\
  \bibnamefont {Chabert}}, \ and\ \bibinfo {author} {\bibfnamefont {Karine}\
  \bibnamefont {Chemla}}} (\bibinfo {year} {1992}),\ \href@noop {} {\emph
  {\bibinfo {title} {Chaos et d\'eterminisme}}}\ (\bibinfo  {publisher} {Points
  Sciences})\BibitemShut {NoStop}%
\bibitem [{\citenamefont {Deutsch}(1983)}]{Deutsch83}%
  \BibitemOpen
  \bibfield  {author} {\bibinfo {author} {\bibnamefont {Deutsch}, \bibfnamefont
  {David}}} (\bibinfo {year} {1983}),\ \bibfield  {title} {\enquote {\bibinfo
  {title} {Uncertainty in quantum measurements},}\ }\href {\doibase
  10.1103/PhysRevLett.50.631} {\bibfield  {journal} {\bibinfo  {journal} {Phys.
  Rev. Lett.}\ }\textbf {\bibinfo {volume} {50}},\ \bibinfo {pages}
  {631--633}}\BibitemShut {NoStop}%
\bibitem [{\citenamefont {Diels}(1906)}]{Diels06}%
  \BibitemOpen
  \bibfield  {author} {\bibinfo {author} {\bibnamefont {Diels}, \bibfnamefont
  {H}}} (\bibinfo {year} {1906}),\ \href@noop {} {\emph {\bibinfo {title}
  {{D}ie fragmente der {V}orsokratiker griechisch und deutsch}}}\ (\bibinfo
  {publisher} {Weidmannsche {B}uchhandlung})\BibitemShut {NoStop}%
\bibitem [{\citenamefont {Dodis}\ \emph {et~al.}(2004)\citenamefont {Dodis},
  \citenamefont {Elbaz}, \citenamefont {Oliveira},\ and\ \citenamefont
  {Raz}}]{Dodis04}%
  \BibitemOpen
  \bibfield  {author} {\bibinfo {author} {\bibnamefont {Dodis}, \bibfnamefont
  {Yevgeniy}}, \bibinfo {author} {\bibfnamefont {Ariel}\ \bibnamefont {Elbaz}},
  \bibinfo {author} {\bibfnamefont {Roberto}\ \bibnamefont {Oliveira}}, \ and\
  \bibinfo {author} {\bibfnamefont {Ran}\ \bibnamefont {Raz}}} (\bibinfo {year}
  {2004}),\ \enquote {\bibinfo {title} {Approximation, randomization, and
  combinatorial optimization. algorithms and techniques: 7th international
  workshop on approximation algorithms for combinatorial optimization problems,
  { APPROX 2004}, and 8th international workshop on randomization and
  computation, { RANDOM 2004, Cambridge, MA, USA, August 22-24, 2004.
  Proceedings}},}\ Chap.\ \bibinfo {chapter} {Improved Randomness Extraction
  from Two Independent Sources}\ (\bibinfo  {publisher} {Springer Berlin
  Heidelberg},\ \bibinfo {address} {Berlin, Heidelberg})\ pp.\ \bibinfo {pages}
  {334--344}\BibitemShut {NoStop}%
\bibitem [{\citenamefont {Einstein}\ \emph {et~al.}(1935)\citenamefont
  {Einstein}, \citenamefont {Podolsky},\ and\ \citenamefont {Rosen}}]{EPR35}%
  \BibitemOpen
  \bibfield  {author} {\bibinfo {author} {\bibnamefont {Einstein},
  \bibfnamefont {A}}, \bibinfo {author} {\bibfnamefont {B.}~\bibnamefont
  {Podolsky}}, \ and\ \bibinfo {author} {\bibfnamefont {N.}~\bibnamefont
  {Rosen}}} (\bibinfo {year} {1935}),\ \bibfield  {title} {\enquote {\bibinfo
  {title} {Can quantum-mechanical description of physical reality be considered
  complete?}}\ }\href {\doibase 10.1103/PhysRev.47.777} {\bibfield  {journal}
  {\bibinfo  {journal} {Phys. Rev.}\ }\textbf {\bibinfo {volume} {47}},\
  \bibinfo {pages} {777--780}}\BibitemShut {NoStop}%
\bibitem [{\citenamefont {Everett}(1957)}]{Everett57}%
  \BibitemOpen
  \bibfield  {author} {\bibinfo {author} {\bibnamefont {Everett}, \bibfnamefont
  {Hugh}}} (\bibinfo {year} {1957}),\ \bibfield  {title} {\enquote {\bibinfo
  {title} {{R}elative state formulation of quantum mechanics},}\ }\href
  {\doibase 10.1103/RevModPhys.29.454} {\bibfield  {journal} {\bibinfo
  {journal} {Rev. Mod. Phys.}\ }\textbf {\bibinfo {volume} {29}},\ \bibinfo
  {pages} {454--462}}\BibitemShut {NoStop}%
\bibitem [{\citenamefont {Falkovich}\ \emph {et~al.}(2001)\citenamefont
  {Falkovich}, \citenamefont {Gaw\ifmmode~\mbox{\c{e}}\else \c{e}\fi{}dzki},\
  and\ \citenamefont {Vergassola}}]{Falkovich01}%
  \BibitemOpen
  \bibfield  {author} {\bibinfo {author} {\bibnamefont {Falkovich},
  \bibfnamefont {G}}, \bibinfo {author} {\bibfnamefont {K.}~\bibnamefont
  {Gaw\ifmmode~\mbox{\c{e}}\else \c{e}\fi{}dzki}}, \ and\ \bibinfo {author}
  {\bibfnamefont {M.}~\bibnamefont {Vergassola}}} (\bibinfo {year} {2001}),\
  \bibfield  {title} {\enquote {\bibinfo {title} {Particles and fields in fluid
  turbulence},}\ }\href {\doibase 10.1103/RevModPhys.73.913} {\bibfield
  {journal} {\bibinfo  {journal} {Rev. Mod. Phys.}\ }\textbf {\bibinfo {volume}
  {73}},\ \bibinfo {pages} {913--975}}\BibitemShut {NoStop}%
\bibitem [{\citenamefont {Farge}(1991)}]{Farge}%
  \BibitemOpen
  \bibfield  {author} {\bibinfo {author} {\bibnamefont {Farge}, \bibfnamefont
  {Marie}}} (\bibinfo {year} {1991}),\ \bibfield  {title} {\enquote {\bibinfo
  {title} {L'evolution des id\'ees sur la turbulence: 1870-1970},}\ }\bibfield
  {booktitle} {\emph {\bibinfo {booktitle} {Un siècle de rapports entre
  mathematiques et physique: 1870-1970}},\ }\href
  {http://wavelets.ens.fr/PUBLICATIONS/publications.htm} {\ \textbf {\bibinfo
  {volume} {40}},\ \bibinfo {pages} {87--96}}\BibitemShut {NoStop}%
\bibitem [{\citenamefont {Fine}(1982)}]{Fine82}%
  \BibitemOpen
  \bibfield  {author} {\bibinfo {author} {\bibnamefont {Fine}, \bibfnamefont
  {Arthur}}} (\bibinfo {year} {1982}),\ \bibfield  {title} {\enquote {\bibinfo
  {title} {Hidden variables, joint probability, and the {B}ell inequalities},}\
  }\href {\doibase 10.1103/PhysRevLett.48.291} {\bibfield  {journal} {\bibinfo
  {journal} {Phys. Rev. Lett.}\ }\textbf {\bibinfo {volume} {48}},\ \bibinfo
  {pages} {291--295}}\BibitemShut {NoStop}%
\bibitem [{\citenamefont {Fletcher}(2012)}]{Fletcher12}%
  \BibitemOpen
  \bibfield  {author} {\bibinfo {author} {\bibnamefont {Fletcher},
  \bibfnamefont {Samuel~Craig}}} (\bibinfo {year} {2012}),\ \bibfield  {title}
  {\enquote {\bibinfo {title} {What counts as a {N}ewtonian system? the view
  from norton's dome},}\ }\href {\doibase 10.1007/s13194-011-0040-8} {\bibfield
   {journal} {\bibinfo  {journal} {European {J}ournal for Philosophy of
  Science}\ }\textbf {\bibinfo {volume} {2}}~(\bibinfo {number} {3}),\ \bibinfo
  {pages} {275--297}}\BibitemShut {NoStop}%
\bibitem [{\citenamefont {Freedman}\ and\ \citenamefont
  {Clauser}(1972)}]{FreedmanPRL1972}%
  \BibitemOpen
  \bibfield  {author} {\bibinfo {author} {\bibnamefont {Freedman},
  \bibfnamefont {Stuart~J}}, \ and\ \bibinfo {author} {\bibfnamefont {John~F.}\
  \bibnamefont {Clauser}}} (\bibinfo {year} {1972}),\ \bibfield  {title}
  {\enquote {\bibinfo {title} {Experimental test of local hidden-variable
  theories},}\ }\href {\doibase 10.1103/PhysRevLett.28.938} {\bibfield
  {journal} {\bibinfo  {journal} {Phys. Rev. Lett.}\ }\textbf {\bibinfo
  {volume} {28}},\ \bibinfo {pages} {938--941}}\BibitemShut {NoStop}%
\bibitem [{\citenamefont {Freeman}(1948)}]{Freeman48}%
  \BibitemOpen
  \bibfield  {author} {\bibinfo {author} {\bibnamefont {Freeman}, \bibfnamefont
  {K}}} (\bibinfo {year} {1948}),\ \href@noop {} {\emph {\bibinfo {title}
  {{A}ncilla to the pre-{S}ocratic philosophers}}}\ (\bibinfo  {publisher}
  {Forgotten Books, Cambridge, Massachusetts})\BibitemShut {NoStop}%
\bibitem [{\citenamefont {Gabizon}\ and\ \citenamefont
  {Shaltiel}(2008)}]{Gabizon08}%
  \BibitemOpen
  \bibfield  {author} {\bibinfo {author} {\bibnamefont {Gabizon}, \bibfnamefont
  {Ariel}}, \ and\ \bibinfo {author} {\bibfnamefont {Ronen}\ \bibnamefont
  {Shaltiel}}} (\bibinfo {year} {2008}),\ \enquote {\bibinfo {title}
  {Approximation, randomization and combinatorial optimization. algorithms and
  techniques: 11th international workshop, approx 2008, and 12th international
  workshop, random 2008, boston, ma, usa, august 25-27, 2008. proceedings},}\
  Chap.\ \bibinfo {chapter} {Increasing the Output Length of Zero-Error
  Dispersers}\ (\bibinfo  {publisher} {Springer Berlin Heidelberg},\ \bibinfo
  {address} {Berlin, Heidelberg})\ pp.\ \bibinfo {pages} {430--443}\BibitemShut
  {NoStop}%
\bibitem [{\citenamefont {Gallego}\ \emph {et~al.}(2012)\citenamefont
  {Gallego}, \citenamefont {W\"urflinger}, \citenamefont {Ac\'\i~n},\ and\
  \citenamefont {Navascu\'es}}]{GallegoPRL2012}%
  \BibitemOpen
  \bibfield  {author} {\bibinfo {author} {\bibnamefont {Gallego}, \bibfnamefont
  {R}}, \bibinfo {author} {\bibfnamefont {L.~E.}\ \bibnamefont {W\"urflinger}},
  \bibinfo {author} {\bibfnamefont {A.}~\bibnamefont {Ac\'\i~n}}, \ and\
  \bibinfo {author} {\bibfnamefont {M.}~\bibnamefont {Navascu\'es}}} (\bibinfo
  {year} {2012}),\ \bibfield  {title} {\enquote {\bibinfo {title} {An
  operational framework for nonlocality},}\ }\href {\doibase
  10.1103/PhysRevLett.109.070401} {\bibfield  {journal} {\bibinfo  {journal}
  {Phys. Rev. Lett.}\ }\textbf {\bibinfo {volume} {109}},\ \bibinfo {pages}
  {070401}}\BibitemShut {NoStop}%
\bibitem [{\citenamefont {Gallego}\ \emph {et~al.}(2013)\citenamefont
  {Gallego}, \citenamefont {Masanes}, \citenamefont {De~La~Torre},
  \citenamefont {Dhara}, \citenamefont {Aolita},\ and\ \citenamefont
  {Acín}}]{Gallego13}%
  \BibitemOpen
  \bibfield  {author} {\bibinfo {author} {\bibnamefont {Gallego}, \bibfnamefont
  {Rodrigo}}, \bibinfo {author} {\bibfnamefont {Lluis}\ \bibnamefont
  {Masanes}}, \bibinfo {author} {\bibfnamefont {Gonzalo}\ \bibnamefont
  {De~La~Torre}}, \bibinfo {author} {\bibfnamefont {Chirag}\ \bibnamefont
  {Dhara}}, \bibinfo {author} {\bibfnamefont {Leandro}\ \bibnamefont {Aolita}},
  \ and\ \bibinfo {author} {\bibfnamefont {Antonio}\ \bibnamefont {Acín}}}
  (\bibinfo {year} {2013}),\ \bibfield  {title} {\enquote {\bibinfo {title}
  {Full randomness from arbitrarily deterministic events},}\ }\href {\doibase
  10.1038/ncomms3654} {\bibfield  {journal} {\bibinfo  {journal} {Nat Commun}\
  }\textbf {\bibinfo {volume} {4}},\ 10.1038/ncomms3654}\BibitemShut {NoStop}%
\bibitem [{\citenamefont {Gawedzki}(2001)}]{Gawedz-rec}%
  \BibitemOpen
  \bibfield  {author} {\bibinfo {author} {\bibnamefont {Gawedzki},
  \bibfnamefont {K}}} (\bibinfo {year} {2001}),\ \href
  {http://www.cambridge.org/gb/academic/subjects/mathematics/fluid-dynamics-and-solid-mechanics/intermittency-turbulent-flows?format=HB&isbn=9780521792219#E81Bf11QWMSMLRQB.97}
  {\emph {\bibinfo {title} {Intermittency in Turbulent Flows, ed. J. C.
  Vassilicos}}}\ (\bibinfo  {publisher} {Cambridge University Press,
  Cambridge})\BibitemShut {NoStop}%
\bibitem [{\citenamefont {Gawedzki}\ and\ \citenamefont
  {Vergassola}(2000)}]{Gawedz2}%
  \BibitemOpen
  \bibfield  {author} {\bibinfo {author} {\bibnamefont {Gawedzki},
  \bibfnamefont {Krzysztof}}, \ and\ \bibinfo {author} {\bibfnamefont
  {Massimo}\ \bibnamefont {Vergassola}}} (\bibinfo {year} {2000}),\ \bibfield
  {title} {\enquote {\bibinfo {title} {Phase transition in the passive scalar
  advection},}\ }\href {\doibase https://doi.org/10.1016/S0167-2789(99)00171-2}
  {\bibfield  {journal} {\bibinfo  {journal} {Physica D: Nonlinear Phenomena}\
  }\textbf {\bibinfo {volume} {138}}~(\bibinfo {number} {1-2}),\ \bibinfo
  {pages} {63--90}}\BibitemShut {NoStop}%
\bibitem [{\citenamefont {Gisin}(2013)}]{gisin13}%
  \BibitemOpen
  \bibfield  {author} {\bibinfo {author} {\bibnamefont {Gisin}, \bibfnamefont
  {Nicolas}}} (\bibinfo {year} {2013}),\ \bibfield  {title} {\enquote {\bibinfo
  {title} {{A}re {T}here {Q}uantum {E}ffects {C}oming from {O}utside
  {S}pace–{T}ime? {N}onlocality, {F}ree {W}ill and “{N}o
  {M}any-{W}orlds”},}\ }in\ \href@noop {} {\emph {\bibinfo {booktitle} {{I}s
  {S}cience {C}ompatible with {F}ree {W}ill? {E}xploring {F}ree {W}ill and
  {C}onsciousness in the {L}ight of {Q}uantum {P}hysics and {N}euroscience}}},\
  Chap.~\bibinfo {chapter} {3}\ (\bibinfo  {publisher} {Springer},\ \bibinfo
  {address} {New York})\ pp.\ \bibinfo {pages} {23--40}\BibitemShut {NoStop}%
\bibitem [{\citenamefont {Giustina}\ \emph {et~al.}(2015)\citenamefont
  {Giustina}, \citenamefont {Versteegh}, \citenamefont {Wengerowsky},
  \citenamefont {Handsteiner}, \citenamefont {Hochrainer}, \citenamefont
  {Phelan}, \citenamefont {Steinlechner}, \citenamefont {Kofler}, \citenamefont
  {Larsson}, \citenamefont {Abell\'an}, \citenamefont {Amaya}, \citenamefont
  {Pruneri}, \citenamefont {Mitchell}, \citenamefont {Beyer}, \citenamefont
  {Gerrits}, \citenamefont {Lita}, \citenamefont {Shalm}, \citenamefont {Nam},
  \citenamefont {Scheidl}, \citenamefont {Ursin}, \citenamefont {Wittmann},\
  and\ \citenamefont {Zeilinger}}]{Giustina15}%
  \BibitemOpen
  \bibfield  {author} {\bibinfo {author} {\bibnamefont {Giustina},
  \bibfnamefont {Marissa}}, \bibinfo {author} {\bibfnamefont {Marijn A.~M.}\
  \bibnamefont {Versteegh}}, \bibinfo {author} {\bibfnamefont {S\"oren}\
  \bibnamefont {Wengerowsky}}, \bibinfo {author} {\bibfnamefont {Johannes}\
  \bibnamefont {Handsteiner}}, \bibinfo {author} {\bibfnamefont {Armin}\
  \bibnamefont {Hochrainer}}, \bibinfo {author} {\bibfnamefont {Kevin}\
  \bibnamefont {Phelan}}, \bibinfo {author} {\bibfnamefont {Fabian}\
  \bibnamefont {Steinlechner}}, \bibinfo {author} {\bibfnamefont {Johannes}\
  \bibnamefont {Kofler}}, \bibinfo {author} {\bibfnamefont {Jan-\AA{}ke}\
  \bibnamefont {Larsson}}, \bibinfo {author} {\bibfnamefont {Carlos}\
  \bibnamefont {Abell\'an}}, \bibinfo {author} {\bibfnamefont {Waldimar}\
  \bibnamefont {Amaya}}, \bibinfo {author} {\bibfnamefont {Valerio}\
  \bibnamefont {Pruneri}}, \bibinfo {author} {\bibfnamefont {Morgan~W.}\
  \bibnamefont {Mitchell}}, \bibinfo {author} {\bibfnamefont {J\"orn}\
  \bibnamefont {Beyer}}, \bibinfo {author} {\bibfnamefont {Thomas}\
  \bibnamefont {Gerrits}}, \bibinfo {author} {\bibfnamefont {Adriana~E.}\
  \bibnamefont {Lita}}, \bibinfo {author} {\bibfnamefont {Lynden~K.}\
  \bibnamefont {Shalm}}, \bibinfo {author} {\bibfnamefont {Sae~Woo}\
  \bibnamefont {Nam}}, \bibinfo {author} {\bibfnamefont {Thomas}\ \bibnamefont
  {Scheidl}}, \bibinfo {author} {\bibfnamefont {Rupert}\ \bibnamefont {Ursin}},
  \bibinfo {author} {\bibfnamefont {Bernhard}\ \bibnamefont {Wittmann}}, \ and\
  \bibinfo {author} {\bibfnamefont {Anton}\ \bibnamefont {Zeilinger}}}
  (\bibinfo {year} {2015}),\ \bibfield  {title} {\enquote {\bibinfo {title}
  {Significant-loophole-free test of {B}ell's theorem with entangled
  photons},}\ }\href {\doibase 10.1103/PhysRevLett.115.250401} {\bibfield
  {journal} {\bibinfo  {journal} {Phys. Rev. Lett.}\ }\textbf {\bibinfo
  {volume} {115}},\ \bibinfo {pages} {250401}}\BibitemShut {NoStop}%
\bibitem [{\citenamefont {Gleason}(1975)}]{Gleason75}%
  \BibitemOpen
  \bibfield  {author} {\bibinfo {author} {\bibnamefont {Gleason}, \bibfnamefont
  {Andrew~M}}} (\bibinfo {year} {1975}),\ \enquote {\bibinfo {title} {The
  logico-algebraic approach to quantum mechanics: Volume i: Historical
  evolution},}\ Chap.\ \bibinfo {chapter} {Measures on the Closed Subspaces of
  a {H}ilbert Space}\ (\bibinfo  {publisher} {Springer Netherlands},\ \bibinfo
  {address} {Dordrecht})\ pp.\ \bibinfo {pages} {123--133}\BibitemShut
  {NoStop}%
\bibitem [{\citenamefont {Gleick}(2008)}]{Gleick08}%
  \BibitemOpen
  \bibfield  {author} {\bibinfo {author} {\bibnamefont {Gleick}, \bibfnamefont
  {James}}} (\bibinfo {year} {2008}),\ \href@noop {} {\emph {\bibinfo {title}
  {{C}haos: {M}aking a {N}ew {S}cience}}}\ (\bibinfo  {publisher} {Penguin
  Books})\BibitemShut {NoStop}%
\bibitem [{\citenamefont {Grudka}\ \emph {et~al.}(2014)\citenamefont {Grudka},
  \citenamefont {Horodecki}, \citenamefont {Horodecki}, \citenamefont
  {Horodecki}, \citenamefont {Paw\l{}owski},\ and\ \citenamefont
  {Ramanathan}}]{Grudka14}%
  \BibitemOpen
  \bibfield  {author} {\bibinfo {author} {\bibnamefont {Grudka}, \bibfnamefont
  {Andrzej}}, \bibinfo {author} {\bibfnamefont {Karol}\ \bibnamefont
  {Horodecki}}, \bibinfo {author} {\bibfnamefont {Micha\l{}}\ \bibnamefont
  {Horodecki}}, \bibinfo {author} {\bibfnamefont {Pawe\l{}}\ \bibnamefont
  {Horodecki}}, \bibinfo {author} {\bibfnamefont {Marcin}\ \bibnamefont
  {Paw\l{}owski}}, \ and\ \bibinfo {author} {\bibfnamefont {Ravishankar}\
  \bibnamefont {Ramanathan}}} (\bibinfo {year} {2014}),\ \bibfield  {title}
  {\enquote {\bibinfo {title} {Free randomness amplification using bipartite
  chain correlations},}\ }\href {\doibase 10.1103/PhysRevA.90.032322}
  {\bibfield  {journal} {\bibinfo  {journal} {Phys. Rev. A}\ }\textbf {\bibinfo
  {volume} {90}},\ \bibinfo {pages} {032322}}\BibitemShut {NoStop}%
\bibitem [{\citenamefont {Gudder}(1970)}]{Gudder70}%
  \BibitemOpen
  \bibfield  {author} {\bibinfo {author} {\bibnamefont {Gudder}, \bibfnamefont
  {Stanley~P}}} (\bibinfo {year} {1970}),\ \bibfield  {title} {\enquote
  {\bibinfo {title} {On {H}idden-{V}ariable {T}heories},}\ }\href {\doibase
  http://dx.doi.org/10.1063/1.1665156} {\bibfield  {journal} {\bibinfo
  {journal} {{J}ournal of Mathematical Physics}\ }\textbf {\bibinfo {volume}
  {11}}~(\bibinfo {number} {2}),\ \bibinfo {pages} {431--436}}\BibitemShut
  {NoStop}%
\bibitem [{\citenamefont {Hall}(2010)}]{Hall10}%
  \BibitemOpen
  \bibfield  {author} {\bibinfo {author} {\bibnamefont {Hall}, \bibfnamefont
  {Michael J~W}}} (\bibinfo {year} {2010}),\ \bibfield  {title} {\enquote
  {\bibinfo {title} {Local deterministic model of singlet state correlations
  based on relaxing measurement independence},}\ }\href {\doibase
  10.1103/PhysRevLett.105.250404} {\bibfield  {journal} {\bibinfo  {journal}
  {Phys. Rev. Lett.}\ }\textbf {\bibinfo {volume} {105}},\ \bibinfo {pages}
  {250404}}\BibitemShut {NoStop}%
\bibitem [{\citenamefont {Halmos}(2013)}]{Halmos13}%
  \BibitemOpen
  \bibfield  {author} {\bibinfo {author} {\bibnamefont {Halmos}, \bibfnamefont
  {Paul~R}}} (\bibinfo {year} {2013}),\ \href@noop {} {\emph {\bibinfo {title}
  {{L}ectures on {E}rgodic {T}heory}}}\ (\bibinfo  {publisher} {Martino Fine
  Books})\BibitemShut {NoStop}%
\bibitem [{\citenamefont {Heisenberg}(1927)}]{Heisenberg27}%
  \BibitemOpen
  \bibfield  {author} {\bibinfo {author} {\bibnamefont {Heisenberg},
  \bibfnamefont {W}}} (\bibinfo {year} {1927}),\ \bibfield  {title} {\enquote
  {\bibinfo {title} {{\"U}ber den anschaulichen inhalt der quantentheoretischen
  kinematik und mechanik},}\ }\href {\doibase 10.1007/BF01397280} {\bibfield
  {journal} {\bibinfo  {journal} {Zeitschrift f{\"u}r Physik}\ }\textbf
  {\bibinfo {volume} {43}}~(\bibinfo {number} {3}),\ \bibinfo {pages}
  {172--198}}\BibitemShut {NoStop}%
\bibitem [{\citenamefont {Hensen}\ \emph {et~al.}(2015)\citenamefont {Hensen},
  \citenamefont {Bernien}, \citenamefont {Dreau}, \citenamefont {Reiserer},
  \citenamefont {Kalb}, \citenamefont {Blok}, \citenamefont {Ruitenberg},
  \citenamefont {Vermeulen}, \citenamefont {Schouten}, \citenamefont {Abellan},
  \citenamefont {Amaya}, \citenamefont {Pruneri}, \citenamefont {Mitchell},
  \citenamefont {Markham}, \citenamefont {Twitchen}, \citenamefont {Elkouss},
  \citenamefont {Wehner}, \citenamefont {Taminiau},\ and\ \citenamefont
  {Hanson}}]{Hensen15}%
  \BibitemOpen
  \bibfield  {author} {\bibinfo {author} {\bibnamefont {Hensen}, \bibfnamefont
  {B}}, \bibinfo {author} {\bibfnamefont {H.}~\bibnamefont {Bernien}}, \bibinfo
  {author} {\bibfnamefont {A.~E.}\ \bibnamefont {Dreau}}, \bibinfo {author}
  {\bibfnamefont {A.}~\bibnamefont {Reiserer}}, \bibinfo {author}
  {\bibfnamefont {N.}~\bibnamefont {Kalb}}, \bibinfo {author} {\bibfnamefont
  {M.~S.}\ \bibnamefont {Blok}}, \bibinfo {author} {\bibfnamefont
  {J.}~\bibnamefont {Ruitenberg}}, \bibinfo {author} {\bibfnamefont {R.~F.~L.}\
  \bibnamefont {Vermeulen}}, \bibinfo {author} {\bibfnamefont {R.~N.}\
  \bibnamefont {Schouten}}, \bibinfo {author} {\bibfnamefont {C.}~\bibnamefont
  {Abellan}}, \bibinfo {author} {\bibfnamefont {W.}~\bibnamefont {Amaya}},
  \bibinfo {author} {\bibfnamefont {V.}~\bibnamefont {Pruneri}}, \bibinfo
  {author} {\bibfnamefont {M.~W.}\ \bibnamefont {Mitchell}}, \bibinfo {author}
  {\bibfnamefont {M.}~\bibnamefont {Markham}}, \bibinfo {author} {\bibfnamefont
  {D.~J.}\ \bibnamefont {Twitchen}}, \bibinfo {author} {\bibfnamefont
  {D.}~\bibnamefont {Elkouss}}, \bibinfo {author} {\bibfnamefont
  {S.}~\bibnamefont {Wehner}}, \bibinfo {author} {\bibfnamefont {T.~H.}\
  \bibnamefont {Taminiau}}, \ and\ \bibinfo {author} {\bibfnamefont
  {R.}~\bibnamefont {Hanson}}} (\bibinfo {year} {2015}),\ \bibfield  {title}
  {\enquote {\bibinfo {title} {Loophole-free {B}ell inequality violation using
  electron spins separated by 1.3 kilometres},}\ }\href {\doibase
  10.1038/nature15759} {\bibfield  {journal} {\bibinfo  {journal} {Nature}\
  }\textbf {\bibinfo {volume} {526}},\ \bibinfo {pages} {682--686}}\BibitemShut
  {NoStop}%
\bibitem [{\citenamefont {Herrero-Collantes}\ and\ \citenamefont
  {Garcia-Escartin}(2017)}]{HerreroARX2016}%
  \BibitemOpen
  \bibfield  {author} {\bibinfo {author} {\bibnamefont {Herrero-Collantes},
  \bibfnamefont {Miguel}}, \ and\ \bibinfo {author} {\bibfnamefont
  {Juan~Carlos}\ \bibnamefont {Garcia-Escartin}}} (\bibinfo {year} {2017}),\
  \bibfield  {title} {\enquote {\bibinfo {title} {Quantum random number
  generators},}\ }\href {\doibase 10.1103/RevModPhys.89.015004} {\bibfield
  {journal} {\bibinfo  {journal} {Rev. Mod. Phys.}\ }\textbf {\bibinfo {volume}
  {89}},\ \bibinfo {pages} {015004}}\BibitemShut {NoStop}%
\bibitem [{\citenamefont {Isham}\ and\ \citenamefont
  {Butterfield}(1998)}]{Isham98}%
  \BibitemOpen
  \bibfield  {author} {\bibinfo {author} {\bibnamefont {Isham}, \bibfnamefont
  {C~J}}, \ and\ \bibinfo {author} {\bibfnamefont {J.}~\bibnamefont
  {Butterfield}}} (\bibinfo {year} {1998}),\ \bibfield  {title} {\enquote
  {\bibinfo {title} {A topos perspective on the {K}ochen-{S}pecker theorem: I.
  quantum states as generalized valuations},}\ }\href
  {http://arxiv.org/abs/quant-ph/9803055v4} {\bibinfo  {journal}
  {arXiv:quant-ph/9803055}\ }\BibitemShut {NoStop}%
\bibitem [{\citenamefont {Isida}\ and\ \citenamefont
  {Ikeda}(1956)}]{Isida1956}%
  \BibitemOpen
\bibfield  {journal} {  }\bibfield  {author} {\bibinfo {author} {\bibnamefont
  {Isida}, \bibfnamefont {Masatugu}}, \ and\ \bibinfo {author} {\bibfnamefont
  {Hiroji}\ \bibnamefont {Ikeda}}} (\bibinfo {year} {1956}),\ \bibfield
  {title} {\enquote {\bibinfo {title} {Random number generator},}\ }\href
  {\doibase 10.1007/BF02863577} {\bibfield  {journal} {\bibinfo  {journal}
  {Annals of the Institute of Statistical Mathematics}\ }\textbf {\bibinfo
  {volume} {8}}~(\bibinfo {number} {1}),\ \bibinfo {pages}
  {119--126}}\BibitemShut {NoStop}%
\bibitem [{\citenamefont {Isley}(2017)}]{Maxwell}%
  \BibitemOpen
  \bibfield  {author} {\bibinfo {author} {\bibnamefont {Isley}, \bibfnamefont
  {Peter}}} (\bibinfo {year} {2017}),\ \href
  {http://www.informationphilosopher.com/solutions/scientists/maxwell/science_and_free_will.html}
  {\enquote {\bibinfo {title} {{E}ssay on {S}cience and {F}ree {W}ill},}\
  }\bibinfo {howpublished} {From Campbell and Garnett, Life of Maxwell, Chapter
  XIV, pp.434-444}\BibitemShut {NoStop}%
\bibitem [{\citenamefont {Ivancevic}\ and\ \citenamefont
  {Ivancevic}(2008)}]{Ivancevic08}%
  \BibitemOpen
  \bibfield  {author} {\bibinfo {author} {\bibnamefont {Ivancevic},
  \bibfnamefont {Vladimir~G}}, \ and\ \bibinfo {author} {\bibfnamefont
  {Tijana~T.}\ \bibnamefont {Ivancevic}}} (\bibinfo {year} {2008}),\ \href@noop
  {} {\emph {\bibinfo {title} {Complex nonlinearity: chaos, phase transitions,
  topology change, and path integrals}}}\ (\bibinfo  {publisher}
  {Springer})\BibitemShut {NoStop}%
\bibitem [{\citenamefont {{Jerger}}\ \emph {et~al.}(2016)\citenamefont
  {{Jerger}}, \citenamefont {{Reshitnyk}}, \citenamefont {{Oppliger}},
  \citenamefont {{Poto{\v c}nik}}, \citenamefont {{Mondal}}, \citenamefont
  {{Wallraff}}, \citenamefont {{Goodenough}}, \citenamefont {{Wehner}},
  \citenamefont {{Juliusson}}, \citenamefont {{Langford}},\ and\ \citenamefont
  {{Fedorov}}}]{JergerARX2016}%
  \BibitemOpen
  \bibfield  {author} {\bibinfo {author} {\bibnamefont {{Jerger}},
  \bibfnamefont {M}}, \bibinfo {author} {\bibfnamefont {Y.}~\bibnamefont
  {{Reshitnyk}}}, \bibinfo {author} {\bibfnamefont {M.}~\bibnamefont
  {{Oppliger}}}, \bibinfo {author} {\bibfnamefont {A.}~\bibnamefont {{Poto{\v
  c}nik}}}, \bibinfo {author} {\bibfnamefont {M.}~\bibnamefont {{Mondal}}},
  \bibinfo {author} {\bibfnamefont {A.}~\bibnamefont {{Wallraff}}}, \bibinfo
  {author} {\bibfnamefont {K.}~\bibnamefont {{Goodenough}}}, \bibinfo {author}
  {\bibfnamefont {S.}~\bibnamefont {{Wehner}}}, \bibinfo {author}
  {\bibfnamefont {K.}~\bibnamefont {{Juliusson}}}, \bibinfo {author}
  {\bibfnamefont {N.~K.}\ \bibnamefont {{Langford}}}, \ and\ \bibinfo {author}
  {\bibfnamefont {A.}~\bibnamefont {{Fedorov}}}} (\bibinfo {year} {2016}),\
  \bibfield  {title} {\enquote {\bibinfo {title} {Contextuality without
  nonlocality in a superconducting quantum system},}\ }\href {\doibase
  10.1038/ncomms12930} {\bibfield  {journal} {\bibinfo  {journal} {Nature
  Communications}\ }\textbf {\bibinfo {volume} {7}},\ \bibinfo {pages}
  {12930}}\BibitemShut {NoStop}%
\bibitem [{\citenamefont {Jofre}\ \emph {et~al.}(2011)\citenamefont {Jofre},
  \citenamefont {Curty}, \citenamefont {Steinlechner}, \citenamefont {Anzolin},
  \citenamefont {Torres}, \citenamefont {Mitchell},\ and\ \citenamefont
  {Pruneri}}]{JofreOE2011b}%
  \BibitemOpen
  \bibfield  {author} {\bibinfo {author} {\bibnamefont {Jofre}, \bibfnamefont
  {M}}, \bibinfo {author} {\bibfnamefont {M.}~\bibnamefont {Curty}}, \bibinfo
  {author} {\bibfnamefont {F.}~\bibnamefont {Steinlechner}}, \bibinfo {author}
  {\bibfnamefont {G.}~\bibnamefont {Anzolin}}, \bibinfo {author} {\bibfnamefont
  {J.~P.}\ \bibnamefont {Torres}}, \bibinfo {author} {\bibfnamefont {M.~W.}\
  \bibnamefont {Mitchell}}, \ and\ \bibinfo {author} {\bibfnamefont
  {V.}~\bibnamefont {Pruneri}}} (\bibinfo {year} {2011}),\ \bibfield  {title}
  {\enquote {\bibinfo {title} {True random numbers from amplified quantum
  vacuum},}\ }\href {\doibase 10.1364/OE.19.020665} {\bibfield  {journal}
  {\bibinfo  {journal} {Opt. Express}\ }\textbf {\bibinfo {volume}
  {19}}~(\bibinfo {number} {21}),\ \bibinfo {pages} {20665--20672}}\BibitemShut
  {NoStop}%
\bibitem [{\citenamefont {Kamp}\ \emph {et~al.}(2011)\citenamefont {Kamp},
  \citenamefont {Rao}, \citenamefont {Vadhan},\ and\ \citenamefont
  {Zuckerman}}]{Kamp11}%
  \BibitemOpen
  \bibfield  {author} {\bibinfo {author} {\bibnamefont {Kamp}, \bibfnamefont
  {Jesse}}, \bibinfo {author} {\bibfnamefont {Anup}\ \bibnamefont {Rao}},
  \bibinfo {author} {\bibfnamefont {Salil}\ \bibnamefont {Vadhan}}, \ and\
  \bibinfo {author} {\bibfnamefont {David}\ \bibnamefont {Zuckerman}}}
  (\bibinfo {year} {2011}),\ \bibfield  {title} {\enquote {\bibinfo {title}
  {Deterministic extractors for small-space sources},}\ }\href {\doibase
  http://dx.doi.org/10.1016/j.jcss.2010.06.014} {\bibfield  {journal} {\bibinfo
   {journal} {Journal of Computer and System Sciences}\ }\textbf {\bibinfo
  {volume} {77}}~(\bibinfo {number} {1}),\ \bibinfo {pages} {191 -- 220}},\
  \bibinfo {note} {celebrating Karp's Kyoto Prize}\BibitemShut {NoStop}%
\bibitem [{\citenamefont {Khinchin}(2014)}]{Khinchin14}%
  \BibitemOpen
  \bibfield  {author} {\bibinfo {author} {\bibnamefont {Khinchin},
  \bibfnamefont {A}}} (\bibinfo {year} {2014}),\ \href@noop {} {\emph {\bibinfo
  {title} {Mathematical {F}oundations of {S}tatistical {M}echanics}}}\
  (\bibinfo  {publisher} {Martino Fine Books})\BibitemShut {NoStop}%
\bibitem [{\citenamefont {Kirchmair}\ \emph {et~al.}(2009)\citenamefont
  {Kirchmair}, \citenamefont {Zahringer}, \citenamefont {Gerritsma},
  \citenamefont {Kleinmann}, \citenamefont {Guhne}, \citenamefont {Cabello},
  \citenamefont {Blatt},\ and\ \citenamefont {Roos}}]{Kirchmair09}%
  \BibitemOpen
  \bibfield  {author} {\bibinfo {author} {\bibnamefont {Kirchmair},
  \bibfnamefont {G}}, \bibinfo {author} {\bibfnamefont {F.}~\bibnamefont
  {Zahringer}}, \bibinfo {author} {\bibfnamefont {R.}~\bibnamefont
  {Gerritsma}}, \bibinfo {author} {\bibfnamefont {M.}~\bibnamefont
  {Kleinmann}}, \bibinfo {author} {\bibfnamefont {O.}~\bibnamefont {Guhne}},
  \bibinfo {author} {\bibfnamefont {A.}~\bibnamefont {Cabello}}, \bibinfo
  {author} {\bibfnamefont {R.}~\bibnamefont {Blatt}}, \ and\ \bibinfo {author}
  {\bibfnamefont {C.~F.}\ \bibnamefont {Roos}}} (\bibinfo {year} {2009}),\
  \bibfield  {title} {\enquote {\bibinfo {title} {State-independent
  experimental test of quantum contextuality},}\ }\href {\doibase
  10.1038/nature08172} {\bibfield  {journal} {\bibinfo  {journal} {Nature}\
  }\textbf {\bibinfo {volume} {460}},\ \bibinfo {pages} {494--497}}\BibitemShut
  {NoStop}%
\bibitem [{\citenamefont {Kochen}\ and\ \citenamefont
  {Specker}(1967)}]{Kochen67}%
  \BibitemOpen
  \bibfield  {author} {\bibinfo {author} {\bibnamefont {Kochen}, \bibfnamefont
  {Simon}}, \ and\ \bibinfo {author} {\bibfnamefont {E.~P.}\ \bibnamefont
  {Specker}}} (\bibinfo {year} {1967}),\ \bibfield  {title} {\enquote {\bibinfo
  {title} {The problem of hidden variables in quantum mechanics},}\ }\href@noop
  {} {\bibfield  {journal} {\bibinfo  {journal} {Journal of Mathematics and
  Mechanics}\ }\textbf {\bibinfo {volume} {17}},\ \bibinfo {pages}
  {59--87}}\BibitemShut {NoStop}%
\bibitem [{\citenamefont {Koh}\ \emph {et~al.}(2012)\citenamefont {Koh},
  \citenamefont {Hall}, \citenamefont {Setiawan}, \citenamefont {Pope},
  \citenamefont {Marletto}, \citenamefont {Kay}, \citenamefont {Scarani},\ and\
  \citenamefont {Ekert}}]{Koh12}%
  \BibitemOpen
  \bibfield  {author} {\bibinfo {author} {\bibnamefont {Koh}, \bibfnamefont
  {Dax~Enshan}}, \bibinfo {author} {\bibfnamefont {Michael J.~W.}\ \bibnamefont
  {Hall}}, \bibinfo {author} {\bibnamefont {Setiawan}}, \bibinfo {author}
  {\bibfnamefont {James~E.}\ \bibnamefont {Pope}}, \bibinfo {author}
  {\bibfnamefont {Chiara}\ \bibnamefont {Marletto}}, \bibinfo {author}
  {\bibfnamefont {Alastair}\ \bibnamefont {Kay}}, \bibinfo {author}
  {\bibfnamefont {Valerio}\ \bibnamefont {Scarani}}, \ and\ \bibinfo {author}
  {\bibfnamefont {Artur}\ \bibnamefont {Ekert}}} (\bibinfo {year} {2012}),\
  \bibfield  {title} {\enquote {\bibinfo {title} {Effects of reduced
  measurement independence on bell-based randomness expansion},}\ }\href
  {\doibase 10.1103/PhysRevLett.109.160404} {\bibfield  {journal} {\bibinfo
  {journal} {Phys. Rev. Lett.}\ }\textbf {\bibinfo {volume} {109}},\ \bibinfo
  {pages} {160404}}\BibitemShut {NoStop}%
\bibitem [{\citenamefont {Kole\.zy\'nski}(2007)}]{Polish}%
  \BibitemOpen
  \bibfield  {author} {\bibinfo {author} {\bibnamefont {Kole\.zy\'nski},
  \bibfnamefont {Andrzej}}} (\bibinfo {year} {2007}),\ \bibfield  {title}
  {\enquote {\bibinfo {title} {Determinizm laplace'a w \'swietle teorii
  fizycznych mechaniki klasycznej},}\ }\href@noop {} {\bibfield  {journal}
  {\bibinfo  {journal} {Zagadnienia Filozoficzne w Nauce}\ }\textbf {\bibinfo
  {volume} {XL}},\ \bibinfo {pages} {59}}\BibitemShut {NoStop}%
\bibitem [{\citenamefont {Korolev}(2007{\natexlab{a}})}]{Korolev07}%
  \BibitemOpen
  \bibfield  {author} {\bibinfo {author} {\bibnamefont {Korolev}, \bibfnamefont
  {Alexandre}}} (\bibinfo {year} {2007}{\natexlab{a}}),\ \bibfield  {title}
  {\enquote {\bibinfo {title} {{I}ndeterminism, asymptotic reasoning, and time
  irreversibility in classical physics},}\ }\href {\doibase 10.1086/525635}
  {\bibfield  {journal} {\bibinfo  {journal} {Philosophy of Science}\ }\textbf
  {\bibinfo {volume} {74}},\ \bibinfo {pages} {943--956}}\BibitemShut {NoStop}%
\bibitem [{\citenamefont {Korolev}(2007{\natexlab{b}})}]{Korolev07a}%
  \BibitemOpen
  \bibfield  {author} {\bibinfo {author} {\bibnamefont {Korolev}, \bibfnamefont
  {Alexandre}}} (\bibinfo {year} {2007}{\natexlab{b}}),\ \bibfield  {title}
  {\enquote {\bibinfo {title} {{T}he {N}orton-{T}ype
  {L}ipschitz-{I}ndeterministic {S}ystems and {E}lastic {P}henomena:
  {I}ndeterminism as an {A}rtefact of {I}nfinite {I}dealizations},}\ \
  }(\bibinfo  {publisher} {Philosophy of Science Assoc. 21st Biennial Mtg.},\
  \bibinfo {address} {Pittsburgh, PA})\BibitemShut {NoStop}%
\bibitem [{\citenamefont {Kosyakov}(2008)}]{Kosyakov08}%
  \BibitemOpen
  \bibfield  {author} {\bibinfo {author} {\bibnamefont {Kosyakov},
  \bibfnamefont {B~P}}} (\bibinfo {year} {2008}),\ \bibfield  {title} {\enquote
  {\bibinfo {title} {{I}s {C}lassical {R}eality {C}ompletely
  {D}eterministic?}}\ }\href {\doibase 10.1007/s10701-007-9185-x} {\bibfield
  {journal} {\bibinfo  {journal} {Foundations of Physics}\ }\textbf {\bibinfo
  {volume} {38}}~(\bibinfo {number} {1}),\ \bibinfo {pages}
  {76--88}}\BibitemShut {NoStop}%
\bibitem [{\citenamefont {Laertius}(1925)}]{Laertius25}%
  \BibitemOpen
  \bibfield  {author} {\bibinfo {author} {\bibnamefont {Laertius},
  \bibfnamefont {Diogenes}}} (\bibinfo {year} {1925}),\ \href@noop {} {\emph
  {\bibinfo {title} {{L}ives of {E}minent {P}hilosophers, translated by {RD}
  {H}icks, {V}ol. 2. {L}oeb {C}lassical {L}ibrary, no. 185}}}\ (\bibinfo
  {publisher} {Harvard University Press, Cambridge, Massachusetts})\BibitemShut
  {NoStop}%
\bibitem [{\citenamefont {Landau}\ and\ \citenamefont
  {Lifshitz}(1960)}]{Landau60}%
  \BibitemOpen
  \bibfield  {author} {\bibinfo {author} {\bibnamefont {Landau}, \bibfnamefont
  {L~D}}, \ and\ \bibinfo {author} {\bibfnamefont {E.~M.}\ \bibnamefont
  {Lifshitz}}} (\bibinfo {year} {1960}),\ \href@noop {} {\emph {\bibinfo
  {title} {{C}lassical mechanics}}}\ (\bibinfo  {publisher} {Pergamon Press,
  Oxford})\BibitemShut {NoStop}%
\bibitem [{\citenamefont {Laplace}(1814)}]{Laplace14}%
  \BibitemOpen
  \bibfield  {author} {\bibinfo {author} {\bibnamefont {Laplace}, \bibfnamefont
  {P~S}}} (\bibinfo {year} {1814}),\ \href@noop {} {\emph {\bibinfo {title}
  {{E}ssai philosophique sur les probabilit\'{e}s}}}\ (\bibinfo  {publisher}
  {Courcier, Paris})\BibitemShut {NoStop}%
\bibitem [{\citenamefont {Laplace}(1951)}]{Laplace51}%
  \BibitemOpen
  \bibfield  {author} {\bibinfo {author} {\bibnamefont {Laplace}, \bibfnamefont
  {P~S}}} (\bibinfo {year} {1951}),\ \href@noop {} {\emph {\bibinfo {title} {A
  {P}hilosophical {E}ssay on {P}robabilities}}}\ (\bibinfo  {publisher}
  {translated into {E}nglish from the original {F}rench 6th ed. by {T}ruscott,
  {F}. {W}. and {Emory, F. L.}, ({D}over {P}ublications, {N}ew {Y}ork,
  1951)})\BibitemShut {NoStop}%
\bibitem [{\citenamefont {Laraudogoitia}(2013)}]{Laraudogoitia13}%
  \BibitemOpen
  \bibfield  {author} {\bibinfo {author} {\bibnamefont {Laraudogoitia},
  \bibfnamefont {Jon~P{\'e}rez}}} (\bibinfo {year} {2013}),\ \bibfield  {title}
  {\enquote {\bibinfo {title} {{O}n {N}orton's dome},}\ }\href {\doibase
  10.1007/s11229-012-0105-z} {\bibfield  {journal} {\bibinfo  {journal}
  {Synthese}\ }\textbf {\bibinfo {volume} {190}}~(\bibinfo {number} {14}),\
  \bibinfo {pages} {2925--2941}}\BibitemShut {NoStop}%
\bibitem [{\citenamefont {Le~Jan}\ and\ \citenamefont
  {Raimond}(1998)}]{Lejan1}%
  \BibitemOpen
  \bibfield  {author} {\bibinfo {author} {\bibnamefont {Le~Jan}, \bibfnamefont
  {Yves}}, \ and\ \bibinfo {author} {\bibfnamefont {Olivier}\ \bibnamefont
  {Raimond}}} (\bibinfo {year} {1998}),\ \bibfield  {title} {\enquote {\bibinfo
  {title} {Solutions statistiques fortes des équations différentielles
  stochastiques},}\ }\href {\doibase
  http://dx.doi.org/10.1016/S0764-4442(99)80039-1} {\bibfield  {journal}
  {\bibinfo  {journal} {Comptes Rendus de l'Académie des Sciences - Series I -
  Mathematics}\ }\textbf {\bibinfo {volume} {327}}~(\bibinfo {number} {10}),\
  \bibinfo {pages} {893 -- 896}}\BibitemShut {NoStop}%
\bibitem [{\citenamefont {Le~Jan}\ and\ \citenamefont
  {Raimond}(2002)}]{Lejan2}%
  \BibitemOpen
  \bibfield  {author} {\bibinfo {author} {\bibnamefont {Le~Jan}, \bibfnamefont
  {Yves}}, \ and\ \bibinfo {author} {\bibfnamefont {Olivier}\ \bibnamefont
  {Raimond}}} (\bibinfo {year} {2002}),\ \bibfield  {title} {\enquote {\bibinfo
  {title} {Integration of {B}rownian vector fields},}\ }\href {\doibase
  10.1214/aop/1023481009} {\bibfield  {journal} {\bibinfo  {journal} {Ann.
  Probab.}\ }\textbf {\bibinfo {volume} {30}}~(\bibinfo {number} {2}),\
  \bibinfo {pages} {826--873}}\BibitemShut {NoStop}%
\bibitem [{\citenamefont {Le~Jan}\ and\ \citenamefont
  {Raimond}(2004)}]{Lejan3}%
  \BibitemOpen
  \bibfield  {author} {\bibinfo {author} {\bibnamefont {Le~Jan}, \bibfnamefont
  {Yves}}, \ and\ \bibinfo {author} {\bibfnamefont {Olivier}\ \bibnamefont
  {Raimond}}} (\bibinfo {year} {2004}),\ \bibfield  {title} {\enquote {\bibinfo
  {title} {Flows, coalescence and noise},}\ }\href {\doibase
  10.1214/009117904000000207} {\bibfield  {journal} {\bibinfo  {journal} {Ann.
  Probab.}\ }\textbf {\bibinfo {volume} {32}}~(\bibinfo {number} {2}),\
  \bibinfo {pages} {1247--1315}}\BibitemShut {NoStop}%
\bibitem [{\citenamefont {L'Ecuyer}\ and\ \citenamefont
  {Simard}(2007)}]{LEcuyerACM2007}%
  \BibitemOpen
  \bibfield  {author} {\bibinfo {author} {\bibnamefont {L'Ecuyer},
  \bibfnamefont {Pierre}}, \ and\ \bibinfo {author} {\bibfnamefont {Richard}\
  \bibnamefont {Simard}}} (\bibinfo {year} {2007}),\ \bibfield  {title}
  {\enquote {\bibinfo {title} {{TestU01}: A c library for empirical testing of
  random number generators},}\ }\href {\doibase 10.1145/1268776.1268777}
  {\bibfield  {journal} {\bibinfo  {journal} {ACM Trans. Math. Softw.}\
  }\textbf {\bibinfo {volume} {33}}~(\bibinfo {number} {4}),\ \bibinfo {pages}
  {22}}\BibitemShut {NoStop}%
\bibitem [{\citenamefont {Liang}\ \emph {et~al.}(2015)\citenamefont {Liang},
  \citenamefont {Rosset}, \citenamefont {Bancal}, \citenamefont {P\"utz},
  \citenamefont {Barnea},\ and\ \citenamefont {Gisin}}]{Liang15}%
  \BibitemOpen
  \bibfield  {author} {\bibinfo {author} {\bibnamefont {Liang}, \bibfnamefont
  {Yeong-Cherng}}, \bibinfo {author} {\bibfnamefont {Denis}\ \bibnamefont
  {Rosset}}, \bibinfo {author} {\bibfnamefont {Jean-Daniel}\ \bibnamefont
  {Bancal}}, \bibinfo {author} {\bibfnamefont {Gilles}\ \bibnamefont {P\"utz}},
  \bibinfo {author} {\bibfnamefont {Tomer~Jack}\ \bibnamefont {Barnea}}, \ and\
  \bibinfo {author} {\bibfnamefont {Nicolas}\ \bibnamefont {Gisin}}} (\bibinfo
  {year} {2015}),\ \bibfield  {title} {\enquote {\bibinfo {title} {Family of
  {B}ell-like inequalities as device-independent witnesses for entanglement
  depth},}\ }\href {\doibase 10.1103/PhysRevLett.114.190401} {\bibfield
  {journal} {\bibinfo  {journal} {Phys. Rev. Lett.}\ }\textbf {\bibinfo
  {volume} {114}},\ \bibinfo {pages} {190401}}\BibitemShut {NoStop}%
\bibitem [{\citenamefont {Lydersen}\ \emph {et~al.}(2010)\citenamefont
  {Lydersen}, \citenamefont {Wiechers}, \citenamefont {Wittmann}, \citenamefont
  {Elser}, \citenamefont {Skaar},\ and\ \citenamefont {Makarov}}]{Lydersen10}%
  \BibitemOpen
  \bibfield  {author} {\bibinfo {author} {\bibnamefont {Lydersen},
  \bibfnamefont {Lars}}, \bibinfo {author} {\bibfnamefont {Carlos}\
  \bibnamefont {Wiechers}}, \bibinfo {author} {\bibfnamefont {Christoffer}\
  \bibnamefont {Wittmann}}, \bibinfo {author} {\bibfnamefont {Dominique}\
  \bibnamefont {Elser}}, \bibinfo {author} {\bibfnamefont {Johannes}\
  \bibnamefont {Skaar}}, \ and\ \bibinfo {author} {\bibfnamefont {Vadim}\
  \bibnamefont {Makarov}}} (\bibinfo {year} {2010}),\ \bibfield  {title}
  {\enquote {\bibinfo {title} {Hacking commercial quantum cryptography systems
  by tailored bright illumination},}\ }\href {\doibase
  10.1038/nphoton.2010.214} {\bibfield  {journal} {\bibinfo  {journal} {Nat
  Photon}\ }\textbf {\bibinfo {volume} {4}},\ \bibinfo {pages}
  {686--689}}\BibitemShut {NoStop}%
\bibitem [{\citenamefont {Maassen}\ and\ \citenamefont
  {Uffink}(1988)}]{Maassen88}%
  \BibitemOpen
  \bibfield  {author} {\bibinfo {author} {\bibnamefont {Maassen}, \bibfnamefont
  {Hans}}, \ and\ \bibinfo {author} {\bibfnamefont {J.~B.~M.}\ \bibnamefont
  {Uffink}}} (\bibinfo {year} {1988}),\ \bibfield  {title} {\enquote {\bibinfo
  {title} {Generalized entropic uncertainty relations},}\ }\href {\doibase
  10.1103/PhysRevLett.60.1103} {\bibfield  {journal} {\bibinfo  {journal}
  {Phys. Rev. Lett.}\ }\textbf {\bibinfo {volume} {60}},\ \bibinfo {pages}
  {1103--1106}}\BibitemShut {NoStop}%
\bibitem [{\citenamefont {Maccone}\ and\ \citenamefont
  {Pati}(2014)}]{Maccone14}%
  \BibitemOpen
  \bibfield  {author} {\bibinfo {author} {\bibnamefont {Maccone}, \bibfnamefont
  {Lorenzo}}, \ and\ \bibinfo {author} {\bibfnamefont {Arun~K.}\ \bibnamefont
  {Pati}}} (\bibinfo {year} {2014}),\ \bibfield  {title} {\enquote {\bibinfo
  {title} {Stronger uncertainty relations for all incompatible observables},}\
  }\href {\doibase 10.1103/PhysRevLett.113.260401} {\bibfield  {journal}
  {\bibinfo  {journal} {Phys. Rev. Lett.}\ }\textbf {\bibinfo {volume} {113}},\
  \bibinfo {pages} {260401}}\BibitemShut {NoStop}%
\bibitem [{\citenamefont {Malament}(2008)}]{Malament08}%
  \BibitemOpen
  \bibfield  {author} {\bibinfo {author} {\bibnamefont {Malament},
  \bibfnamefont {David~B}}} (\bibinfo {year} {2008}),\ \bibfield  {title}
  {\enquote {\bibinfo {title} {Norton’s slippery slope},}\ }\href
  {http://www.jstor.org/stable/10.1086/594525} {\bibfield  {journal} {\bibinfo
  {journal} {{P}hilosophy of {S}cience}\ }\textbf {\bibinfo {volume}
  {75}}~(\bibinfo {number} {5}),\ \bibinfo {pages} {799--816}}\BibitemShut
  {NoStop}%
\bibitem [{\citenamefont {Marsaglia}(2008)}]{Marsaglia08}%
  \BibitemOpen
  \bibfield  {author} {\bibinfo {author} {\bibnamefont {Marsaglia},
  \bibfnamefont {G}}} (\bibinfo {year} {2008}),\ \href
  {http://www.stat.fsu.edu/pub/diehard/} {\emph {\bibinfo {title} {The
  {M}arsaglia Random Number CDROM including the {D}iehard Battery of Tests of
  Randomness}}}\BibitemShut {NoStop}%
\bibitem [{\citenamefont {Menezes}\ \emph {et~al.}(1996)\citenamefont
  {Menezes}, \citenamefont {van Oorschot},\ and\ \citenamefont
  {Vanstone}}]{Menezes96}%
  \BibitemOpen
  \bibfield  {author} {\bibinfo {author} {\bibnamefont {Menezes}, \bibfnamefont
  {Alfred~J}}, \bibinfo {author} {\bibfnamefont {Paul~C.}\ \bibnamefont {van
  Oorschot}}, \ and\ \bibinfo {author} {\bibfnamefont {Scott~A.}\ \bibnamefont
  {Vanstone}}} (\bibinfo {year} {1996}),\ \href {http://cacr.uwaterloo.ca/hac/}
  {\emph {\bibinfo {title} {Handbook of Applied Cryptography}}}\ (\bibinfo
  {publisher} {CRC Press},\ \bibinfo {address} {Boca Raton, FL,
  USA})\BibitemShut {NoStop}%
\bibitem [{\citenamefont {Messiah}(2014)}]{Messiah14}%
  \BibitemOpen
  \bibfield  {author} {\bibinfo {author} {\bibnamefont {Messiah}, \bibfnamefont
  {Albert}}} (\bibinfo {year} {2014}),\ \href@noop {} {\emph {\bibinfo {title}
  {{Q}uantum {M}echnaics}}}\ (\bibinfo  {publisher} {Dover
  Publications})\BibitemShut {NoStop}%
\bibitem [{\citenamefont {Michael~Mueller}(2015)}]{Mueller15}%
  \BibitemOpen
  \bibfield  {author} {\bibinfo {author} {\bibnamefont {Michael~Mueller},
  \bibfnamefont {Thomas}}} (\bibinfo {year} {2015}),\ \bibfield  {title}
  {\enquote {\bibinfo {title} {{T}he {B}oussinesq {D}ebate: {R}eversibility,
  {I}nstability, and {F}ree {W}ill},}\ }\href {\doibase
  10.1017/S0269889715000290} {\bibfield  {journal} {\bibinfo  {journal}
  {Science in Context}\ }\textbf {\bibinfo {volume} {28}}~(\bibinfo {number}
  {4}),\ \bibinfo {pages} {613–635}}\BibitemShut {NoStop}%
\bibitem [{\citenamefont {Miller}\ and\ \citenamefont
  {Shi}(2014)}]{miller_shi2}%
  \BibitemOpen
  \bibfield  {author} {\bibinfo {author} {\bibnamefont {Miller}, \bibfnamefont
  {C~A}}, \ and\ \bibinfo {author} {\bibfnamefont {Y.}~\bibnamefont {Shi}}}
  (\bibinfo {year} {2014}),\ \bibfield  {title} {\enquote {\bibinfo {title}
  {{Universal security for randomness expansion from the spot-checking
  protocol}},}\ }\href@noop {} {\bibfield  {journal} {\bibinfo  {journal}
  {ArXiv e-prints}\ }}\Eprint {http://arxiv.org/abs/1411.6608} {arXiv:1411.6608
  [quant-ph]} \BibitemShut {NoStop}%
\bibitem [{\citenamefont {Miller}\ and\ \citenamefont
  {Shi}(2016)}]{miller_shi}%
  \BibitemOpen
  \bibfield  {author} {\bibinfo {author} {\bibnamefont {Miller}, \bibfnamefont
  {Carl~A}}, \ and\ \bibinfo {author} {\bibfnamefont {Yaoyun}\ \bibnamefont
  {Shi}}} (\bibinfo {year} {2016}),\ \bibfield  {title} {\enquote {\bibinfo
  {title} {Robust protocols for securely expanding randomness and distributing
  keys using untrusted quantum devices},}\ }\href {\doibase 10.1145/2885493}
  {\bibfield  {journal} {\bibinfo  {journal} {J. ACM}\ }\textbf {\bibinfo
  {volume} {63}}~(\bibinfo {number} {4}),\ \bibinfo {pages}
  {33:1--33:63}}\BibitemShut {NoStop}%
\bibitem [{\citenamefont {Mironowicz}\ \emph {et~al.}(2015)\citenamefont
  {Mironowicz}, \citenamefont {Gallego},\ and\ \citenamefont
  {Paw\l{}owski}}]{Mironowicz15}%
  \BibitemOpen
  \bibfield  {author} {\bibinfo {author} {\bibnamefont {Mironowicz},
  \bibfnamefont {Piotr}}, \bibinfo {author} {\bibfnamefont {Rodrigo}\
  \bibnamefont {Gallego}}, \ and\ \bibinfo {author} {\bibfnamefont {Marcin}\
  \bibnamefont {Paw\l{}owski}}} (\bibinfo {year} {2015}),\ \bibfield  {title}
  {\enquote {\bibinfo {title} {Robust amplification of {S}antha-{V}azirani
  sources with three devices},}\ }\href {\doibase 10.1103/PhysRevA.91.032317}
  {\bibfield  {journal} {\bibinfo  {journal} {Phys. Rev. A}\ }\textbf {\bibinfo
  {volume} {91}},\ \bibinfo {pages} {032317}}\BibitemShut {NoStop}%
\bibitem [{\citenamefont {Mitchell}\ \emph {et~al.}(2015)\citenamefont
  {Mitchell}, \citenamefont {Abellan},\ and\ \citenamefont
  {Amaya}}]{MitchellPRA2015}%
  \BibitemOpen
  \bibfield  {author} {\bibinfo {author} {\bibnamefont {Mitchell},
  \bibfnamefont {Morgan~W}}, \bibinfo {author} {\bibfnamefont {Carlos}\
  \bibnamefont {Abellan}}, \ and\ \bibinfo {author} {\bibfnamefont {Waldimar}\
  \bibnamefont {Amaya}}} (\bibinfo {year} {2015}),\ \bibfield  {title}
  {\enquote {\bibinfo {title} {Strong experimental guarantees in ultrafast
  quantum random number generation},}\ }\href {\doibase
  10.1103/PhysRevA.91.012314} {\bibfield  {journal} {\bibinfo  {journal} {Phys.
  Rev. A}\ }\textbf {\bibinfo {volume} {91}},\ \bibinfo {pages}
  {012314}}\BibitemShut {NoStop}%
\bibitem [{\citenamefont {Motwani}\ and\ \citenamefont
  {Raghavan}(1995)}]{Motwani95}%
  \BibitemOpen
  \bibfield  {author} {\bibinfo {author} {\bibnamefont {Motwani}, \bibfnamefont
  {Rajeev}}, \ and\ \bibinfo {author} {\bibfnamefont {Prabhakar}\ \bibnamefont
  {Raghavan}}} (\bibinfo {year} {1995}),\ \href@noop {} {\emph {\bibinfo
  {title} {Randomized Algorithms}}}\ (\bibinfo  {publisher} {Cambridge
  University Press},\ \bibinfo {address} {New York, NY, USA})\BibitemShut
  {NoStop}%
\bibitem [{\citenamefont {Myerson}\ \emph {et~al.}(2008)\citenamefont
  {Myerson}, \citenamefont {Szwer}, \citenamefont {Webster}, \citenamefont
  {Allcock}, \citenamefont {Curtis}, \citenamefont {Imreh}, \citenamefont
  {Sherman}, \citenamefont {Stacey}, \citenamefont {Steane},\ and\
  \citenamefont {Lucas}}]{MyersonPRL2008}%
  \BibitemOpen
  \bibfield  {author} {\bibinfo {author} {\bibnamefont {Myerson}, \bibfnamefont
  {A~H}}, \bibinfo {author} {\bibfnamefont {D.~J.}\ \bibnamefont {Szwer}},
  \bibinfo {author} {\bibfnamefont {S.~C.}\ \bibnamefont {Webster}}, \bibinfo
  {author} {\bibfnamefont {D.~T.~C.}\ \bibnamefont {Allcock}}, \bibinfo
  {author} {\bibfnamefont {M.~J.}\ \bibnamefont {Curtis}}, \bibinfo {author}
  {\bibfnamefont {G.}~\bibnamefont {Imreh}}, \bibinfo {author} {\bibfnamefont
  {J.~A.}\ \bibnamefont {Sherman}}, \bibinfo {author} {\bibfnamefont {D.~N.}\
  \bibnamefont {Stacey}}, \bibinfo {author} {\bibfnamefont {A.~M.}\
  \bibnamefont {Steane}}, \ and\ \bibinfo {author} {\bibfnamefont {D.~M.}\
  \bibnamefont {Lucas}}} (\bibinfo {year} {2008}),\ \bibfield  {title}
  {\enquote {\bibinfo {title} {High-fidelity readout of trapped-ion qubits},}\
  }\href {\doibase 10.1103/PhysRevLett.100.200502} {\bibfield  {journal}
  {\bibinfo  {journal} {Phys. Rev. Lett.}\ }\textbf {\bibinfo {volume} {100}},\
  \bibinfo {pages} {200502}}\BibitemShut {NoStop}%
\bibitem [{\citenamefont {{National Institutes of Standards and
  Technology}}(2011)}]{NISTBeacon}%
  \BibitemOpen
  \bibfield  {author} {\bibinfo {author} {\bibnamefont {{National Institutes of
  Standards and Technology}},}} (\bibinfo {year} {2011}),\ \href
  {http://www.nist.gov/itl/csd/ct/nist_beacon.cfm} {\enquote {\bibinfo {title}
  {{NIST} {R}andomness {B}eacon},}\ }\Eprint
  {http://arxiv.org/abs/{http://www.nist.gov/itl/csd/ct/nist{\textunderscore}beacon.cfm}}
  {{http://www.nist.gov/itl/csd/ct/nist{\textunderscore}beacon.cfm}}
  \BibitemShut {NoStop}%
\bibitem [{\citenamefont {Navascués}\ \emph {et~al.}(2008)\citenamefont
  {Navascués}, \citenamefont {Pironio},\ and\ \citenamefont
  {Acín}}]{Navascues08}%
  \BibitemOpen
  \bibfield  {author} {\bibinfo {author} {\bibnamefont {Navascués},
  \bibfnamefont {Miguel}}, \bibinfo {author} {\bibfnamefont {Stefano}\
  \bibnamefont {Pironio}}, \ and\ \bibinfo {author} {\bibfnamefont {Antonio}\
  \bibnamefont {Acín}}} (\bibinfo {year} {2008}),\ \bibfield  {title}
  {\enquote {\bibinfo {title} {A convergent hierarchy of semidefinite programs
  characterizing the set of quantum correlations},}\ }\href
  {http://stacks.iop.org/1367-2630/10/i=7/a=073013} {\bibfield  {journal}
  {\bibinfo  {journal} {New Journal of Physics}\ }\textbf {\bibinfo {volume}
  {10}}~(\bibinfo {number} {7}),\ \bibinfo {pages} {073013}}\BibitemShut
  {NoStop}%
\bibitem [{\citenamefont {von Neumann}(1951)}]{VonNeumannAMS1951}%
  \BibitemOpen
  \bibfield  {author} {\bibinfo {author} {\bibnamefont {von Neumann},
  \bibfnamefont {John}}} (\bibinfo {year} {1951}),\ \bibfield  {title}
  {\enquote {\bibinfo {title} {Various techniques used in connection with
  random digits},}\ }\href@noop {} {\bibfield  {journal} {\bibinfo  {journal}
  {Applied Math Series}\ }\textbf {\bibinfo {volume} {12}},\ \bibinfo {pages}
  {36}}\BibitemShut {NoStop}%
\bibitem [{\citenamefont {Neumann}(1955)}]{Neumann55}%
  \BibitemOpen
  \bibfield  {author} {\bibinfo {author} {\bibnamefont {Neumann}, \bibfnamefont
  {John~Von}}} (\bibinfo {year} {1955}),\ \href
  {http://press.princeton.edu/titles/2113.html} {\emph {\bibinfo {title}
  {Mathematical Foundations of Quantum Mechanics}}}\ (\bibinfo  {publisher}
  {Princeton University Press})\BibitemShut {NoStop}%
\bibitem [{\citenamefont {Newman}(1956)}]{Newman56}%
  \BibitemOpen
  \bibfield  {author} {\bibinfo {author} {\bibnamefont {Newman}, \bibfnamefont
  {J~R}}} (\bibinfo {year} {1956}),\ \href@noop {} {\emph {\bibinfo {title}
  {{T}he world of mathematics}}}\ (\bibinfo  {publisher} {Simon and Schuster,
  New York})\BibitemShut {NoStop}%
\bibitem [{\citenamefont {Nie}\ \emph {et~al.}(2015)\citenamefont {Nie},
  \citenamefont {Huang}, \citenamefont {Liu}, \citenamefont {Payne},
  \citenamefont {Zhang},\ and\ \citenamefont {Pan}}]{Nie2015}%
  \BibitemOpen
  \bibfield  {author} {\bibinfo {author} {\bibnamefont {Nie}, \bibfnamefont
  {You-Qi}}, \bibinfo {author} {\bibfnamefont {Leilei}\ \bibnamefont {Huang}},
  \bibinfo {author} {\bibfnamefont {Yang}\ \bibnamefont {Liu}}, \bibinfo
  {author} {\bibfnamefont {Frank}\ \bibnamefont {Payne}}, \bibinfo {author}
  {\bibfnamefont {Jun}\ \bibnamefont {Zhang}}, \ and\ \bibinfo {author}
  {\bibfnamefont {Jian-Wei}\ \bibnamefont {Pan}}} (\bibinfo {year} {2015}),\
  \bibfield  {title} {\enquote {\bibinfo {title} {The generation of 68 gbps
  quantum random number by measuring laser phase fluctuations},}\ }\href
  {http://scitation.aip.org/content/aip/journal/rsi/86/6/10.1063/1.4922417}
  {\bibfield  {journal} {\bibinfo  {journal} {Review of Scientific
  Instruments}\ }\textbf {\bibinfo {volume} {86}}~(\bibinfo {number} {6}),\
  \bibinfo {pages} {063105}}\BibitemShut {NoStop}%
\bibitem [{\citenamefont {Nisan}\ and\ \citenamefont
  {Ta-Shma}(1999)}]{Nisan99}%
  \BibitemOpen
  \bibfield  {author} {\bibinfo {author} {\bibnamefont {Nisan}, \bibfnamefont
  {Noam}}, \ and\ \bibinfo {author} {\bibfnamefont {Amnon}\ \bibnamefont
  {Ta-Shma}}} (\bibinfo {year} {1999}),\ \bibfield  {title} {\enquote {\bibinfo
  {title} {Extracting randomness: A survey and new constructions},}\ }\href
  {\doibase http://dx.doi.org/10.1006/jcss.1997.1546} {\bibfield  {journal}
  {\bibinfo  {journal} {Journal of Computer and System Sciences}\ }\textbf
  {\bibinfo {volume} {58}}~(\bibinfo {number} {1}),\ \bibinfo {pages} {148 --
  173}}\BibitemShut {NoStop}%
\bibitem [{\citenamefont {Norton}(2007)}]{Norton07}%
  \BibitemOpen
  \bibfield  {author} {\bibinfo {author} {\bibnamefont {Norton}, \bibfnamefont
  {J~D}}} (\bibinfo {year} {2007}),\ \enquote {\bibinfo {title} {Causation,
  physics, and the constitution of reality: Russell's republic revisited},}\
  Chap.\ \bibinfo {chapter} {{C}ausation as folk science}\ (\bibinfo
  {publisher} {Oxford University Press},\ \bibinfo {address} {Oxford})\ pp.\
  \bibinfo {pages} {11--44}\BibitemShut {NoStop}%
\bibitem [{\citenamefont {Norton}(2008)}]{Norton08}%
  \BibitemOpen
  \bibfield  {author} {\bibinfo {author} {\bibnamefont {Norton}, \bibfnamefont
  {J~D}}} (\bibinfo {year} {2008}),\ \bibfield  {title} {\enquote {\bibinfo
  {title} {{T}he {D}ome: {A}n {U}nexpectedly {S}imple {F}ailure of
  {D}eterminism},}\ }\href {\doibase 10.1086/594524} {\bibfield  {journal}
  {\bibinfo  {journal} {Philosophy of Science}\ }\textbf {\bibinfo {volume}
  {75}}~(\bibinfo {number} {5}),\ \bibinfo {pages} {786--798}}\BibitemShut
  {NoStop}%
\bibitem [{\citenamefont {Papineau}(2010)}]{Papineau2010}%
  \BibitemOpen
  \bibfield  {author} {\bibinfo {author} {\bibnamefont {Papineau},
  \bibfnamefont {David}}} (\bibinfo {year} {2010}),\ \bibfield  {title}
  {\enquote {\bibinfo {title} {{A Fair Deal for Everettians}},}\ }in\ \href
  {http://www.oxfordscholarship.com/view/10.1093/acprof:oso/9780199560561.001.0001/acprof-9780199560561-chapter-9}
  {\emph {\bibinfo {booktitle} {{Many Worlds?: Everett, Quantum Theory, and
  Reality}}}},\ \bibinfo {editor} {edited by\ \bibinfo {editor} {\bibfnamefont
  {Simon}\ \bibnamefont {Saunders}}, \bibinfo {editor} {\bibfnamefont
  {Jonathan}\ \bibnamefont {Barrett}}, \bibinfo {editor} {\bibfnamefont
  {Adrian}\ \bibnamefont {Kent}}, \ and\ \bibinfo {editor} {\bibfnamefont
  {David}\ \bibnamefont {Wallace}}},\ Chap.~\bibinfo {chapter} {9}\ (\bibinfo
  {publisher} {Oxford Publisher},\ \bibinfo {address} {Oxford})\ pp.\ \bibinfo
  {pages} {206--226}\BibitemShut {NoStop}%
\bibitem [{\citenamefont {Penrose}(1979)}]{Penrose79}%
  \BibitemOpen
  \bibfield  {author} {\bibinfo {author} {\bibnamefont {Penrose}, \bibfnamefont
  {O}}} (\bibinfo {year} {1979}),\ \bibfield  {title} {\enquote {\bibinfo
  {title} {{F}oundations of statistical mechanics},}\ }\href
  {http://stacks.iop.org/0034-4885/42/i=12/a=002} {\bibfield  {journal}
  {\bibinfo  {journal} {Reports on Progress in Physics}\ }\textbf {\bibinfo
  {volume} {42}}~(\bibinfo {number} {12}),\ \bibinfo {pages}
  {1937}}\BibitemShut {NoStop}%
\bibitem [{\citenamefont {Peres}(1995)}]{Peres95}%
  \BibitemOpen
  \bibfield  {author} {\bibinfo {author} {\bibnamefont {Peres}, \bibfnamefont
  {A}}} (\bibinfo {year} {1995}),\ \href {\doibase 10.1007/0-306-47120-5}
  {\emph {\bibinfo {title} {Quantum Theory: Concepts and Methods}}}\ (\bibinfo
  {publisher} {Springer Netherlands})\BibitemShut {NoStop}%
\bibitem [{\citenamefont {{Pironio}}(2015)}]{PironioARX2015}%
  \BibitemOpen
  \bibfield  {author} {\bibinfo {author} {\bibnamefont {{Pironio}},
  \bibfnamefont {S}}} (\bibinfo {year} {2015}),\ \bibfield  {title} {\enquote
  {\bibinfo {title} {{Random 'choices' and the locality loophole}},}\
  }\href@noop {} {\bibfield  {journal} {\bibinfo  {journal} {ArXiv e-prints}\
  }}\Eprint {http://arxiv.org/abs/1510.00248} {arXiv:1510.00248 [quant-ph]}
  \BibitemShut {NoStop}%
\bibitem [{\citenamefont {Pironio}\ \emph {et~al.}(2010)\citenamefont
  {Pironio}, \citenamefont {Acín}, \citenamefont {Massar}, \citenamefont
  {de~la Giroday}, \citenamefont {Matsukevich}, \citenamefont {Maunz},
  \citenamefont {Olmschenk}, \citenamefont {Hayes}, \citenamefont {Luo},
  \citenamefont {Manning},\ and\ \citenamefont {Monroe}}]{Pironio10}%
  \BibitemOpen
  \bibfield  {author} {\bibinfo {author} {\bibnamefont {Pironio}, \bibfnamefont
  {S}}, \bibinfo {author} {\bibfnamefont {A.}~\bibnamefont {Acín}}, \bibinfo
  {author} {\bibfnamefont {S.}~\bibnamefont {Massar}}, \bibinfo {author}
  {\bibfnamefont {A.~Boyer}\ \bibnamefont {de~la Giroday}}, \bibinfo {author}
  {\bibfnamefont {D.~N.}\ \bibnamefont {Matsukevich}}, \bibinfo {author}
  {\bibfnamefont {P.}~\bibnamefont {Maunz}}, \bibinfo {author} {\bibfnamefont
  {S.}~\bibnamefont {Olmschenk}}, \bibinfo {author} {\bibfnamefont
  {D.}~\bibnamefont {Hayes}}, \bibinfo {author} {\bibfnamefont
  {L.}~\bibnamefont {Luo}}, \bibinfo {author} {\bibfnamefont {T.~A.}\
  \bibnamefont {Manning}}, \ and\ \bibinfo {author} {\bibfnamefont
  {C.}~\bibnamefont {Monroe}}} (\bibinfo {year} {2010}),\ \bibfield  {title}
  {\enquote {\bibinfo {title} {Random numbers certified by {B}ell’s
  theorem},}\ }\href {\doibase 10.1038/nature09008} {\bibfield  {journal}
  {\bibinfo  {journal} {Nature}\ }\textbf {\bibinfo {volume} {23}},\ \bibinfo
  {pages} {1021--1024}}\BibitemShut {NoStop}%
\bibitem [{\citenamefont {Pironio}\ \emph {et~al.}(2015)\citenamefont
  {Pironio}, \citenamefont {Scarani},\ and\ \citenamefont
  {Vidick}}]{Pironio15}%
  \BibitemOpen
  \bibfield  {author} {\bibinfo {author} {\bibnamefont {Pironio}, \bibfnamefont
  {Stefano}}, \bibinfo {author} {\bibfnamefont {Valerio}\ \bibnamefont
  {Scarani}}, \ and\ \bibinfo {author} {\bibfnamefont {Thomas}\ \bibnamefont
  {Vidick}}} (\bibinfo {year} {2015}),\ \bibfield  {title} {\enquote {\bibinfo
  {title} {Focus on device independent quantum information},}\ }\href
  {http://iopscience.iop.org/1367-2630/focus/Focus-on-Device-Independent-Quantum-Information}
  {\bibinfo  {journal} {New Journal of Physics}\ }\BibitemShut {NoStop}%
\bibitem [{\citenamefont {Pivoluska}\ and\ \citenamefont
  {Plesch}(2014)}]{Pivoluska14}%
  \BibitemOpen
\bibfield  {journal} {  }\bibfield  {author} {\bibinfo {author} {\bibnamefont
  {Pivoluska}, \bibfnamefont {Matej}}, \ and\ \bibinfo {author} {\bibfnamefont
  {Martin}\ \bibnamefont {Plesch}}} (\bibinfo {year} {2014}),\ \bibfield
  {title} {\enquote {\bibinfo {title} {Device independent random number
  generation},}\ }\href {\doibase 10.2478/apsrt-2014-0006} {\bibfield
  {journal} {\bibinfo  {journal} {Acta Physica Slovaca}\ }\textbf {\bibinfo
  {volume} {64}},\ \bibinfo {pages} {600 -- 663}}\BibitemShut {NoStop}%
\bibitem [{\citenamefont {Poincar\'e}(1912)}]{Poincare12}%
  \BibitemOpen
  \bibfield  {author} {\bibinfo {author} {\bibnamefont {Poincar\'e},
  \bibfnamefont {H}}} (\bibinfo {year} {1912}),\ \href@noop {} {\emph {\bibinfo
  {title} {{C}alcul des probabilit\'es}}}\ (\bibinfo  {publisher}
  {Gauthier-Villars, Paris})\BibitemShut {NoStop}%
\bibitem [{\citenamefont {Popescu}\ and\ \citenamefont
  {Rohrlich}(1992)}]{Popescu92}%
  \BibitemOpen
  \bibfield  {author} {\bibinfo {author} {\bibnamefont {Popescu}, \bibfnamefont
  {Sandu}}, \ and\ \bibinfo {author} {\bibfnamefont {Daniel}\ \bibnamefont
  {Rohrlich}}} (\bibinfo {year} {1992}),\ \bibfield  {title} {\enquote
  {\bibinfo {title} {Generic quantum nonlocality},}\ }\href {\doibase
  http://dx.doi.org/10.1016/0375-9601(92)90711-T} {\bibfield  {journal}
  {\bibinfo  {journal} {Physics Letters A}\ }\textbf {\bibinfo {volume}
  {166}}~(\bibinfo {number} {5}),\ \bibinfo {pages} {293 -- 297}}\BibitemShut
  {NoStop}%
\bibitem [{\citenamefont {Popescu}\ and\ \citenamefont {Rohrlich}(1994)}]{PR}%
  \BibitemOpen
  \bibfield  {author} {\bibinfo {author} {\bibnamefont {Popescu}, \bibfnamefont
  {Sandu}}, \ and\ \bibinfo {author} {\bibfnamefont {Daniel}\ \bibnamefont
  {Rohrlich}}} (\bibinfo {year} {1994}),\ \bibfield  {title} {\enquote
  {\bibinfo {title} {Quantum nonlocality as an axiom},}\ }\href {\doibase
  10.1007/BF02058098} {\bibfield  {journal} {\bibinfo  {journal} {Foundations
  of Physics}\ }\textbf {\bibinfo {volume} {24}}~(\bibinfo {number} {3}),\
  \bibinfo {pages} {379--385}}\BibitemShut {NoStop}%
\bibitem [{\citenamefont {Popper}(1982)}]{Popper82}%
  \BibitemOpen
  \bibfield  {author} {\bibinfo {author} {\bibnamefont {Popper}, \bibfnamefont
  {K~R}}} (\bibinfo {year} {1982}),\ \href@noop {} {\emph {\bibinfo {title}
  {{T}he {O}pen {U}niverse. {A}n {A}rgument for {I}ndeterminism}}}\ (\bibinfo
  {publisher} {Routledge, London and New York})\BibitemShut {NoStop}%
\bibitem [{\citenamefont {Rajasekar}\ and\ \citenamefont
  {Sanjuan}(2016)}]{Sanjuan16}%
  \BibitemOpen
  \bibfield  {author} {\bibinfo {author} {\bibnamefont {Rajasekar},
  \bibfnamefont {Shanmuganathan}}, \ and\ \bibinfo {author} {\bibfnamefont
  {Miguel A.~F.}\ \bibnamefont {Sanjuan}}} (\bibinfo {year} {2016}),\ \href
  {\doibase 10.1007/978-3-319-24886-8} {\emph {\bibinfo {title} {Nonlinear
  {R}esonances}}}\ (\bibinfo  {publisher} {Springer International
  Publishing})\BibitemShut {NoStop}%
\bibitem [{\citenamefont {Ramanathan}\ \emph {et~al.}(2016)\citenamefont
  {Ramanathan}, \citenamefont {Brand\~ao}, \citenamefont {Horodecki},
  \citenamefont {Horodecki}, \citenamefont {Horodecki},\ and\ \citenamefont
  {Wojew\'odka}}]{Ravi16a}%
  \BibitemOpen
  \bibfield  {author} {\bibinfo {author} {\bibnamefont {Ramanathan},
  \bibfnamefont {Ravishankar}}, \bibinfo {author} {\bibfnamefont {Fernando G.
  S.~L.}\ \bibnamefont {Brand\~ao}}, \bibinfo {author} {\bibfnamefont {Karol}\
  \bibnamefont {Horodecki}}, \bibinfo {author} {\bibfnamefont {Micha\l{}}\
  \bibnamefont {Horodecki}}, \bibinfo {author} {\bibfnamefont {Pawe\l{}}\
  \bibnamefont {Horodecki}}, \ and\ \bibinfo {author} {\bibfnamefont {Hanna}\
  \bibnamefont {Wojew\'odka}}} (\bibinfo {year} {2016}),\ \bibfield  {title}
  {\enquote {\bibinfo {title} {Randomness amplification under minimal
  fundamental assumptions on the devices},}\ }\href {\doibase
  10.1103/PhysRevLett.117.230501} {\bibfield  {journal} {\bibinfo  {journal}
  {Phys. Rev. Lett.}\ }\textbf {\bibinfo {volume} {117}},\ \bibinfo {pages}
  {230501}}\BibitemShut {NoStop}%
\bibitem [{\citenamefont {Raz}(2005)}]{Raz05}%
  \BibitemOpen
  \bibfield  {author} {\bibinfo {author} {\bibnamefont {Raz}, \bibfnamefont
  {Ran}}} (\bibinfo {year} {2005}),\ \bibfield  {title} {\enquote {\bibinfo
  {title} {Extractors with weak random seeds},}\ }in\ \href {\doibase
  10.1145/1060590.1060593} {\emph {\bibinfo {booktitle} {Proceedings of the
  Thirty-seventh Annual ACM Symposium on Theory of Computing}}},\ \bibinfo
  {series and number} {STOC '05}\ (\bibinfo  {publisher} {ACM},\ \bibinfo
  {address} {New York, NY, USA})\ pp.\ \bibinfo {pages} {11--20}\BibitemShut
  {NoStop}%
\bibitem [{\citenamefont {Roberts}(2009)}]{Roberts09}%
  \BibitemOpen
  \bibfield  {author} {\bibinfo {author} {\bibnamefont {Roberts}, \bibfnamefont
  {B~W}}} (\bibinfo {year} {2009}),\ \bibfield  {title} {\enquote {\bibinfo
  {title} {{W}ilson's case against the dome: {N}ot necessary, not
  sufficient},}\ }\href@noop {} {\bibinfo  {journal} {Unpublished manuscript}\
  }\BibitemShut {NoStop}%
\bibitem [{\citenamefont {Robertson}(1929)}]{Robertson29}%
  \BibitemOpen
\bibfield  {journal} {  }\bibfield  {author} {\bibinfo {author} {\bibnamefont
  {Robertson}, \bibfnamefont {H~P}}} (\bibinfo {year} {1929}),\ \bibfield
  {title} {\enquote {\bibinfo {title} {The uncertainty principle},}\ }\href
  {\doibase 10.1103/PhysRev.34.163} {\bibfield  {journal} {\bibinfo  {journal}
  {Phys. Rev.}\ }\textbf {\bibinfo {volume} {34}},\ \bibinfo {pages}
  {163--164}}\BibitemShut {NoStop}%
\bibitem [{\citenamefont {Rosenfeld}\ \emph {et~al.}(2011)\citenamefont
  {Rosenfeld}, \citenamefont {Hofmann}, \citenamefont {Ortegel}, \citenamefont
  {Krug}, \citenamefont {Henkel}, \citenamefont {Kurtsiefer}, \citenamefont
  {Weber},\ and\ \citenamefont {Weinfurter}}]{RosenfeldOS2011}%
  \BibitemOpen
  \bibfield  {author} {\bibinfo {author} {\bibnamefont {Rosenfeld},
  \bibfnamefont {W}}, \bibinfo {author} {\bibfnamefont {J.}~\bibnamefont
  {Hofmann}}, \bibinfo {author} {\bibfnamefont {N.}~\bibnamefont {Ortegel}},
  \bibinfo {author} {\bibfnamefont {M.}~\bibnamefont {Krug}}, \bibinfo {author}
  {\bibfnamefont {F.}~\bibnamefont {Henkel}}, \bibinfo {author} {\bibfnamefont
  {Ch.}\ \bibnamefont {Kurtsiefer}}, \bibinfo {author} {\bibfnamefont
  {M.}~\bibnamefont {Weber}}, \ and\ \bibinfo {author} {\bibfnamefont
  {H.}~\bibnamefont {Weinfurter}}} (\bibinfo {year} {2011}),\ \bibfield
  {title} {\enquote {\bibinfo {title} {Towards high-fidelity interference of
  photons emitted by two remotely trapped rb-87 atoms},}\ }\href {\doibase
  10.1134/S0030400X11110233} {\bibfield  {journal} {\bibinfo  {journal} {Optics
  and Spectroscopy}\ }\textbf {\bibinfo {volume} {111}}~(\bibinfo {number}
  {4}),\ \bibinfo {pages} {535--539}}\BibitemShut {NoStop}%
\bibitem [{\citenamefont {Rosicka}\ \emph {et~al.}(2016)\citenamefont
  {Rosicka}, \citenamefont {Ramanathan}, \citenamefont {Gnaciński},
  \citenamefont {Horodecki}, \citenamefont {Horodecki}, \citenamefont
  {Horodecki},\ and\ \citenamefont {Severini}}]{Rosicka16}%
  \BibitemOpen
  \bibfield  {author} {\bibinfo {author} {\bibnamefont {Rosicka}, \bibfnamefont
  {M}}, \bibinfo {author} {\bibfnamefont {R}~\bibnamefont {Ramanathan}},
  \bibinfo {author} {\bibfnamefont {P}~\bibnamefont {Gnaciński}}, \bibinfo
  {author} {\bibfnamefont {K}~\bibnamefont {Horodecki}}, \bibinfo {author}
  {\bibfnamefont {M}~\bibnamefont {Horodecki}}, \bibinfo {author}
  {\bibfnamefont {P}~\bibnamefont {Horodecki}}, \ and\ \bibinfo {author}
  {\bibfnamefont {S}~\bibnamefont {Severini}}} (\bibinfo {year} {2016}),\
  \bibfield  {title} {\enquote {\bibinfo {title} {Linear game non-contextuality
  and {B}ell inequalities—a graph-theoretic approach},}\ }\href
  {http://stacks.iop.org/1367-2630/18/i=4/a=045020} {\bibfield  {journal}
  {\bibinfo  {journal} {New Journal of Physics}\ }\textbf {\bibinfo {volume}
  {18}}~(\bibinfo {number} {4}),\ \bibinfo {pages} {045020}}\BibitemShut
  {NoStop}%
\bibitem [{\citenamefont {Rukhin}\ \emph {et~al.}(2010)\citenamefont {Rukhin},
  \citenamefont {Soto}, \citenamefont {Nechvatal}, \citenamefont {Smid},
  \citenamefont {Barker}, \citenamefont {Leigh}, \citenamefont {Levenson},
  \citenamefont {Vangel}, \citenamefont {Banks}, \citenamefont {Heckert},
  \citenamefont {Dray},\ and\ \citenamefont {Vo}}]{RukhinNIST2010}%
  \BibitemOpen
  \bibfield  {author} {\bibinfo {author} {\bibnamefont {Rukhin}, \bibfnamefont
  {Andrew}}, \bibinfo {author} {\bibfnamefont {Juan}\ \bibnamefont {Soto}},
  \bibinfo {author} {\bibfnamefont {James}\ \bibnamefont {Nechvatal}}, \bibinfo
  {author} {\bibfnamefont {Miles}\ \bibnamefont {Smid}}, \bibinfo {author}
  {\bibfnamefont {Elaine}\ \bibnamefont {Barker}}, \bibinfo {author}
  {\bibfnamefont {Stefan}\ \bibnamefont {Leigh}}, \bibinfo {author}
  {\bibfnamefont {Mark}\ \bibnamefont {Levenson}}, \bibinfo {author}
  {\bibfnamefont {Mark}\ \bibnamefont {Vangel}}, \bibinfo {author}
  {\bibfnamefont {David}\ \bibnamefont {Banks}}, \bibinfo {author}
  {\bibfnamefont {Alan}\ \bibnamefont {Heckert}}, \bibinfo {author}
  {\bibfnamefont {James}\ \bibnamefont {Dray}}, \ and\ \bibinfo {author}
  {\bibfnamefont {San}\ \bibnamefont {Vo}}} (\bibinfo {year} {2010}),\ \href
  {http://csrc.nist.gov/publications/PubsSPs.html#800-22} {\emph {\bibinfo
  {title} {A Statistical Test Suite for Random and Pseudorandom Number
  Generators for Cryptographic Applications}}},\ \bibinfo {type} {Tech. Rep.}\
  \bibinfo {number} {800-22}\ (\bibinfo  {institution} {National Institute of
  Standards and Technology})\BibitemShut {NoStop}%
\bibitem [{\citenamefont {Sanju\'an}(2009{\natexlab{a}})}]{blog2}%
  \BibitemOpen
  \bibfield  {author} {\bibinfo {author} {\bibnamefont {Sanju\'an},
  \bibfnamefont {Miguel A~F}}} (\bibinfo {year} {2009}{\natexlab{a}}),\
  \bibfield  {title} {\enquote {\bibinfo {title} {{J}ames {C}lerk {M}axwell,
  caos y determinismo},}\ }\href
  {http://www.madrimasd.org/blogs/complejidad/2009/10/24/127508} {\bibinfo
  {journal} {BLOG madri+d}\ }\BibitemShut {NoStop}%
\bibitem [{\citenamefont {Sanju\'an}(2009{\natexlab{b}})}]{blog1}%
  \BibitemOpen
\bibfield  {journal} {  }\bibfield  {author} {\bibinfo {author} {\bibnamefont
  {Sanju\'an}, \bibfnamefont {Miguel A~F}}} (\bibinfo {year}
  {2009}{\natexlab{b}}),\ \bibfield  {title} {\enquote {\bibinfo {title} {{M}ax
  {B}orn y el determinismo cl\'asico},}\ }\href
  {http://www.madrimasd.org/blogs/complejidad/2009/11/06/128252} {\bibinfo
  {journal} {BLOG madri+d}\ }\BibitemShut {NoStop}%
\bibitem [{\citenamefont {Sanju\'an}(2009{\natexlab{c}})}]{blog3}%
  \BibitemOpen
\bibfield  {journal} {  }\bibfield  {author} {\bibinfo {author} {\bibnamefont
  {Sanju\'an}, \bibfnamefont {Miguel A~F}}} (\bibinfo {year}
  {2009}{\natexlab{c}}),\ \bibfield  {title} {\enquote {\bibinfo {title}
  {¿{C}onocía {F}eynman la teoría del caos?}}\ }\href
  {http://www.madrimasd.org/blogs/complejidad/2009/11/06/128257} {\bibinfo
  {journal} {BLOG madri+d}\ }\BibitemShut {NoStop}%
\bibitem [{\citenamefont {Santha}\ and\ \citenamefont
  {Vazirani}(1986)}]{Santha86}%
  \BibitemOpen
\bibfield  {journal} {  }\bibfield  {author} {\bibinfo {author} {\bibnamefont
  {Santha}, \bibfnamefont {Miklos}}, \ and\ \bibinfo {author} {\bibfnamefont
  {Umesh~V.}\ \bibnamefont {Vazirani}}} (\bibinfo {year} {1986}),\ \bibfield
  {title} {\enquote {\bibinfo {title} {Generating quasi-random sequences from
  semi-random sources},}\ }\href {\doibase
  http://dx.doi.org/10.1016/0022-0000(86)90044-9} {\bibfield  {journal}
  {\bibinfo  {journal} {Journal of Computer and System Sciences}\ }\textbf
  {\bibinfo {volume} {33}}~(\bibinfo {number} {1}),\ \bibinfo {pages} {75 --
  87}}\BibitemShut {NoStop}%
\bibitem [{\citenamefont {Saunders}(1998)}]{Saunders1998}%
  \BibitemOpen
  \bibfield  {author} {\bibinfo {author} {\bibnamefont {Saunders},
  \bibfnamefont {Simon}}} (\bibinfo {year} {1998}),\ \bibfield  {title}
  {\enquote {\bibinfo {title} {{Time, Quantum Mechanics, and Probability}},}\
  }\href {\doibase 10.1023/A:1005079904008} {\bibfield  {journal} {\bibinfo
  {journal} {Synthese}\ }\textbf {\bibinfo {volume} {114}}~(\bibinfo {number}
  {3}),\ \bibinfo {pages} {373--404}}\BibitemShut {NoStop}%
\bibitem [{\citenamefont {Saunders}(2010)}]{Saunders2010}%
  \BibitemOpen
  \bibfield  {author} {\bibinfo {author} {\bibnamefont {Saunders},
  \bibfnamefont {Simon}}} (\bibinfo {year} {2010}),\ \bibfield  {title}
  {\enquote {\bibinfo {title} {{Chance in the Everett Interpretation}},}\ }in\
  \href
  {http://www.oxfordscholarship.com/view/10.1093/acprof:oso/9780199560561.001.0001/acprof-9780199560561-chapter-8}
  {\emph {\bibinfo {booktitle} {{Many Worlds?: Everett, Quantum Theory, and
  Reality}}}},\ \bibinfo {editor} {edited by\ \bibinfo {editor} {\bibfnamefont
  {Simon}\ \bibnamefont {Saunders}}, \bibinfo {editor} {\bibfnamefont
  {Jonathan}\ \bibnamefont {Barrett}}, \bibinfo {editor} {\bibfnamefont
  {Adrian}\ \bibnamefont {Kent}}, \ and\ \bibinfo {editor} {\bibfnamefont
  {David}\ \bibnamefont {Wallace}}},\ Chap.~\bibinfo {chapter} {8}\ (\bibinfo
  {publisher} {Oxford Publisher},\ \bibinfo {address} {Oxford})\ pp.\ \bibinfo
  {pages} {181--205}\BibitemShut {NoStop}%
\bibitem [{\citenamefont {Schawlow}\ and\ \citenamefont
  {Townes}(1958)}]{SchawlowPR1958}%
  \BibitemOpen
  \bibfield  {author} {\bibinfo {author} {\bibnamefont {Schawlow},
  \bibfnamefont {A~L}}, \ and\ \bibinfo {author} {\bibfnamefont {C.~H.}\
  \bibnamefont {Townes}}} (\bibinfo {year} {1958}),\ \bibfield  {title}
  {\enquote {\bibinfo {title} {Infrared and optical masers},}\ }\href {\doibase
  10.1103/PhysRev.112.1940} {\bibfield  {journal} {\bibinfo  {journal} {Phys.
  Rev.}\ }\textbf {\bibinfo {volume} {112}},\ \bibinfo {pages}
  {1940--1949}}\BibitemShut {NoStop}%
\bibitem [{\citenamefont {Schr\"odinger}(1930)}]{Schrodinger30}%
  \BibitemOpen
  \bibfield  {author} {\bibinfo {author} {\bibnamefont {Schr\"odinger},
  \bibfnamefont {E}}} (\bibinfo {year} {1930}),\ \bibfield  {title} {\enquote
  {\bibinfo {title} {About {H}eisenberg {U}ncertainty {R}elation},}\ }in\
  \href@noop {} {\emph {\bibinfo {booktitle} {Proc. Prussian Acad. Sci., Phys.
  Math. Section}}},\ Vol.\ \bibinfo {volume} {XIX},\ p.\ \bibinfo {pages}
  {293}\BibitemShut {NoStop}%
\bibitem [{\citenamefont {Schr\"odinger}(1989)}]{Schrodinger89}%
  \BibitemOpen
  \bibfield  {author} {\bibinfo {author} {\bibnamefont {Schr\"odinger},
  \bibfnamefont {Erwin}}} (\bibinfo {year} {1989}),\ \href@noop {} {\emph
  {\bibinfo {title} {{S}tatistical {T}hermodynamics}}}\ (\bibinfo  {publisher}
  {Dover Publications})\BibitemShut {NoStop}%
\bibitem [{\citenamefont {Scully}\ and\ \citenamefont
  {W{\'o}dkiewicz}(1989)}]{Scully89}%
  \BibitemOpen
  \bibfield  {author} {\bibinfo {author} {\bibnamefont {Scully}, \bibfnamefont
  {Marlan~O}}, \ and\ \bibinfo {author} {\bibfnamefont {K.}~\bibnamefont
  {W{\'o}dkiewicz}}} (\bibinfo {year} {1989}),\ \enquote {\bibinfo {title}
  {{Coherence and Quantum Optics VI: Proceedings of the Sixth Rochester
  Conference on Coherence and Quantum Optics held at the {U}niversity of
  {R}ochester, June 26--28, 1989}},}\ Chap.\ \bibinfo {chapter} {On the Quantum
  Malus Law for Photon and Spin Quantum Correlations}\ (\bibinfo  {publisher}
  {Springer US},\ \bibinfo {address} {Boston, MA})\ pp.\ \bibinfo {pages}
  {1047--1050}\BibitemShut {NoStop}%
\bibitem [{\citenamefont {Shalm}\ \emph {et~al.}(2015)\citenamefont {Shalm},
  \citenamefont {Meyer-Scott}, \citenamefont {Christensen}, \citenamefont
  {Bierhorst}, \citenamefont {Wayne}, \citenamefont {Stevens}, \citenamefont
  {Gerrits}, \citenamefont {Glancy}, \citenamefont {Hamel}, \citenamefont
  {Allman}, \citenamefont {Coakley}, \citenamefont {Dyer}, \citenamefont
  {Hodge}, \citenamefont {Lita}, \citenamefont {Verma}, \citenamefont
  {Lambrocco}, \citenamefont {Tortorici}, \citenamefont {Migdall},
  \citenamefont {Zhang}, \citenamefont {Kumor}, \citenamefont {Farr},
  \citenamefont {Marsili}, \citenamefont {Shaw}, \citenamefont {Stern},
  \citenamefont {Abell\'an}, \citenamefont {Amaya}, \citenamefont {Pruneri},
  \citenamefont {Jennewein}, \citenamefont {Mitchell}, \citenamefont {Kwiat},
  \citenamefont {Bienfang}, \citenamefont {Mirin}, \citenamefont {Knill},\ and\
  \citenamefont {Nam}}]{Shalm15}%
  \BibitemOpen
  \bibfield  {author} {\bibinfo {author} {\bibnamefont {Shalm}, \bibfnamefont
  {Lynden~K}}, \bibinfo {author} {\bibfnamefont {Evan}\ \bibnamefont
  {Meyer-Scott}}, \bibinfo {author} {\bibfnamefont {Bradley~G.}\ \bibnamefont
  {Christensen}}, \bibinfo {author} {\bibfnamefont {Peter}\ \bibnamefont
  {Bierhorst}}, \bibinfo {author} {\bibfnamefont {Michael~A.}\ \bibnamefont
  {Wayne}}, \bibinfo {author} {\bibfnamefont {Martin~J.}\ \bibnamefont
  {Stevens}}, \bibinfo {author} {\bibfnamefont {Thomas}\ \bibnamefont
  {Gerrits}}, \bibinfo {author} {\bibfnamefont {Scott}\ \bibnamefont {Glancy}},
  \bibinfo {author} {\bibfnamefont {Deny~R.}\ \bibnamefont {Hamel}}, \bibinfo
  {author} {\bibfnamefont {Michael~S.}\ \bibnamefont {Allman}}, \bibinfo
  {author} {\bibfnamefont {Kevin~J.}\ \bibnamefont {Coakley}}, \bibinfo
  {author} {\bibfnamefont {Shellee~D.}\ \bibnamefont {Dyer}}, \bibinfo {author}
  {\bibfnamefont {Carson}\ \bibnamefont {Hodge}}, \bibinfo {author}
  {\bibfnamefont {Adriana~E.}\ \bibnamefont {Lita}}, \bibinfo {author}
  {\bibfnamefont {Varun~B.}\ \bibnamefont {Verma}}, \bibinfo {author}
  {\bibfnamefont {Camilla}\ \bibnamefont {Lambrocco}}, \bibinfo {author}
  {\bibfnamefont {Edward}\ \bibnamefont {Tortorici}}, \bibinfo {author}
  {\bibfnamefont {Alan~L.}\ \bibnamefont {Migdall}}, \bibinfo {author}
  {\bibfnamefont {Yanbao}\ \bibnamefont {Zhang}}, \bibinfo {author}
  {\bibfnamefont {Daniel~R.}\ \bibnamefont {Kumor}}, \bibinfo {author}
  {\bibfnamefont {William~H.}\ \bibnamefont {Farr}}, \bibinfo {author}
  {\bibfnamefont {Francesco}\ \bibnamefont {Marsili}}, \bibinfo {author}
  {\bibfnamefont {Matthew~D.}\ \bibnamefont {Shaw}}, \bibinfo {author}
  {\bibfnamefont {Jeffrey~A.}\ \bibnamefont {Stern}}, \bibinfo {author}
  {\bibfnamefont {Carlos}\ \bibnamefont {Abell\'an}}, \bibinfo {author}
  {\bibfnamefont {Waldimar}\ \bibnamefont {Amaya}}, \bibinfo {author}
  {\bibfnamefont {Valerio}\ \bibnamefont {Pruneri}}, \bibinfo {author}
  {\bibfnamefont {Thomas}\ \bibnamefont {Jennewein}}, \bibinfo {author}
  {\bibfnamefont {Morgan~W.}\ \bibnamefont {Mitchell}}, \bibinfo {author}
  {\bibfnamefont {Paul~G.}\ \bibnamefont {Kwiat}}, \bibinfo {author}
  {\bibfnamefont {Joshua~C.}\ \bibnamefont {Bienfang}}, \bibinfo {author}
  {\bibfnamefont {Richard~P.}\ \bibnamefont {Mirin}}, \bibinfo {author}
  {\bibfnamefont {Emanuel}\ \bibnamefont {Knill}}, \ and\ \bibinfo {author}
  {\bibfnamefont {Sae~Woo}\ \bibnamefont {Nam}}} (\bibinfo {year} {2015}),\
  \bibfield  {title} {\enquote {\bibinfo {title} {Strong loophole-free test of
  local realism*},}\ }\href {\doibase 10.1103/PhysRevLett.115.250402}
  {\bibfield  {journal} {\bibinfo  {journal} {Phys. Rev. Lett.}\ }\textbf
  {\bibinfo {volume} {115}},\ \bibinfo {pages} {250402}}\BibitemShut {NoStop}%
\bibitem [{\citenamefont {Shaltiel}(2002)}]{Shaltiel02}%
  \BibitemOpen
  \bibfield  {author} {\bibinfo {author} {\bibnamefont {Shaltiel},
  \bibfnamefont {Ronen}}} (\bibinfo {year} {2002}),\ \bibfield  {title}
  {\enquote {\bibinfo {title} {Recent developments in explicit constructions of
  extractors},}\ }\href@noop {} {\bibfield  {journal} {\bibinfo  {journal}
  {Bulletin of the {EATCS}}\ }\textbf {\bibinfo {volume} {77}},\ \bibinfo
  {pages} {67--95}}\BibitemShut {NoStop}%
\bibitem [{\citenamefont {Smoluchowski}(1918)}]{Smoluchowski18}%
  \BibitemOpen
  \bibfield  {author} {\bibinfo {author} {\bibnamefont {Smoluchowski},
  \bibfnamefont {M~V}}} (\bibinfo {year} {1918}),\ \bibfield  {title} {\enquote
  {\bibinfo {title} {{\"U}ber den {B}egriff des {Z}ufalls und den {U}rsprung
  der {W}ahrscheinlichkeitsgesetze in der {P}hysik},}\ }\href@noop {}
  {\bibfield  {journal} {\bibinfo  {journal} {Naturwissenschaften}\ }\textbf
  {\bibinfo {volume} {6}},\ \bibinfo {pages} {253--263}}\BibitemShut {NoStop}%
\bibitem [{\citenamefont {Suarez}(2013)}]{suarez13}%
  \BibitemOpen
  \bibfield  {author} {\bibinfo {author} {\bibnamefont {Suarez}, \bibfnamefont
  {Antoine}}} (\bibinfo {year} {2013}),\ \bibfield  {title} {\enquote {\bibinfo
  {title} {{F}ree {W}ill and {N}onlocality at {D}etection as {B}asic
  {P}rinciples of {Q}uantum {P}hysics},}\ }in\ \href@noop {} {\emph {\bibinfo
  {booktitle} {{I}s {S}cience {C}ompatible with {F}ree {W}ill? {E}xploring
  {F}ree {W}ill and {C}onsciousness in the {L}ight of {Q}uantum {P}hysics and
  {N}euroscience}}},\ Chap.~\bibinfo {chapter} {4}\ (\bibinfo  {publisher}
  {Springer},\ \bibinfo {address} {New York})\ pp.\ \bibinfo {pages}
  {63--79}\BibitemShut {NoStop}%
\bibitem [{\citenamefont {Suarez}\ and\ \citenamefont
  {Adams}(2013)}]{suarezbook13}%
  \BibitemOpen
  \bibinfo {editor} {\bibnamefont {Suarez}, \bibfnamefont {Antoine}}, \ and\
  \bibinfo {editor} {\bibfnamefont {Peter}\ \bibnamefont {Adams}},\ Eds.
  (\bibinfo {year} {2013}),\ \href@noop {} {\emph {\bibinfo {title} {{I}s
  {S}cience {C}ompatible with {F}ree {W}ill? {E}xploring {F}ree {W}ill and
  {C}onsciousness in the {L}ight of {Q}uantum {P}hysics and {N}euroscience}}}\
  (\bibinfo  {publisher} {Springer},\ \bibinfo {address} {New
  York})\BibitemShut {NoStop}%
\bibitem [{\citenamefont {Tolman}(2010)}]{Tolman10}%
  \BibitemOpen
  \bibfield  {author} {\bibinfo {author} {\bibnamefont {Tolman}, \bibfnamefont
  {Richard~C}}} (\bibinfo {year} {2010}),\ \href@noop {} {\emph {\bibinfo
  {title} {The {P}rinciples of {S}tatistical {M}echanics ({D}over {B}ooks on
  Physics)}}}\ (\bibinfo  {publisher} {Dover Publications})\BibitemShut
  {NoStop}%
\bibitem [{\citenamefont {de~la Torre}\ \emph {et~al.}(2015)\citenamefont
  {de~la Torre}, \citenamefont {Hoban}, \citenamefont {Dhara}, \citenamefont
  {Prettico},\ and\ \citenamefont {Ac\'{\i}n}}]{delatorre}%
  \BibitemOpen
  \bibfield  {author} {\bibinfo {author} {\bibnamefont {de~la Torre},
  \bibfnamefont {Gonzalo}}, \bibinfo {author} {\bibfnamefont {Matty~J.}\
  \bibnamefont {Hoban}}, \bibinfo {author} {\bibfnamefont {Chirag}\
  \bibnamefont {Dhara}}, \bibinfo {author} {\bibfnamefont {Giuseppe}\
  \bibnamefont {Prettico}}, \ and\ \bibinfo {author} {\bibfnamefont {Antonio}\
  \bibnamefont {Ac\'{\i}n}}} (\bibinfo {year} {2015}),\ \bibfield  {title}
  {\enquote {\bibinfo {title} {Maximally nonlocal theories cannot be maximally
  random},}\ }\href {\doibase 10.1103/PhysRevLett.114.160502} {\bibfield
  {journal} {\bibinfo  {journal} {Phys. Rev. Lett.}\ }\textbf {\bibinfo
  {volume} {114}},\ \bibinfo {pages} {160502}}\BibitemShut {NoStop}%
\bibitem [{\citenamefont {Trevisan}(2001)}]{randextr}%
  \BibitemOpen
  \bibfield  {author} {\bibinfo {author} {\bibnamefont {Trevisan},
  \bibfnamefont {L}}} (\bibinfo {year} {2001}),\ \bibfield  {title} {\enquote
  {\bibinfo {title} {Extractors and pseudorandom generators},}\ }\href
  {\doibase 10.1145/502090.502099} {\bibfield  {journal} {\bibinfo  {journal}
  {J. ACM}\ }\textbf {\bibinfo {volume} {48}}~(\bibinfo {number} {4}),\
  \bibinfo {pages} {860--879}}\BibitemShut {NoStop}%
\bibitem [{\citenamefont {Tura}\ \emph
  {et~al.}(2014{\natexlab{a}})\citenamefont {Tura}, \citenamefont {Augusiak},
  \citenamefont {Sainz}, \citenamefont {V{\'e}rtesi}, \citenamefont
  {Lewenstein},\ and\ \citenamefont {Ac{\'\i}n}}]{Tura14a}%
  \BibitemOpen
  \bibfield  {author} {\bibinfo {author} {\bibnamefont {Tura}, \bibfnamefont
  {J}}, \bibinfo {author} {\bibfnamefont {R.}~\bibnamefont {Augusiak}},
  \bibinfo {author} {\bibfnamefont {A.~B.}\ \bibnamefont {Sainz}}, \bibinfo
  {author} {\bibfnamefont {T.}~\bibnamefont {V{\'e}rtesi}}, \bibinfo {author}
  {\bibfnamefont {M.}~\bibnamefont {Lewenstein}}, \ and\ \bibinfo {author}
  {\bibfnamefont {A.}~\bibnamefont {Ac{\'\i}n}}} (\bibinfo {year}
  {2014}{\natexlab{a}}),\ \bibfield  {title} {\enquote {\bibinfo {title}
  {Detecting nonlocality in many-body quantum states},}\ }\href {\doibase
  10.1126/science.1247715} {\bibfield  {journal} {\bibinfo  {journal}
  {Science}\ }\textbf {\bibinfo {volume} {344}}~(\bibinfo {number} {6189}),\
  \bibinfo {pages} {1256--1258}}\BibitemShut {NoStop}%
\bibitem [{\citenamefont {Tura}\ \emph {et~al.}(2015)\citenamefont {Tura},
  \citenamefont {Augusiak}, \citenamefont {Sainz}, \citenamefont {Lücke},
  \citenamefont {Klempt}, \citenamefont {Lewenstein},\ and\ \citenamefont
  {Acín}}]{Tura15}%
  \BibitemOpen
  \bibfield  {author} {\bibinfo {author} {\bibnamefont {Tura}, \bibfnamefont
  {J}}, \bibinfo {author} {\bibfnamefont {R.}~\bibnamefont {Augusiak}},
  \bibinfo {author} {\bibfnamefont {A.B.}\ \bibnamefont {Sainz}}, \bibinfo
  {author} {\bibfnamefont {B.}~\bibnamefont {Lücke}}, \bibinfo {author}
  {\bibfnamefont {C.}~\bibnamefont {Klempt}}, \bibinfo {author} {\bibfnamefont
  {M.}~\bibnamefont {Lewenstein}}, \ and\ \bibinfo {author} {\bibfnamefont
  {A.}~\bibnamefont {Acín}}} (\bibinfo {year} {2015}),\ \bibfield  {title}
  {\enquote {\bibinfo {title} {Nonlocality in many-body quantum systems
  detected with two-body correlators},}\ }\href {\doibase
  http://dx.doi.org/10.1016/j.aop.2015.07.021} {\bibfield  {journal} {\bibinfo
  {journal} {Annals of Physics}\ }\textbf {\bibinfo {volume} {362}},\ \bibinfo
  {pages} {370 -- 423}}\BibitemShut {NoStop}%
\bibitem [{\citenamefont {Tura}\ \emph
  {et~al.}(2014{\natexlab{b}})\citenamefont {Tura}, \citenamefont {Sainz},
  \citenamefont {Vértesi}, \citenamefont {Acín}, \citenamefont {Lewenstein},\
  and\ \citenamefont {Augusiak}}]{Tura14}%
  \BibitemOpen
  \bibfield  {author} {\bibinfo {author} {\bibnamefont {Tura}, \bibfnamefont
  {J}}, \bibinfo {author} {\bibfnamefont {A.~B.}\ \bibnamefont {Sainz}},
  \bibinfo {author} {\bibfnamefont {T.}~\bibnamefont {Vértesi}}, \bibinfo
  {author} {\bibfnamefont {A.}~\bibnamefont {Acín}}, \bibinfo {author}
  {\bibfnamefont {M.}~\bibnamefont {Lewenstein}}, \ and\ \bibinfo {author}
  {\bibfnamefont {R.}~\bibnamefont {Augusiak}}} (\bibinfo {year}
  {2014}{\natexlab{b}}),\ \bibfield  {title} {\enquote {\bibinfo {title}
  {Translationally invariant multipartite bell inequalities involving only
  two-body correlators},}\ }\href
  {http://stacks.iop.org/1751-8121/47/i=42/a=424024} {\bibfield  {journal}
  {\bibinfo  {journal} {Journal of Physics A: Mathematical and Theoretical}\
  }\textbf {\bibinfo {volume} {47}}~(\bibinfo {number} {42}),\ \bibinfo {pages}
  {424024}}\BibitemShut {NoStop}%
\bibitem [{\citenamefont {Tylec}\ and\ \citenamefont {Kuś}(2015)}]{Tylec15}%
  \BibitemOpen
  \bibfield  {author} {\bibinfo {author} {\bibnamefont {Tylec}, \bibfnamefont
  {T~I}}, \ and\ \bibinfo {author} {\bibfnamefont {M.}~\bibnamefont {Kuś}}}
  (\bibinfo {year} {2015}),\ \bibfield  {title} {\enquote {\bibinfo {title}
  {Non-signaling boxes and quantum logics},}\ }\href
  {http://stacks.iop.org/1751-8121/48/i=50/a=505303} {\bibfield  {journal}
  {\bibinfo  {journal} {Journal of Physics A: Mathematical and Theoretical}\
  }\textbf {\bibinfo {volume} {48}}~(\bibinfo {number} {50}),\ \bibinfo {pages}
  {505303}}\BibitemShut {NoStop}%
\bibitem [{\citenamefont {Vaidman}(2014)}]{Vaidman14}%
  \BibitemOpen
  \bibfield  {author} {\bibinfo {author} {\bibnamefont {Vaidman}, \bibfnamefont
  {Lev}}} (\bibinfo {year} {2014}),\ \bibfield  {title} {\enquote {\bibinfo
  {title} {Quantum theory and determinism},}\ }\href {\doibase
  10.1007/s40509-014-0008-4} {\bibfield  {journal} {\bibinfo  {journal}
  {Quantum Studies: Mathematics and Foundations}\ }\textbf {\bibinfo {volume}
  {1}}~(\bibinfo {number} {1}),\ \bibinfo {pages} {5--38}}\BibitemShut
  {NoStop}%
\bibitem [{\citenamefont {Vallejo}\ and\ \citenamefont
  {Sanjuan}(2017)}]{Sanjuan17}%
  \BibitemOpen
  \bibfield  {author} {\bibinfo {author} {\bibnamefont {Vallejo}, \bibfnamefont
  {Juan~C}}, \ and\ \bibinfo {author} {\bibfnamefont {Miguel A.~F.}\
  \bibnamefont {Sanjuan}}} (\bibinfo {year} {2017}),\ \href {\doibase
  10.1007/978-3-319-51893-0} {\emph {\bibinfo {title} {Predictability of
  Chaotic Dynamics}}}\ (\bibinfo  {publisher} {Springer International
  Publishing})\BibitemShut {NoStop}%
\bibitem [{\citenamefont {Vanden~Eijnden}\ and\ \citenamefont
  {Vanden~Eijnden}(2000)}]{Vanden1}%
  \BibitemOpen
  \bibfield  {author} {\bibinfo {author} {\bibnamefont {Vanden~Eijnden},
  \bibfnamefont {Weinan~E}}, \ and\ \bibinfo {author} {\bibfnamefont {Eric}\
  \bibnamefont {Vanden~Eijnden}}} (\bibinfo {year} {2000}),\ \bibfield  {title}
  {\enquote {\bibinfo {title} {Generalized flows, intrinsic stochasticity, and
  turbulent transport},}\ }\href {\doibase 10.1073/pnas.97.15.8200} {\bibfield
  {journal} {\bibinfo  {journal} {Proceedings of the National Academy of
  Sciences}\ }\textbf {\bibinfo {volume} {97}}~(\bibinfo {number} {15}),\
  \bibinfo {pages} {8200--8205}}\BibitemShut {NoStop}%
\bibitem [{\citenamefont {Vazirani}\ and\ \citenamefont
  {Vidick}(2012)}]{Vazirani12}%
  \BibitemOpen
  \bibfield  {author} {\bibinfo {author} {\bibnamefont {Vazirani},
  \bibfnamefont {Umesh}}, \ and\ \bibinfo {author} {\bibfnamefont {Thomas}\
  \bibnamefont {Vidick}}} (\bibinfo {year} {2012}),\ \bibfield  {title}
  {\enquote {\bibinfo {title} {Certifiable quantum dice: Or, true random number
  generation secure against quantum adversaries},}\ }in\ \href {\doibase
  10.1145/2213977.2213984} {\emph {\bibinfo {booktitle} {Proceedings of the
  Forty-fourth Annual ACM Symposium on Theory of Computing}}},\ \bibinfo
  {series and number} {STOC '12}\ (\bibinfo  {publisher} {ACM},\ \bibinfo
  {address} {New York, NY, USA})\ pp.\ \bibinfo {pages} {61--76}\BibitemShut
  {NoStop}%
\bibitem [{\citenamefont {Vazirani}(1987)}]{Vazirani87}%
  \BibitemOpen
  \bibfield  {author} {\bibinfo {author} {\bibnamefont {Vazirani},
  \bibfnamefont {Umesh~V}}} (\bibinfo {year} {1987}),\ \bibfield  {title}
  {\enquote {\bibinfo {title} {Strong communication complexity or generating
  quasirandom sequences form two communicating semi-random sources},}\ }\href
  {\doibase 10.1007/BF02579325} {\bibfield  {journal} {\bibinfo  {journal}
  {Combinatorica}\ }\textbf {\bibinfo {volume} {7}}~(\bibinfo {number} {4}),\
  \bibinfo {pages} {375--392}}\BibitemShut {NoStop}%
\bibitem [{\citenamefont {Vicente}(2014)}]{Vicente14}%
  \BibitemOpen
  \bibfield  {author} {\bibinfo {author} {\bibnamefont {Vicente}, \bibfnamefont
  {Julio I~de}}} (\bibinfo {year} {2014}),\ \bibfield  {title} {\enquote
  {\bibinfo {title} {On nonlocality as a resource theory and nonlocality
  measures},}\ }\href {http://stacks.iop.org/1751-8121/47/i=42/a=424017}
  {\bibfield  {journal} {\bibinfo  {journal} {Journal of Physics A:
  Mathematical and Theoretical}\ }\textbf {\bibinfo {volume} {47}}~(\bibinfo
  {number} {42}),\ \bibinfo {pages} {424017}}\BibitemShut {NoStop}%
\bibitem [{\citenamefont {Weinan}\ and\ \citenamefont
  {Vanden~Eijnden}(2001)}]{Vanden2}%
  \BibitemOpen
  \bibfield  {author} {\bibinfo {author} {\bibnamefont {Weinan}, \bibfnamefont
  {E}}, \ and\ \bibinfo {author} {\bibfnamefont {Eric}\ \bibnamefont
  {Vanden~Eijnden}}} (\bibinfo {year} {2001}),\ \bibfield  {title} {\enquote
  {\bibinfo {title} {{T}urbulent {P}randtl number effect on passive scalar
  advection},}\ }\href {\doibase https://doi.org/10.1016/S0167-2789(01)00196-8}
  {\bibfield  {journal} {\bibinfo  {journal} {Physica D: Nonlinear Phenomena}\
  }\textbf {\bibinfo {volume} {152–153}},\ \bibinfo {pages} {636 --
  645}}\BibitemShut {NoStop}%
\bibitem [{\citenamefont {Wheeler}\ and\ \citenamefont
  {Zurek}(1983)}]{WheelerZurek83}%
  \BibitemOpen
  \bibfield  {author} {\bibinfo {author} {\bibnamefont {Wheeler}, \bibfnamefont
  {John~Archibald}}, \ and\ \bibinfo {author} {\bibfnamefont {Wojciech~Hubert}\
  \bibnamefont {Zurek}}} (\bibinfo {year} {1983}),\ \href
  {http://www.jstor.org/stable/j.ctt7ztxn5} {\emph {\bibinfo {title} {Quantum
  Theory and Measurement}}}\ (\bibinfo  {publisher} {Princeton University
  Press})\BibitemShut {NoStop}%
\bibitem [{\citenamefont {Wilson}(2009)}]{Wilson09}%
  \BibitemOpen
  \bibfield  {author} {\bibinfo {author} {\bibnamefont {Wilson}, \bibfnamefont
  {Mark}}} (\bibinfo {year} {2009}),\ \bibfield  {title} {\enquote {\bibinfo
  {title} {{D}eterminism and the {M}ystery of the {M}issing {P}hysics},}\
  }\href {\doibase 10.1093/bjps/axn052} {\bibfield  {journal} {\bibinfo
  {journal} {The British {J}ournal for the Philosophy of Science}\ }\textbf
  {\bibinfo {volume} {60}}~(\bibinfo {number} {1}),\ \bibinfo {pages}
  {173--193}}\BibitemShut {NoStop}%
\bibitem [{\citenamefont {W\'odkiewicz}(1995)}]{Wodkiewicz95}%
  \BibitemOpen
  \bibfield  {author} {\bibinfo {author} {\bibnamefont {W\'odkiewicz},
  \bibfnamefont {Krzysztof}}} (\bibinfo {year} {1995}),\ \bibfield  {title}
  {\enquote {\bibinfo {title} {Classical and quantum {M}alus laws},}\ }\href
  {\doibase 10.1103/PhysRevA.51.2785} {\bibfield  {journal} {\bibinfo
  {journal} {Phys. Rev. A}\ }\textbf {\bibinfo {volume} {51}},\ \bibinfo
  {pages} {2785--2788}}\BibitemShut {NoStop}%
\bibitem [{\citenamefont {Wojew\'odka}\ \emph {et~al.}(2016)\citenamefont
  {Wojew\'odka}, \citenamefont {Brand\~ao}, \citenamefont {Grudka},
  \citenamefont {Horodecki}, \citenamefont {Horodecki}, \citenamefont
  {Horodecki}, \citenamefont {Pw{\l}owski},\ and\ \citenamefont
  {Ramanathan}}]{WBGHHHPR16}%
  \BibitemOpen
  \bibfield  {author} {\bibinfo {author} {\bibnamefont {Wojew\'odka},
  \bibfnamefont {H}}, \bibinfo {author} {\bibfnamefont {F.~G.~S.~L.}\
  \bibnamefont {Brand\~ao}}, \bibinfo {author} {\bibfnamefont {A.}~\bibnamefont
  {Grudka}}, \bibinfo {author} {\bibfnamefont {M.}~\bibnamefont {Horodecki}},
  \bibinfo {author} {\bibfnamefont {K.}~\bibnamefont {Horodecki}}, \bibinfo
  {author} {\bibnamefont {Horodecki}}, \bibinfo {author} {\bibfnamefont
  {M.}~\bibnamefont {Pw{\l}owski}}, \ and\ \bibinfo {author} {\bibfnamefont
  {R.}~\bibnamefont {Ramanathan}}} (\bibinfo {year} {2016}),\ \bibfield
  {title} {\enquote {\bibinfo {title} {{Amplifying the randomness of weak
  sources correlated with devices}},}\ }\href@noop {} {\bibfield  {journal}
  {\bibinfo  {journal} {ArXiv e-prints}\ }}\Eprint
  {http://arxiv.org/abs/1601.06455} {arXiv:1601.06455 [quant-ph]} \BibitemShut
  {NoStop}%
\bibitem [{\citenamefont {Wódkiewicz}(1985)}]{Wodkiewicz85}%
  \BibitemOpen
  \bibfield  {author} {\bibinfo {author} {\bibnamefont {Wódkiewicz},
  \bibfnamefont {K}}} (\bibinfo {year} {1985}),\ \bibfield  {title} {\enquote
  {\bibinfo {title} {Quantum {M}alu's law},}\ }\href {\doibase
  http://dx.doi.org/10.1016/0375-9601(85)90339-1} {\bibfield  {journal}
  {\bibinfo  {journal} {Physics Letters A}\ }\textbf {\bibinfo {volume}
  {112}}~(\bibinfo {number} {6}),\ \bibinfo {pages} {276--278}}\BibitemShut
  {NoStop}%
\bibitem [{\citenamefont {Xu}\ \emph {et~al.}(2012)\citenamefont {Xu},
  \citenamefont {Qi}, \citenamefont {Ma}, \citenamefont {Xu}, \citenamefont
  {Zheng},\ and\ \citenamefont {Lo}}]{XuOE2012}%
  \BibitemOpen
  \bibfield  {author} {\bibinfo {author} {\bibnamefont {Xu}, \bibfnamefont
  {Feihu}}, \bibinfo {author} {\bibfnamefont {Bing}\ \bibnamefont {Qi}},
  \bibinfo {author} {\bibfnamefont {Xiongfeng}\ \bibnamefont {Ma}}, \bibinfo
  {author} {\bibfnamefont {He}~\bibnamefont {Xu}}, \bibinfo {author}
  {\bibfnamefont {Haoxuan}\ \bibnamefont {Zheng}}, \ and\ \bibinfo {author}
  {\bibfnamefont {Hoi-Kwong}\ \bibnamefont {Lo}}} (\bibinfo {year} {2012}),\
  \bibfield  {title} {\enquote {\bibinfo {title} {Ultrafast quantum random
  number generation based on quantum phase fluctuations},}\ }\href {\doibase
  10.1364/OE.20.012366} {\bibfield  {journal} {\bibinfo  {journal} {Opt.
  Express}\ }\textbf {\bibinfo {volume} {20}}~(\bibinfo {number} {11}),\
  \bibinfo {pages} {12366--12377}}\BibitemShut {NoStop}%
\bibitem [{\citenamefont {Yuan}\ \emph {et~al.}(2014)\citenamefont {Yuan},
  \citenamefont {Lucamarini}, \citenamefont {Dynes}, \citenamefont
  {Fr{\"o}hlich}, \citenamefont {Plews},\ and\ \citenamefont
  {Shields}}]{YuanAPL2014}%
  \BibitemOpen
  \bibfield  {author} {\bibinfo {author} {\bibnamefont {Yuan}, \bibfnamefont
  {Z~L}}, \bibinfo {author} {\bibfnamefont {M.}~\bibnamefont {Lucamarini}},
  \bibinfo {author} {\bibfnamefont {J.~F.}\ \bibnamefont {Dynes}}, \bibinfo
  {author} {\bibfnamefont {B.}~\bibnamefont {Fr{\"o}hlich}}, \bibinfo {author}
  {\bibfnamefont {A.}~\bibnamefont {Plews}}, \ and\ \bibinfo {author}
  {\bibfnamefont {A.~J.}\ \bibnamefont {Shields}}} (\bibinfo {year} {2014}),\
  \bibfield  {title} {\enquote {\bibinfo {title} {Robust random number
  generation using steady-state emission of gain-switched laser diodes},}\
  }\href {\doibase http://dx.doi.org/10.1063/1.4886761} {\bibfield  {journal}
  {\bibinfo  {journal} {Applied Physics Letters}\ }\textbf {\bibinfo {volume}
  {104}}~(\bibinfo {number} {26}),\ \bibinfo {eid} {261112},\
  http://dx.doi.org/10.1063/1.4886761}\BibitemShut {NoStop}%
\bibitem [{\citenamefont {Zinkernagel}(2010)}]{Zinkernagel10}%
  \BibitemOpen
  \bibfield  {author} {\bibinfo {author} {\bibnamefont {Zinkernagel},
  \bibfnamefont {Henrik}}} (\bibinfo {year} {2010}),\ \enquote {\bibinfo
  {title} {{{EPSA} Philosophical Issues in the Sciences: Launch of the European
  Philosophy of Science Association}},}\ Chap.\ \bibinfo {chapter} {Causal
  Fundamentalism in Physics}\ (\bibinfo  {publisher} {Springer Netherlands},\
  \bibinfo {address} {Dordrecht})\ pp.\ \bibinfo {pages} {311--322}\BibitemShut
  {NoStop}%
\bibitem [{\citenamefont {Zurek}(2003)}]{Zurek03}%
  \BibitemOpen
  \bibfield  {author} {\bibinfo {author} {\bibnamefont {Zurek}, \bibfnamefont
  {Wojciech~Hubert}}} (\bibinfo {year} {2003}),\ \bibfield  {title} {\enquote
  {\bibinfo {title} {Decoherence, einselection, and the quantum origins of the
  classical},}\ }\href {\doibase 10.1103/RevModPhys.75.715} {\bibfield
  {journal} {\bibinfo  {journal} {Rev. Mod. Phys.}\ }\textbf {\bibinfo {volume}
  {75}},\ \bibinfo {pages} {715--775}}\BibitemShut {NoStop}%
\bibitem [{\citenamefont {Zurek}(2009)}]{Zurek09}%
  \BibitemOpen
  \bibfield  {author} {\bibinfo {author} {\bibnamefont {Zurek}, \bibfnamefont
  {Wojciech~Hubert}}} (\bibinfo {year} {2009}),\ \bibfield  {title} {\enquote
  {\bibinfo {title} {Quantum {D}arwinism},}\ }\href {\doibase
  10.1038/nphys1202} {\bibfield  {journal} {\bibinfo  {journal} {Nat Phys}\
  }\textbf {\bibinfo {volume} {5}},\ \bibinfo {pages} {181--188}}\BibitemShut
  {NoStop}%
\end{thebibliography}

%merlin.mbs apsrmp4-1.bst 2010-07-25 4.21a (PWD, AO, DPC) hacked
%Control: author (3) reversed first dotless

\end{document}